\documentclass[12pt]{article}
\usepackage{amssymb}
\usepackage{graphicx}
\usepackage{amsmath}
\def\ee{\end{eqnarray}}
\def\Lag{\mathcal{L}}
\newcommand{\nn}{\nonumber}
\def\p{\partial}
\def\ra{\rightarrow}
\def\D{\mathcal{D}}

\def\=:{=\hspace{-.7em}\raisebox{1.1ex}{.}\hspace{.1em}\raisebox{-0.2ex}{.} }
\newcommand{\NF}{N_{\rm F}}
\newcommand{\NC}{N_{\rm C}}
\newcommand{\hs}[1]{\hspace{#1 mm}}
\newcommand {\1}[1]{\frac{1}{#1}}

\makeatletter
\@addtoreset{equation}{section}

\renewcommand{\thefootnote}{\fnsymbol{footnote}}
\makeatother

\begin{document}
\thispagestyle{empty}
\begin{flushright}
TIT/HEP--524 \\
{\tt hep-th/0405194} \\
May, 2004 \\
\end{flushright}
\vspace{3mm}

\begin{center}
{\Large \bf 
Non-Abelian Walls in Supersymmetric Gauge Theories} 
\\[12mm]
\vspace{5mm}

\normalsize
  {\large \bf 
Youichi~Isozumi}
\footnote{\it  e-mail address: 
isozumi@th.phys.titech.ac.jp
}, 
  {\large \bf 
Muneto~Nitta}
\footnote{\it  e-mail address: 
nitta@th.phys.titech.ac.jp
}, 
  {\large \bf 
 Keisuke~Ohashi 
}\footnote{\it  e-mail address: 
keisuke@th.phys.titech.ac.jp
}, 
~and~~  {\large \bf 
Norisuke~Sakai}
\footnote{\it  e-mail address: 
nsakai@th.phys.titech.ac.jp
} 

\vskip 1.5em

{ \it Department of Physics, Tokyo Institute of 
Technology \\
Tokyo 152-8551, JAPAN  
 }
\vspace{15mm}

{\bf Abstract}\\[5mm]
{\parbox{13cm}{\hspace{5mm}
The Bogomol'nyi-Prasad-Sommerfield (BPS) 
multi-wall solutions are constructed in 
supersymmetric 
$U(N_{\rm C})$ gauge theories in five dimensions with 
$N_{\rm F}(>N_{\rm C})$ hypermultiplets in the fundamental representation. 
Exact solutions are obtained with full generic moduli 
for infinite gauge coupling and with partial moduli 
for finite gauge coupling. 
The generic wall solutions require nontrivial configurations 
for either gauge fields or 
off-diagonal components of adjoint scalars 
depending on the gauge. 
Effective theories of moduli fields are 
constructed 
as world-volume gauge theories. 
Nambu-Goldstone and 
quasi-Nambu-Goldstone scalars 
are distinguished and worked out.  
Total moduli space of the BPS non-Abelian walls 
including all topological sectors 
is found to be the complex Grassmann manifold 
$SU(N_{\rm F}) / 
[SU(N_{\rm C})\times SU(N_{\rm F}-{N}_{\rm C}) \times U(1)]$ 
endowed with a deformed metric. 
}}
\end{center}
\vfill
\newpage
\setcounter{page}{1}
\setcounter{footnote}{0}
\renewcommand{\thefootnote}{\arabic{footnote}}

\section{Introduction}\label{INTRO}

In constructing unified theories with extra 
dimensions~\cite{HoravaWitten}--\cite{RandallSundrum}, 
it is crucial to obtain topological defects and localization 
of massless or nearly massless modes on the defect. 
Walls in five-dimensional theories are 
the simplest of the topological defects leading to 
the four-dimensional world-volume. 
In constructing topological defects, supersymmetric (SUSY) 
theories are helpful, since partial preservation of 
SUSY automatically gives a solution of equations of 
motion~\cite{WittenOlive}. These states are called 
BPS states. 
The simplest of these BPS states is the 
wall~\cite{Cvetic:1991vp,AT}. 
The resulting theory tend to produce an ${\cal N}=1$ 
SUSY theory on the world volume, 
which can provide realistic unified models with the 
desirable properties~\cite{DGSW}. 
Although scalars and spinors 
can be obtained as localized modes on the 
wall~\cite{Rubakov}, 
it has been difficult to obtain localized massless gauge 
bosons in five dimensions, in spite of many 
interesting proposals, especially in lower 
dimensions~\cite{DvaliShifman}--\cite{SY2}. 
Recently a model of the localized massless gauge 
bosons on the wall has 
been obtained for Abelian gauge theories 
using SUSY QED interacting with 
tensor multiplets~\cite{IOS1,IOS2}. 
Walls in non-Abelian gauge theories are called non-Abelian 
walls. 
They are expected to help 
obtaining non-Abelian gauge bosons localized 
on the world volume. 
Moreover, non-Abelian wall solutions 
have rich structures and are interesting in its own right. 
BPS walls in a non-Abelian SUSY gauge theories 
have recently been studied in lower dimensions in a 
particular context~\cite{SY2}. 

The purpose of this paper is to construct BPS walls 
in five-dimensional non-Abelian gauge theories with 
eight supercharges and to obtain the effective theories of 
moduli on the four-dimensional world-volume. 
In particular we study the $U(N_{\rm C})$ gauge theory 
with $N_{\rm F}(>N_{\rm C})$ flavors of hypermultiplets in the 
fundamental representation. 
To obtain discrete vacua, we consider non-degenerate 
masses for hypermultiplets, and the Fayet-Iliopoulos 
(FI) parameter is introduced~\cite{ANS}. 
By taking the limit of infinite gauge coupling, 
we obtain exact BPS multi-wall solutions with generic 
moduli parameters covering the complete moduli space. 
For a restricted class of moduli parameters called 
$U(1)$-factorizable moduli, 
we also obtain exact BPS multi-wall solutions for 
certain values of finite gauge coupling. 
We find that the total moduli space 
is a compact complex manifold, the Grassmann manifold 
$G_{N_{\rm F},N_{\rm C}}\equiv {SU(N_{\rm F}) \over 
SU(N_{\rm C})\times SU(N_{\rm F}-N_{\rm C}) \times U(1)}$ 
as reported in \cite{INOS}. 
Each moduli parameter 
provides 
a massless field for the effective field theory 
on the world volume of walls. 
We find explicitly Nambu-Goldstone scalars associated 
with the spontaneously broken global symmetry. 
We also identify those massless scalars that are not 
explained by the spontaneously broken symmetry 
and are called quasi-Nambu-Goldstone scalars. 
We find it convenient to introduce a matrix function 
$S(y)$ as a function of the extra-dimensional coordinate 
and constant moduli matrices $H_0^i$ to describe 
the solution. 
The redundancy of the description is expressed as 
a global symmetry $GL(N_{\rm C}, {\bf C})$ 
of these data $(S, H_0^i)$. 
This symmetry turns out to be very useful and 
eventually be promoted to a local gauge symmetry 
when we consider effective theories 
on the world volume of walls.\footnote{
Our gauge symmetry on the world volume 
seems to be different from that obtained 
previously for effective theories of moduli 
fields using the brane constructions 
where the $U(k)$ gauge symmetry emerges 
for the $k$ solitons~\cite{Wi,brane-monopole,HT}.}
Therefore we call the symmetry the 
world-volume symmetry. 
We also obtain a general formula for the metric 
in moduli space which gives the effective theory 
of moduli fields on the world volume. 
The formula can be reduced to an explicit integral 
representation in the case of infinite gauge 
coupling. 
We also establish a duality between 
BPS wall solutions with $U(N_{\rm C})$ color 
and $N_{\rm F}$ flavor and those with 
$U(\tilde N_{\rm C}\equiv N_{\rm F}-N_{\rm C})$ color 
and $N_{\rm F}$ flavor.

Our solutions and their moduli space are unchanged 
under dimensional reduction to two, three and four 
space-time dimensions. 
In particular, in four space-time dimensions,  
there exists a long history for construction of  
BPS solitons and their moduli space 
in the gauge-Higgs system. 
A beautiful method for construction of 
instantons was given by 
Atiyah, Hitchin, Drinfeld and Manin (ADHM)~\cite{ADHM}.
It was modified by Nahm to the one for BPS monopoles~\cite{Nahm}. 
Recently the moduli space for non-Abelian vortices 
has been constructed by Hanany and Tong~\cite{HT}.
However, a systematic method for construction of 
walls in non-Abelian gauge theories 
has not been obtained although there exist 
some for walls in  Abelian gauge theories and/or 
nonlinear sigma models derivable from Abelian 
gauge theories~\cite{GTT2,To,To2,gravity2,IOS1}.  
Our method presents the last gap for 
the construction of solitons and moduli space 
in the gauge-Higgs system. 
Our wall moduli space as well as the moduli space of vortices 
are constructed by the K\"ahler quotient 
while moduli spaces of instantons and monopoles are constructed by 
the hyper-K\"ahler quotient. 
One interesting feature for  
non-Abelian walls may be that the total moduli space 
is finite dimensional in contrast to 
total moduli spaces for other solitons 
which are infinite dimensional.

Since we are interested in wall solutions with Poincar\'e 
invariance in the wall world-volume, only the 
extra-dimensional 
component $W_y$ may be nontrivial for the gauge field. 
One can always choose a gauge of the original local gauge symmetry to 
eliminate the extra-dimensional component $W_y$ of gauge field in the case 
of $U(1)$ gauge theories.
Therefore all the explicit wall solutions so far 
obtained have vanishing gauge field 
configurations~\cite{To,ShifmanYung,IOS1}. 
In the case of the non-Abelian gauge group, 
it is usually convenient to eliminate 
all the vector multiplet scalars 
for generators outside of the Cartan 
subalgebra ${\cal H}$, and all the gauge fields 
for generators in the Cartan 
subalgebra: 
$\Sigma^{I\not\in{\cal H}}=0$, 
$W_y^{I\in {\cal H}}=0$. 
We find that our BPS multi-wall solutions 
for generic moduli have nontrivial gauge field 
configurations: $W_y^{I\not\in {\cal H}}\not=0$. 
We will also give a gauge invariant description of 
these nontrivial vector multiplet configurations and evaluate 
these gauge invariant quantities for explicit examples.

The SUSY vacua in our $U(N_{\rm C})$ model 
are found to be the color-flavor 
locking form specified by the non-vanishing flavor 
$A_r$ for each color component $r$, such as 
$\langle A_1 \cdots A_{N_{\rm C}} \rangle$ 
abbreviated as $\langle A \rangle$. 
BPS multi-wall solutions interpolate between 
two SUSY vacua which are specified by 
boundary conditions: 
a SUSY vacuum 
$\langle A_1 \cdots A_{N_{\rm C}} \rangle$ 
at $y=\infty$ and 
another SUSY vacuum 
$\langle B_1 \cdots B_{N_{\rm C}} \rangle$ 
at $y=-\infty$. 
The boundary condition at $\pm \infty$ 
defines a topological sector denoted as 
$\langle A_1 \cdots A_{N_{\rm C}} \rangle 
\leftarrow 
\langle B_1 \cdots B_{N_{\rm C}} \rangle$. 
The total moduli space is defined by a sum 
over $k$ of the moduli spaces of $k$-walls 
${\cal M}^k_{N_{\rm F}, N_{\rm C}}$, 
but may also be expressed as a sum over the 
topological 
sectors 
${\cal M}^{\langle A\rangle \leftarrow \langle B\rangle}
_{N_{\rm F}, N_{\rm C}}$ defined by boundary conditions 
at $y=\pm \infty$: 
\begin{equation}
G_{N_{\rm F}, N_{\rm C}} =
\sum_k {\cal M}^k _{N_{\rm F}, N_{\rm C}}= 
\sum_{\langle A\rangle \leftarrow \langle B\rangle} 
{\cal M}^{\langle A\rangle \leftarrow \langle B\rangle}
_{N_{\rm F}, N_{\rm C}}. 
\label{eq:total-moduli-space}
\end{equation}
Among various BPS walls, there are 
walls interpolating between two vacua with identical 
labels except one label that have adjacent 
flavors: 
$\langle A_1 \cdots A_{N_{\rm C}} \rangle 
\leftarrow 
\langle B_1 \cdots B_{N_{\rm C}} \rangle$ 
with $A_j = B_j, \, j\not=i$, and $A_i + 1=B_i$. 
These walls are building blocks of 
multi-walls and are called elementary walls. 
We find that a quantum number $(A_i,A_i+
1)$ 
can be ascribed to the elementary wall with 
$A_i +
1=B_i$ and a matrix algebra can be formulated 
to describe the non-Abelian walls. 
Composite walls made of several elementary walls 
can be represented by a product of matrices 
corresponding to constituent elementary walls.
If the matrices do not commute, the commutator gives a 
single wall made by compressing the two walls. 
We call such a wall compressed wall. 
This is the situation for Abelian walls. 
On the other hand, we can also have commuting matrices for 
non-Abelian walls. 
If the matrices are commuting, 
the two elementary walls are called penetrable, 
since the intermediate 
vacuum changes character while the constituent walls 
go through each other maintaining their identities by changing from one sign of the relative position 
to the other sign. 

In Sec.~\ref{sc:model-vacua-BPSeq}, 
we introduce our model, 
work out SUSY vacua with a 
convenient diagrammatic representation, 
and obtain $1/2$ BPS equations. 
In Sec.~\ref{BPSWS}, 
exact solutions of the BPS equations 
are obtained both for infinite and for 
finite gauge couplings, 
by introducing moduli matrices and the world-volume 
symmetry. 
In Sec.~\ref{CEWSIC}, 
explicit solutions at 
infinite coupling are presented 
for a number of illustrative examples. 
In Sec.~\ref{MSFNAW}, 
the topology and metric of the moduli space 
of the non-Abelian BPS wall solutions are studied. 
In Sec.~\ref{sc:discussion}, we discuss the 
implications of our 
results and future directions of research. 
A number of useful details are 
described in several Appendices.

\section{The Model, SUSY Vacua and BPS Equations}
\label{sc:model-vacua-BPSeq}
\subsection{The Model}
\label{sc:model}
Since we are interested in theories in five dimensions, 
we need eight supercharges. 
With this minimum number of supersymmetry (SUSY), 
simple building blocks are 
vector multiplets and hypermultiplets. 
Wall solutions require discrete vacua, which can be 
obtained by considering $U(1)$ factors besides 
semi-simple gauge group~\cite{ANS}. 
We denote the gauge group suffix 
and flavor group suffix in our fundamental theory 
by the uppercase letters G and F, respectively. 
The $U(1)_{\rm G}$ vector multiplet with coupling 
constant $g_0$ 
consists of a $U(1)_{\rm G}$ gauge field $W_M^0$, a real 
scalar field $\Sigma^0$, a $SU(2)_R$ triplet of real 
auxiliary field $Y^{a0}$, and  
an $SU(2)_R$ doublet of gauginos $\lambda^{i0}$. 
We denote space-time indices by 
$M,N, \cdots=0,1,2,3,4$, and $SU(2)_R$ triplet, 
doublet indices by $a,i$ respectively. 
The $U(1)_{\rm G}$ part of vector multiplets allows 
us to introduce the FI term which gives rise to 
discrete vacua once mass terms for hypermultiplets 
are introduced~\cite{ANS}. 

We also have a non-Abelian vector multiplet 
for a semi-simple gauge group $G$ with coupling 
constant $g$. 
It consists of a gauge field $W_M$, a scalar $\Sigma$, 
auxiliary fields $Y^{a}$, and gauginos $\lambda^{i}$, 
which are now in the adjoint representation of $G$. 
We use a matrix notation for 
these component fields, such as $\Sigma=\Sigma^I T_I$. 
We denote the Hermitian generators in the 
Lie algebra ${\cal G}$ of the gauge group $G$ 
as $T^I\in {\cal G} \ (I = 1,2,\cdots, {\rm dim}(G))$, 
which satisfy the following normalization condition 
and commutation relation
\begin{eqnarray}
{\rm Tr}(T_I T_J) 
= T({\cal R})\delta_{IJ}, \ \ \ [ T_I , T_J]
=i f_{IJ}{}^{K} T_K,\label{norm-comm}
\end{eqnarray}
where $f_{IJ}{}^{K}$ are the structure constants 
of the gauge group $G$, 
and $T({\cal R})$ is the normalization constant 
for the representation ${\cal R}$. 
Furthermore, we denote 
the generators in the Cartan subalgebra 
${\cal H}$ of ${\cal G}$ 
by a suffix $x$ as 
$T^x\in {\cal H}$. 
For later convenience, we denote the generator of 
the $U(1)$ factor group as $T^0$ with the same 
normalization 
as the non-Abelian group generators (\ref{norm-comm}). 
Moreover, we collectively denote 
generators as 
$T^I$ with $I$ running over $I=0$ for the $U(1)$ 
and $I=1, \cdots, 
{\rm dim}({\cal G})$ for the non-Abelian group. 
We also denote 
gauge couplings 
as $g_I$ ($I=0, 1, \cdots, 
{\rm dim}({\cal G})$), with $g_I \equiv g$ for 
$I=1, \cdots, {\rm dim}({\cal G})$. 
Similarly we also combine the $U(1)$ generator 
with those in the Cartan subalgebra to denote diagonal generators: 
$T^x$ with $x=0,1, \cdots, {\rm dim}({\cal H})$.

We have hypermultiplets as matter fields, 
consisting of $SU(2)_R$ doublet of complex scalar 
quark fields $H^{irA}$, 
$SU(2)_R$ doublet of auxiliary fields $F_i^{rA}$, 
and Dirac fields $\psi^{rA}$. 
Color indices $r,s,\cdots$ run over 
$1,2,\cdots, {\cal R}$ where ${\cal R}$ denotes 
the dimension of the representation of the hypermultiplet, 
whereas $A,B, \cdots =1,2,\cdots, N_{\rm F}$ stand 
for flavor indices. 
We consider $N_{\rm F}>N_{\rm C}$ to obtain 
disconnected SUSY vacua appropriate for constructing walls. 

We shall consider a model with minimal 
kinetic terms for vector and hypermultiplets. 
The eight supercharges 
allow only a few parameters in our model: 
gauge coupling constants $g_0$  for $U(1)_{\rm G}$, 
and $g$ for the non-Abelian semi-simple gauge group $G$, 
the masses of $A$-th hypermultiplet $m_A$, and 
the FI parameters $\zeta^a$ for the $U(1)_{\rm G}$ 
vector multiplet. 
Then the bosonic part of our Lagrangian reads 
\begin{eqnarray}
\Lag_{\rm bosonic} 
&\!\!\!=&\!\!\! 
-\sum_{I\ge 0}\frac{1}{4g_I^2}
F_{MN}^I(W)F^{IMN}(W) \nn\\
&\!\!\!&\!\!\!
+\sum_{I\ge 0}\frac{1}{2g_I^2}
\D_M \Sigma^I  \D^M \Sigma^I\nn\\  
&\!\!\!&\!\!\!
{}- \zeta^a Y^{a0}
+\sum_{I\ge 0}\frac{1}{2g_I^2}(Y^{aI})^2\nn\\
&\!\!\!&\!\!\!
{}+ (\D_M H^{irA})^* \D^M H^{irA} 
-(H^{irA})^*[(
\Sigma-m_A )^2]^r{}_s H^{isA}\nn\\
&\!\!\!&\!\!\!
{}+(H^{irA})^*  (\sigma^a)^i{}_j 
(Y^a)^r{}_s H^{jsA}
+(F_i^{rA})^* F_i^{rA},
\label{fundamental-Lag}
\end{eqnarray}
where the summation over group indices $I$ is explicitly 
denoted. 
In the following, however, we will suppress 
the summation with the understanding 
that the sum over repeated indices $I$ should be done 
including $I=0$, unless stated otherwise. 
Summation over repeated indices is also implied for 
other indices. 
The covariant derivatives are defined as 
$\D_M H^{irA}=(\p_M \delta_s^r + i(W_M^I)^r{}_s)H^{isA} 
\ (I=0,1,\cdots, {\rm dim}(G))$, 
$\D_M \Sigma = \p_M \Sigma + i[ W_M , \Sigma ]$, 
and field strength is defined as 
$F_{MN}=\frac{1}{i}[\D_M , \D_N]
=\p_M W_N -\p_N W_M + i[W_M, W_N]$ 
and our convention of metric is 
$\eta_{MN}={\rm diag}(+1,-1,-1,-1,-1)$. 

In this paper, we assume non-degenerate 
mass parameters $m_A$ unless stated otherwise. 
Then the flavor symmetry reduces to 
\begin{eqnarray}
 G_{\rm F} = U(1)_{\rm F}^{N_{\rm F}-1},
 \label{break-flavor}
\end{eqnarray} 
where 
$U(1)_{\rm F}$ corresponding to common phase is
gauged by $U(1)_{\rm G}$ local gauge symmetry. 
We choose the order of the mass parameters 
as $m_A > m_{A+1}$ for all $A$.

\subsection{SUSY Vacua and its Diagrammatic Representation}
\label{sc:SusyVac}
SUSY vacua can be obtained by requiring vanishing vacuum 
energy. 
Let us first write down equations of motion 
for auxiliary fields 
\begin{eqnarray}
Y^{a0}&\!\!\!=&\!\!
g_0^2 [ \zeta^a - (H^{irA})^* 
(\sigma^a)^i{}_j(T_0)^r{}_s H^{jsA}  ],
\label{EOM-aux1} \\
Y^{aI}&\!\!\!=&\!\!\!
-g^2 (H^{irA})^* (\sigma^a)^i{}_j (T_I)^r{}_s H^{jsA}, \ \ (I\not = 0),
\label{EOM-aux2} \\
F_i^{rA}&\!\!\!=&\!\!\!0.
\label{EOM-aux3}
\end{eqnarray}
After eliminating auxiliary fields, we obtain 
the on-shell version of the bosonic part of the 
Lagrangian 
\begin{eqnarray}
\Lag_{\rm bosonic} 
&\!\!\!=&\!\!\! 
-\frac{1}{4g_I^2}
F_{MN}^I(W)F^{IMN}(W)
{}+\frac{1}{2g_I^2}
\D_M \Sigma^I \D^M \Sigma^I 
\nn\\
&\!\!\!&\!\!\!
{}+ (\D_M H^{irA})^*  \D^M H^{irA} -V,
\label{fundamental-Lag2}
\end{eqnarray}
where the scalar potential $V$ is given by
\begin{eqnarray}
V&\!\!\!=&\!\!\! 
\frac{1}{2g_I^2}(Y^{aI})^2
+(F_i^{rA})^* F_i^{rA} 
+ (H^{irA})^* [(
\Sigma -m_A )^2]^r{}_s H^{isA}
\nn\\
&\!\!\!=&\!\!\!
\frac{g_I^2}{2}[ \zeta^a\delta_{0I} - (H^{irA})^* 
(\sigma^a)^i{}_j  (T^I)^r{}_s H^{jsA} ]^2
\nn\\&\!\!\!&\!\!\!\hspace{2cm} 
{}+(H^{irA})^* [(
\Sigma -m_A)^2]^r{}_s H^{isA}.
\end{eqnarray}
The vanishing vacuum energy requires 
both contributions from vector and hypermultiplets 
to vanish. 
Conditions of vanishing contribution from vector 
multiplet can be summarized to one equation as
\begin{eqnarray}
(H^{irA})^* (\sigma^a)^i{}_j (T_I)^r{}_s H^{jrA} 
=\zeta^a \delta_{0I}, \  (I=0,1,\cdots ,{\rm dim}(G) ).\quad 
\label{eq:susy-cod}
\end{eqnarray}
The $SU(2)_R$ symmetry allows us to 
choose the FI parameters to lie in the third direction 
without loss of generality 
\begin{eqnarray}
\zeta^a = (0, \ 0, \ \zeta), \ \ \ \zeta>0. 
\label{eq:FIparameter}
\end{eqnarray}
Then the SUSY condition (\ref{eq:susy-cod}) 
for the vector multiplets is reduced to 
\begin{eqnarray}
a=3:&&\!\! \ (H^{1rA})^*
(T_I)^r{}_s H^{1sA} - (H^{2rA})^* 
(T_I)^r{}_sH^{2sA}=\zeta \delta_{0I},\qquad 
\label{D-term-cond++1}\\
a=1,2:&&\!\! \ (H^{1rA})^*
(T_I)^r{}_s H^{2sA}=0,
\label{D-term-cond++2}
\end{eqnarray}
where $r,s=1,2,\cdots, {\cal R}$, $A=1,2,\cdots, N_{\rm F}$ 
and $I=0,1,2,\cdots, {\rm dim}(G)$. 
Requiring the vanishing contribution to vacuum energy 
from hypermultiplets 
gives the SUSY condition for hypermultiplets as 
\begin{eqnarray}
(\Sigma^0 T_0 + \sum_{I\ge 1}^{{\rm dim}(G)}\Sigma^I 
T_I -m_A \mathbf{1}_{{\cal R}
})^r{}_s H^{isA}=0 , \  
\label{susy-cond-H}
\end{eqnarray}
for each index $A$.
By local gauge transformations of $G$, we can always 
choose $\Sigma^{I\not\in{\cal H}}=0$. 
To parametrize the remaining vector multiplet scalars 
belonging to the Cartan subalgebra ${\cal H}$, 
we introduce orthogonal matrices $T_x$ as 
\begin{eqnarray}
(T_x)^r{}_s&\equiv &\delta ^r{}_sn_{x r}({\cal R}), 
\quad {\rm for~} x=0,1,2,\dots, {\rm dim}({\cal H}),\qquad 
\end{eqnarray}
where $n_{xr}$ is the $U(1)_x$-charge of the scalar 
carrying the color index $r$, 
and note that $n_{xr}$ have the following properties 
due to the traceless condition of $T_x$ 
and the normalization (\ref{norm-comm}) 
\begin{eqnarray}
&&n_{0 r}({\cal R})=\sqrt{ T ( \cal R ) \over \cal R}, 
\quad \sum_{r=1}^{\cal R}n_{x(\not=0) r}({\cal R})=0,\quad 
\nn\\
&&\sum_{r=1}^{\cal R}n_{x r}({\cal R})n_{y r}({\cal R})
=T({\cal R })\,\delta _{x y}. 
\label{property-nvectors}
\end{eqnarray}  
Rewriting the condition (\ref{susy-cond-H}) with $n_{x r}({\cal R})$, 
we obtain
\begin{eqnarray}
&&\left\{ \left({\Sigma^0\over \sqrt{2\cal R}} - m_A\right)
 + \sum_{x=1}^{{\rm dim}({\cal H})}
\Sigma^x n_{x r}({\cal R}) \right\}
H^{irA}\nn\\
&&=
\left\{\sum_{x=0}^{{\rm dim}
({\cal H})}\Sigma^x n_{x r}({\cal R})-m_A \right\} H^{irA}
=0 \label{F-term-cond++}
\end{eqnarray}
for each index $A$. 
In order to have a non-vanishing hypermultiplet scalar 
$H^{irA}$ 
with the color $r$ and the flavor $A$, 
we need to require the corresponding coefficient 
in Eq.~(\ref{F-term-cond++}) to vanish: 
\begin{eqnarray}
(\Sigma )^r{}_r=\sum_{x=0}^{{\rm dim}
({\cal H})}\Sigma^x n_{x r}({\cal R})=m_A .
\label{Sig-cond}
\end{eqnarray}
Let us consider a $({\rm dim}({\cal H})+1)$-dimensional 
space $\vec \Sigma \equiv (\Sigma ^0,\Sigma ^{x=1},\cdots,
\Sigma^{x={\rm dim}({\cal H})})$ 
of the vector multiplet scalars. 
The condition (\ref{Sig-cond}) implies that 
the region in 
$\vec \Sigma$ for a non-vanishing hypermultiplet scalar 
$H^{irA}$ should be contained in a 
${\rm dim}({\cal H})$-dimensional hyperplane, 
which contains 
a point $\vec \Sigma =(\sqrt{2\cal R}m_A,0,\cdots,0)$ 
and is orthogonal to the vector 
$\vec n_r({\cal R})$ with component 
$(\vec{n}_r)_x \equiv n_{xr}$. 
Obviously, two scalars $H^{irA}$ and $H^{irB}$ 
with the same color index $r$ can be non-vanishing 
only if $m_A=m_B$. 
Vacua with the $n$ non-vanishing 
scalars should lie in the 
$({\rm dim}({\cal H})+1-n)$-dimensional hyperplane 
in $\vec \Sigma $. 
These hyperplanes can easily be visualized 
diagrammatically in 
$\vec \Sigma $ space. 
These diagrams are quite useful to 
understand the structure of the vacua intuitively, 
and to construct the 
domain walls interpolating between these vacua 
as we see below. 

We shall discuss mainly the cases where there are 
non-vanishing scalars carrying flavor $A_r$ 
for the $r$-th color component, ($H^{irA_r}\not=0$). 
In these cases, Eqs.~(\ref{Sig-cond}) and 
(\ref{property-nvectors}) determine 
the scalar $\Sigma^x$ in terms of 
$n_{x r}(\cal R)$ as 
\begin{eqnarray}
\Sigma ^x =\frac{1}{T({\cal R})}\sum_{r=1}^{\cal R}n_{x r}({\cal R})\,m_{A_r}.
\end{eqnarray}
In particular 
$\Sigma ^0$ is given by 
an average value of the mass 
parameters,
\begin{eqnarray}
 \Sigma ^0=\frac{1}{\sqrt{T({\cal R}) {\cal R}}}\sum_{r=1}^{\cal R}\,m_{A_r},
\end{eqnarray}
which is independent of gauge-choices.

\subsection{SUSY Vacua for $U(N_{\rm C})$ Gauge Group with $N_{\rm F}$ 
Flavors
}
\label{sc:vacua-UN}

The procedure to solve the SUSY conditions 
(\ref{D-term-cond++1}) and (\ref{D-term-cond++2}) 
for the vector multiplets 
depends on details of the system. 
In this paper, we mostly consider 
a simple example of the 
$U(1)\times G=U(N_{\rm C})$ gauge group, 
and $N_{\rm F}$ hypermultiplets in the fundamental 
representation of $U(N_{\rm C})$, for which 
we choose $T({\cal R})=1/2$. 
We assume  non-degenerate mass parameters\footnote{
Almost all of our discussions are also 
applicable to the degenerate mass case apart from 
some subtleties associated with global symmetry 
which we hope to return in other publications. }
 : 
with the ordering $m_A > m_{A+1}$ for all $A$ 
as was mentioned below Eq.~(\ref{break-flavor}).

\begin{figure}[thb]
\begin{center}
\includegraphics[width=7cm,clip]{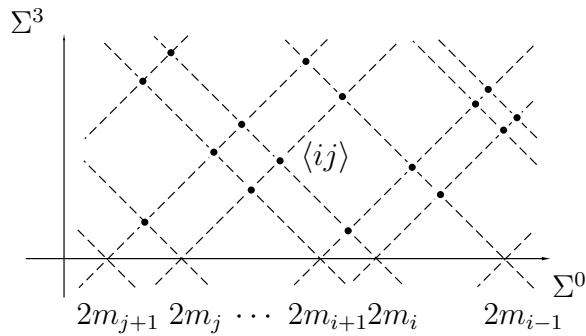}
\put(-205,100){$\Sigma^3$}
\put(0,0){$\Sigma^0$}
\put(-95,50){$\langle ij\rangle$}
\put(-180,-10){$2m_{j+1}$}
\put(-145,-10){$2m_j$}
\put(-120,-10){$\cdots$}
\put(-100,-10){$2m_{i+1}$}
\put(-70,-10){$2m_{i}$}
\put(-26,-10){$2m_{i-1}$}
\end{center}
\caption{
Diagrammatic representation of vacua for $N_{\rm C}=2$ case. 
Dashed lines are defined by $\Sigma^0+\Sigma^3 =2m_A$
and $\Sigma^0 - \Sigma^3 =2m_A$. 
Vacua are given as intersection 
points of these lines except for $\Sigma ^3=0$, 
because of Eq.~(\ref{D-term-cond-b}). 
}
\label{su2nf}
\end{figure}

It is convenient to combine the $N_{\rm F}$ 
hypermultiplets in the fundamental representation 
into the following $N_{\rm C}\times N_{\rm F}$ matrix 
\begin{eqnarray}
H^i\equiv 
\left( 
\begin{array}{cccc}
H^{i11}&H^{i12} &\cdots &H^{i1N_{\rm F}} \\
H^{i21}&H^{i22}& \cdots & H^{i2N_{\rm F}} \\
\vdots&\vdots &\ddots & \vdots\\
H^{iN_{\rm C} 1}&H^{iN_{\rm C} 2}& \cdots 
& H^{iN_{\rm C} N_{\rm F}} 
\end{array}
\right) 
.
\end{eqnarray}
In the following, we will denote this matrix as $H^i$, 
while its $r A$ components are denoted as 
$H^{irA}$. 
We also use $N_{\rm F} \times N_{\rm C}$ matrix $H^i{}^\dagger$ 
whose components are $(H^i{}^\dagger)_{Ar} \equiv (H^{irA})^*$.  
The SUSY condition (\ref{D-term-cond++1}) for 
vector multiplets 
can be rewritten in terms of this matrix as 
\begin{eqnarray} 
H^{1}  H^{1\dagger}  - H^{2} H^{2\dagger} 
=2\zeta T_0 =c\mathbf{1}_{N_{\rm C}},
\label{D-term-cond-a}
\end{eqnarray}
where we rescaled the FI parameter $\zeta$ to define $c$ 
\begin{equation}
c\equiv \zeta \sqrt{2/N_{\rm C}}. 
\end{equation}
Another SUSY condition for vector multiplets, 
(\ref{D-term-cond++2}) becomes 
\begin{eqnarray}
 H^2 H^{1\dagger}= 0.
\label{D-term-cond-b}
\end{eqnarray}
Since we assume non-degenerate masses for hypermultiplets, 
we find from the conditions 
(\ref{F-term-cond++}), (\ref{D-term-cond-a}) and 
(\ref{D-term-cond-b}) 
that only one flavor $A=A_r$ 
can be non-vanishing 
for each color component $r$ of hypermultiplet scalars 
$H^{irA}$ 
with 
\begin{eqnarray}
 H^{1rA}=\sqrt{c}\,\delta ^{A_r}{}_A,\quad H^{2rA}=0,
 \label{eq:hyper-vacuum}
\end{eqnarray}
since $c=\zeta \sqrt{2/N_{\rm C}}>0$ as defined 
in Eq.~(\ref{eq:FIparameter}). 
Here we used global gauge transformations 
to eliminate possible phase factors. 
This is often called the color-flavor locking vacuum. 
The vector multiplet scalars $\Sigma^x$ is 
determined in $\vec \Sigma$ as intersection 
points 
of $N_{\rm C}$ hyperplanes defined by (\ref{Sig-cond}), 
as illustrated in Fig.~\ref{su2nf}
\begin{eqnarray}
{1\over \sqrt{2N_{\rm C}}}\Sigma ^0
+{1\over 2}\Sigma ^3+{1\over \sqrt{3}}\Sigma ^8+\cdots
&=&m_{A_1}\nn,\\
{1\over \sqrt{2N_{\rm C}}}\Sigma ^0
-{1\over 2}\Sigma ^3+{1\over \sqrt{3}}\Sigma ^8+\cdots
&=&m_{A_2}\nn,\\
{1\over \sqrt{2N_{\rm C}}}\Sigma ^0\quad \qquad 
-{2\over \sqrt{3}}\Sigma ^8+\cdots&=&m_{A_3},\nn\\
&\vdots& . \label{NC-sheets}
\end{eqnarray}
These discrete vacua are equivalently expressed 
in the matrix notation as 
\begin{eqnarray}
\Sigma ={\rm diag.}(m_{A_1},\,m_{A_2},\,\cdots,\,
m_{A_{N_{\rm C}}}).
\end{eqnarray}
We denote a SUSY vacuum specified by a set of 
non-vanishing hypermultiplet 
scalars with the flavor $\{A_r\}$ for each color 
component $r$ as 
\begin{eqnarray}
 \langle A_1\,A_2\,\cdots\,A_{N_{\rm C}}\rangle .
\end{eqnarray}    
Since global gauge transformations can exchange flavors 
$A_r$ and $A_s$ for the color component $r$ and $s$, 
respectively, 
the ordering of the flavors $A_1, \cdots, A_{N_{\rm C}}$ 
does not matter in considering only vacua: 
$\langle 123 \rangle
=\langle 213 \rangle
$. 
Thus a number of SUSY 
vacua is given by~\cite{ANS} 
\begin{eqnarray}
 {}_{N_{\rm F}}C_{N_{\rm C}} 
 = {N_{\rm F}! \over N_{\rm C}!  (N_{\rm F} - N_{\rm C})!} 
   \label{vacnumber}
\end{eqnarray} 
and 
we usually take $A_1<A_2<\cdots<A_{N_{\rm C}}$. 

Walls interpolate between two vacua at $y=\infty$ 
and $y=-\infty$. 
These boundary conditions at $y=\pm \infty$ 
define topological sectors, such as 
$\langle 1 2 \rangle \leftarrow \langle 3 4 \rangle$. 
(Multi-)walls are classified by the topological 
sectors. 
Clearly  $\langle 1 2 \rangle 
\leftarrow \langle 3 4 \rangle$ is 
identical to 
$\langle 12 \rangle \leftarrow \langle 43 \rangle$. 

When we consider walls, however, it is often 
convenient to fix a gauge in presenting solutions. 
The gauge transformations allow us to eliminate 
all the vector multiplet scalars 
$\Sigma^{I\not\in{\cal H}}$ 
for generators outside of the Cartan 
subalgebra ${\cal H}$, and all the gauge fields 
$W_y^{I\in {\cal H}}$ in  the Cartan 
subalgebra ${\cal H}$. 
In this gauge, gauge fields $W_y^{I\not\in{\cal H}}$ 
can no longer be 
eliminated, since gauge is completely fixed. 
We shall usually use this gauge 
\begin{equation}
\Sigma^{I\not\in{\cal H}}=0, \qquad 
W_y^{I\in {\cal H}}=0 
\end{equation}
in this paper unless 
otherwise stated. 
If we wish, we can choose another gauge where 
the extra dimension component $W_y^I$ 
of the gauge field vanishes for all the generators. 
Then all components of 
vector multiplet scalars $\Sigma^I$ 
including those out of Cartan subalgebra become 
nontrivial. 
In that gauge, our BPS multi-wall solutions are 
expressed by nontrivial vector multiplet scalars 
$\Sigma^I$ for all the generators, instead of gauge 
fields $W_y^I$. 

When gauge is fixed in any one of these gauge choices, 
the ordering of flavors have physical 
significance, since changing one side of 
the boundary condition ($y=+\infty$) while 
keeping the other side ($y=-\infty$) 
requires local gauge transformations which will no longer 
be allowed. 
For example, we denote the wall connecting two vacua 
labeled by $\langle12\rangle$ 
at $y=+\infty$ and $\langle34\rangle$ at $y=-\infty$ 
as $\langle 12 \leftarrow 34 \rangle$ in the gauge 
fixed representation. 
The wall connecting vacua 
$\langle12\rangle$ and $\langle34\rangle$ 
is different from 
$\langle12\rangle$ and $\langle43\rangle$ 
in the gauge-fixed representation : 
$\langle 12 \leftarrow 34 \rangle\not=
\langle 12 \leftarrow 43 \rangle$. 

\subsection{Half BPS Equations for Domain Walls}\label{BPSEqDW}
We assume $g_0=g$ in the following. 
Let us obtain the BPS equations for domain walls 
interpolating between two SUSY vacua. 
The SUSY transformation laws of fermions of vector 
multiplets and hypermultiplets are given by\footnote{
In this paper, our gamma matrices are defined as 
$\{ \gamma^M, \gamma^N \}=2\eta^{MN}$, 
$\gamma^{MN}\equiv \frac{1}{2}[\gamma^M, \gamma^N]$,
$\gamma^5\equiv i\gamma^0\gamma^1\gamma^2\gamma^3 =-i\gamma^4$,
$\epsilon ^{12}=\epsilon _{12}=1$, 
and $P_\pm = \frac{1\pm\gamma^5}{2}$.
} 
\begin{eqnarray}
\delta_\varepsilon \lambda^i 
&\!\!\!=\!\!\!& \Bigl\{ \frac{1}{2}\gamma^{MN}F_{MN}(W) 
+ \gamma^M {\cal D}_M \Sigma \Bigl\} \varepsilon^i 
+ i (Y^a \sigma^a)^i{}_j \varepsilon^j,\quad \ \
\label{sutra-lambda}\\
\delta_\varepsilon \psi 
&\!\!\!=\!\!\!& \sqrt{2}\Bigl\{ -i\gamma^M {\cal D}_M H^{i} 
+\!  \Sigma H^{i}-H^{i}M \Bigl\} \epsilon_{ij}\varepsilon^j 
+\! \sqrt{2}F_i \varepsilon^i,\quad \ \
\label{sutra-psi}
\end{eqnarray}
where we use $N_{\rm F}\times N_{\rm F}$ Hermitian 
mass matrix $M$ defined by  
\begin{equation}
(M)^A{}_B\equiv m_A\delta ^A{}_B
.
\end{equation}
To obtain wall solutions, we assume that 
all fields depend only on the coordinate of one extra 
dimension $x^4$ which we denote as $y$. 
We also assume the Poincar\'e invariance 
on the four-dimensional world volume of the wall, 
implying 
\begin{eqnarray}
F_{MN}(W) =0, \ \ W_\mu =0,
\label{eq:gauge-field-wall}
\end{eqnarray}
where we take $x^\mu=(x^0, x^1,x^2,x^3)$ as 
four-dimensional world-volume coordinates. 
Note that $W_y$ need not vanish. 
We demand that half of supercharges defined by 
\begin{eqnarray}
P_+ \varepsilon^1 =0,  \ P_- \varepsilon^2 =0, 
\quad (\gamma ^4\varepsilon ^i
=-i(\sigma ^3)^i{}_j\varepsilon ^j),
\label{eq:halfBPS-cond}
\end{eqnarray}
to be conserved~\cite{IOS1}.  
By using these wall ansatz (\ref{eq:gauge-field-wall}) 
and unbroken supercharges (\ref{eq:halfBPS-cond}), 
the transformation laws (\ref{sutra-lambda}), 
(\ref{sutra-psi}) 
on the background of the 1/2 BPS state reduce to 
\begin{eqnarray}
\delta_\varepsilon \lambda^i \Big|_{\rm background} &=&
-i(\sigma ^3)^i{}_j\left(\D_y \Sigma -Y^3\right)\varepsilon^j 
+ i (Y^1 \sigma^1+Y^2\sigma ^2)^i{}_j \varepsilon^j,
\label{eq:SUSYtrans-half1} 
\\
\delta_\varepsilon \psi\Big|_{\rm background} &=& 
\sqrt{2}\Bigl\{ (\sigma _3)^i{}_k \D_y H^{k} 
+ \Sigma H^{i}-H^{i}M \Bigl\} \epsilon_{ij}\varepsilon^j 
+ \sqrt{2}F_i \varepsilon^i.
\label{eq:SUSYtrans-half2} 
\end{eqnarray} 
Preservation of the half of supercharges requires these 
transformations (\ref{eq:SUSYtrans-half1})
and (\ref{eq:SUSYtrans-half2})
to vanish. 
Using Eqs.~(\ref{EOM-aux1})-(\ref{EOM-aux3}), 
we find the following BPS equations for domain walls 
in the matrix-notation 
\begin{eqnarray}
\D_y \Sigma &=& Y^3 
={g^2\over 2}\left(c{\bf 1}_{N_{\rm C}}-H^1H^1{}^\dagger 
+H^2H^2{}^\dagger \right),
\label{BPSeq-Sigma}\\
0&=&Y^1+iY^2= - g^2 H^2H^1{}^\dagger ,\label{BPSeq-Y12}\\
\D_y H^1 &=& -\Sigma H^1 + H^1 M,\nonumber\\ 
\D_y H^2 &=& \Sigma H^2 -H^2 M.\label{BPSeq-H}
\end{eqnarray}

The Bogomol'nyi completion of the energy density of our 
system can be performed as 
\begin{eqnarray}
{\cal E}
&=&\frac{1}{g^2}{\rm Tr}(\D_y \Sigma -Y^3)^2
+\frac{1}{g^2} {\rm Tr}[(Y^1)^2 + (Y^2)^2 ]
\nn\\
&&+ {\rm Tr}|\D_y H^1 + \Sigma H^1 -H^1M|^2
\nn \\ && {}
+ {\rm Tr}|\D_y H^2 - \Sigma H^2 +H^2M|^2\nn\\
&&{}+ c \p_y{\rm Tr}\Sigma
- \p_y \left\{{\rm Tr}
\left[
\left(\Sigma H^1 - H^1 M\right) H^1{}^\dagger \right.\right.\nn\\ 
&& \hspace{8em} {}\left.\left.
+ \left(-\Sigma H^2 
+H^2 M\right)H^2{}^\dagger\right]\right\}.
\label{eq:bogomolnyi}
\end{eqnarray}
Let us consider a configuration approaching to 
a SUSY vacuum labeled by 
$\langle A_1A_2\cdots A_{N_{\rm C}}\rangle $ 
at the boundary 
of positive infinity $y=+\infty$, and to 
a vacuum 
$\langle B_1B_2\cdots B_{N_{\rm C}}\rangle $ 
at the boundary 
of negative infinity $y=-\infty$. 
If the SUSY condition $\Sigma H^i -H^iM =0$ is satisfied 
at $y=\pm \infty$, 
the second term of the last line of Eq.~(\ref{eq:bogomolnyi}) 
vanishes. 
Therefore the minimum energy is achieved by 
the configuration satisfying the BPS 
Eqs.~(\ref{BPSeq-Sigma})-(\ref{BPSeq-H}), 
and the energy (per unit world-volume of the wall) 
for the BPS saturated configuration 
is given by 
\begin{eqnarray}
T_{\rm w}&\!\!=\!\!&\!\!\int^{+\infty}_{-\infty}\hspace{-1.5em}dy{\cal E}
=c 
\left[{\rm Tr}\Sigma \right]^{+\infty}_{-\infty}
\!\!=\!c \left(\sum_{k=1}^{N_{\rm C}}m_{A_k}
-\sum_{k=1}^{N_{\rm C}}m_{B_k}\right).\qquad 
\label{eq:tension}
\end{eqnarray}
If a wall connects two SUSY vacua 
with identical labels except for 
a single label which are adjacent, 
such as 
$\langle B_1,\cdots,B_k,A-1,B_{k+2},\cdots,B_{N_{\rm C}} 
\leftarrow  B_1,\cdots,B_k,A,B_{k+2},\cdots,B_{N_{\rm C}} 
\rangle$, 
its tension is given by 
\begin{eqnarray}
T_{\langle B_1,\cdots,B_k,A-1,B_{k+2},\cdots,B_{N_{\rm C}} 
\leftarrow  B_1,\cdots,B_k,A,B_{k+2},\cdots,B_{N_{\rm C}} 
\rangle}
=c \,(m_{A-1}-m_{A})>0.
\end{eqnarray}
Since the tension depends only on the two labels 
which are different in the two vacua, we denote 
it as $T_{\langle A-1 \leftarrow A \rangle }$. 
Note that the tension $T_{\rm w}$ 
of general BPS multi-walls 
can be expressed by a sum of 
these minimal units. 
In this sense, these walls can be thought of building blocks 
of various walls. 
Therefore we call these walls 
$\langle B_1,\cdots,B_k,A-1,B_{k+2},\cdots,B_{N_{\rm C}} 
\leftarrow  B_1,\cdots,B_k,A,B_{k+2},\cdots,B_{N_{\rm C}} 
\rangle$ as elementary walls.

For non-BPS walls, we obtain a lower bound for 
their tension by 
\begin{eqnarray}
\int^{+\infty}_{-\infty}{\cal E}dy > \Big[c {\rm Tr} \Sigma 
-{\rm Tr}\left\{ 
\left(\Sigma H^1 - H^1 M\right)H^1{}^\dagger\right.\nn\\
\left.{}+ \left(-\Sigma H^2 
+H^2 M\right)H^2{}^\dagger\right\}\Big]_{-\infty}^{\infty}. 
\end{eqnarray}

\section{BPS Wall Solutions}
\label{BPSWS}

In this section, we construct solutions for BPS 
Eqs.~(\ref{BPSeq-Sigma})-(\ref{BPSeq-H}) 
and examine their properties in detail. 

\subsection{The BPS Equations for Arbitrary Gauge Coupling}
\label{TBPSEABC}

It is convenient to introduce an 
$N_{\rm C}\times N_{\rm C}$ invertible 
complex matrix function $S(y)$ defined by
\begin{eqnarray}
\Sigma + iW_y \equiv S^{-1}\p_y S.
\label{def-S}
\end{eqnarray}
Note that this 
differential equation\footnote{
In Abelian $U(1)_{\rm G}$ case, a complex function $\psi$ 
was used to solve the BPS equation for 
walls~\cite{To,IOS1}. 
It is related to $S$ through 
$S= e^{\psi}$. 
Our matrix function $S$ is 
its generalization to non-Abelian cases. 
}
determines the function $S(y)$ with 
$N_{\rm C}^2$ arbitrary complex integration constants, 
from which the world-volume symmetry emerges as we see 
later. 
Let us change variables from $H^1,\,H^2$ to 
$N_{\rm C}\times N_{\rm F}$ matrix functions 
$f^1,\,f^2$ by using $S$ 
\begin{eqnarray}
H^1\equiv S^{-1}f^1,\quad H^2\equiv S^\dagger f^2. 
\label{def-f}
\end{eqnarray}
Substituting (\ref{def-S}), (\ref{def-f}) 
to the BPS Eq.~(\ref{BPSeq-H}) 
for $H^i$, we obtain 
\begin{eqnarray}
\p_y f^1 = f^1 M,\quad \partial _yf^2 = -f^2 M
\end{eqnarray}
which can be easily solved as
\begin{eqnarray}
f^1=H_0^1 \, e^{My},\quad f^2=H^2_0 e^{-My} 
\end{eqnarray}
with the $N_{\rm C}\times N_{\rm F}$ constant 
complex matrices $H_0^1,\,H^2_0$ as 
integration constants, which we call moduli matrices. 
Therefore $H^i$ can be solved completely in terms of $S$ as
\begin{eqnarray}
H^1=S^{-1}H_0^1 e^{My},\quad H^2
=S^\dagger H_0^2 e^{-My}.
\label{sol-H}
\end{eqnarray}
The definitions (\ref{def-S}),\,(\ref{def-f}) show 
that a set 
$(S, H_0^1, H_0^2)$ and 
another set $(S', H_0^1{}', H_0^2{}')$ give 
the same original fields $\Sigma ,\,W_y, H^i$, 
provided they are related by 
\begin{eqnarray}
S' = VS,\quad 
H_0^1{}'=VH_0^1,\quad 
H_0^2{}'
=(V^\dagger )^{-1}H_0^2,
\label{art-sym}
\end{eqnarray} 
where  
$V\in GL (N_{\rm C},{\bf C})$. 
This transformation $V$ defines an equivalence class 
among sets of the matrix function and moduli matrices 
$(S, H_0^1, H_0^2)$ which represents physically 
equivalent results. 
This symmetry  comes from 
the $N_{\rm C}^2$ integration constants in solving 
(\ref{def-S}), and represents 
the redundancy of describing the wall solution 
in terms of $(S, H_0^1, H_0^2)$. 
We call this `world-volume symmetry', 
since this symmetry will eventually be promoted to a 
local gauge symmetry in the world-volume of walls 
when we consider the effective action on the walls. 
It will turn out to play an important role to study 
moduli of solutions for domain walls. 

Another BPS equation (\ref{BPSeq-Y12}) 
reduces to the following condition for the moduli matrices 
\begin{eqnarray}
 H^1_0H_0^2{}^\dagger =0. \label{cond-H01H02}
\end{eqnarray}
With our choice of the direction of the FI parameter 
(\ref{eq:FIparameter}), 
$H^2$ vanishes in any SUSY vacuum as given in 
Eq.~(\ref{eq:hyper-vacuum}) corresponding to 
non-degenerate masses, 
which we consider here. 
Thus we expect that the moduli matrix for domain walls 
$H^2_0$ corresponding to the field $H^2$ also vanishes. 
Consequently the field $H^2$ for domain wall solutions 
vanishes identically in the extra dimension.  
In Appendix \ref{PH02=0}, we prove 
this expectation with the aid of 
Eqs.~(\ref{sol-H}) and 
(\ref{cond-H01H02}), 
by requiring that the scalar fields $H^1,\,H^2$ should 
converge at the boundaries. 
We also show that $H^2$ can be 
non-vanishing only as constant vacuum values 
fixed by boundary conditions, even in the case of 
degenerate mass parameters for hypermultiplets. 
Therefore we take
\begin{eqnarray}
 H^2_0=0,\qquad (H^2=0), \qquad H_0 \equiv H^1_0 . 
\end{eqnarray} 

Since the BPS equations for hypermultiplets are solved 
by means of the matrix function $S$ as in Eq.~(\ref{sol-H}), 
the remaining BPS equations for the vector multiplets 
can be written in terms of the matrix $S$ and the moduli 
matrix $H_0$.   
Since the matrix function $S$ originates from the 
vector multiplet scalars $\Sigma$ and the fifth 
component of the gauge fields $W_y$ as in 
Eq.~(\ref{def-S}), the gauge transformations on the original 
fields $\Sigma ,\,W_y,\,H^1,\,H^2$
\begin{eqnarray}
H^1  &\rightarrow & H^1{}' = U H^1,
\qquad 
H^2  \rightarrow  H^2{}' = U H^2,
\nn\\
\Sigma +iW_y &\ra &\Sigma' +iW_y' 
= U\left(\Sigma +iW_y\right)U^\dagger  + U\p_y U^\dagger ,
\qquad 
\end{eqnarray}
can be obtained by multiplying a unitary matrix 
$U^\dagger(y)$ 
from the right of $S$:  
\begin{eqnarray}
 S\,\rightarrow \,S'=SU^\dagger ,\quad U^\dagger U=1, 
\label{eq:gauge-tr-S}
\end{eqnarray}
without causing any transformations on the moduli 
matrices $H_0$. 
Thus we define $\Omega$ out of 
$S$ 
\begin{eqnarray}
 \Omega \equiv SS^\dagger , 
\label{def-Omega}
\end{eqnarray}
which is invariant under the gauge transformations 
(\ref{eq:gauge-tr-S}) of the fundamental theory. 
Note that this $\Omega$ is not invariant under the 
world-volume symmetry transformations 
(\ref{art-sym}):
\begin{eqnarray}
 \Omega \ \rightarrow \ \Omega' = V \Omega V^\dagger .
\label{WV-transf-Omega}
\end{eqnarray}
Together with 
the gauge invariant moduli matrix $H_0$, 
the BPS equations (\ref{BPSeq-Sigma}) for vector 
multiplets can be rewritten in the following 
gauge invariant form 
\begin{eqnarray}
\p_y^2 \Omega -\p_y \Omega \Omega^{-1} \p_y \Omega = g^2 
\left(c\, \Omega - H_0 \,e^{2My} H_0{}^\dagger 
\right),\label{diff-eq-S}
\end{eqnarray}
where, we used the following equality
\begin{eqnarray}
 {\cal D}_y\Sigma ={1\over 2}S^{-1}\left(\partial _y^2\Omega -\p_y \Omega \Omega^{-1} \p_y \Omega \right)
(S^\dagger )^{-1}.
\end{eqnarray} 
Needless to say, we can calculate uniquely the 
$N_{\rm C}\times N_{\rm C}$ complex matrix $S$ 
from the $N_{\rm C}\times N_{\rm C}$ Hermitian 
matrix $\Omega $ with a 
suitable gauge choice.\footnote{
For instance, 
we can take a gauge choice where $S$ is 
an upper (lower) triangular matrix whose 
diagonal elements are positive real, 
then Eq.~(\ref{def-Omega}) determines the non-vanishing 
components of the matrix $S$ straightforwardly 
from the lower-right (upper-left) components 
to the upper-left (lower-right) components.
}
Therefore, once a solution of $\Omega $ for 
Eq.~(\ref{diff-eq-S}) with a given moduli matrix $H_0$ is 
obtained, 
the matrix $S$ can be determined and then, 
all the quantities, $\Sigma ,\,W_y,\,H^1$ and $H^2$ 
are obtained by Eqs.~(\ref{def-S}) and (\ref{sol-H}). 

The remaining task for us to obtain the 
general solutions of the BPS equations is only to solve 
Eq.~(\ref{diff-eq-S}) with given boundary 
conditions.\footnote{
We need to translate the boundary conditions 
for the 
original fields, $\Sigma ,\,(W_y)$ and $H^1$ 
to those for $\Omega $ with a 
given $H_0$ to solve the equation (\ref{diff-eq-S}). 
} 
Since we are going to impose two boundary conditions 
at $y=\infty$ and at $y=-\infty$ to the second order 
differential equation (\ref{diff-eq-S}), the number of 
necessary boundary conditions precisely matches to 
obtain the unique solution. 
From this reason we expect that the nonlinear 
differential equation (\ref{diff-eq-S}) 
supplemented by the boundary conditions 
determines the 
solution uniquely with no additional integration constants. 
Therefore there should be no more moduli parameters 
in addition to the moduli 
matrix $H_0$. 
This point will become obvious when we consider the case of 
infinite coupling in Sec.~\ref{BPSWSIGC}. 
For finite coupling, a detailed analysis of 
the nonlinear differential 
equation with boundary conditions at infinity become 
rather complicated. 
However, we have analyzed in detail 
the almost analogous nonlinear 
differential equation in the case of the Abelian gauge 
theory at finite gauge coupling in order to obtain BPS 
wall solutions \cite{IOS1}. 
We have worked out an iterative approximation scheme 
to solve the nonlinear differential 
equation, say from $y=\infty$, by imposing the boundary condition, 
and found that a series of exponential terms are obtained 
with just a single arbitrary parameter to fix the solution. 
This freedom of the arbitrary parameter can be used 
to satisfy the boundary condition at the other side 
$y=-\infty$. 
The only subtlety lies in the fact that the iterative scheme 
does not seem to converge uniformly in $y$, so that we 
need to do sufficiently large numbers of iterations 
to obtain a good approximation as we go to smaller and 
smaller values of $y$. 
For the case of non-Abelian gauge theories at 
finite gauge coupling, there is no reason to believe 
a behavior different from the Abelian counterpart. 
However, it is more desirable to show it rigorously, 
for instance by index theorems, one of 
which was given for the $N_{\rm C}=1$ case~\cite{Lee}. 
Thus we believe that we should consider only the moduli 
contained in the moduli matrix $H_0$, 
in order to discuss the moduli space of domain walls.
\footnote{
If we consider 
degenerate masses for hypermultiplets, 
we have cases with non-vanishing $H_0^2$, 
which are determined by boundary conditions 
without giving any additional moduli 
as explained in Appendix~\ref{PH02=0}. 
}

For an arbitrary gauge coupling $g$, an arbitrary mass 
matrix $M$ and 
an arbitrary moduli matrix $H_0$, 
it seems, however, quite difficult to solve the nonlinear 
differential equation (\ref{diff-eq-S}) explicitly. 
In Sec.~\ref{BPSWSIGC}, 
we consider the case of the infinite gauge coupling, 
$g^2\rightarrow \infty $, where exact multi-wall 
solutions can be constructed explicitly for generic 
moduli and with arbitrary masses for hypermultiplets. 
In Sec.~\ref{sc:finite-coupl}, we obtain solutions for 
finite, but particular gauge couplings and with particular 
masses for hypermultiplets. 
This class of solution exploits the 
previously solved cases with finite gauge coupling 
for Abelian gauge theories\cite{IOS1} 
and covers only restricted subspaces of the full 
moduli space. 

\subsection{Gauge Invariant Observables}
\label{GIO}
Here, we give some useful 
identities to obtain gauge invariant quantities. 
It is tedious to calculate the gauge-variant matrix 
function $S$ from the gauge invariant matrix 
function $\Omega =SS^\dagger $.   
However, we can obtain the gauge invariant quantities without 
determining the explicit expression of the 
gauge-variant matrix function $S$. 
In almost all situations, we are interested in 
gauge invariant informations which can be obtained 
from the gauge invariant matrix $\Omega $. 
Thus we only give an 
explicit form of gauge invariant quantities without 
giving the matrix $S$ in most part of this paper. 
The Weyl invariants made of the scalar $\Sigma $ 
are given by 
\begin{eqnarray}
{\rm Tr}((\Sigma )^n)
&=&{1\over 2^n}{\rm Tr}
\left((\Omega ^{-1}\partial _y\Omega )^n\right), 
\end{eqnarray} 
where we used 
\begin{eqnarray}
 \Sigma ={\rm Re}\left(S^{-1}\partial _yS\right)
={1\over 2}S^{-1}(\partial _y\Omega )\Omega ^{-1}S.
\end{eqnarray}
In particular, the Weyl invariant for $n=1$ 
\begin{eqnarray}
\Sigma ^{I=0}
=\sqrt{2\over N_{\rm C}}{\rm Tr}\left(\Sigma \right)
={1\over \sqrt{2N_{\rm C}}}
\partial _y\left(\log(\det \Omega )\right),
\label{eq:sigm-Omega}
\end{eqnarray}
is important to obtain the tension of the walls. 
Information of the number of walls and their locations 
can be extracted from the profile of the function 
$\det\Omega $, as we explain in Appendix~\ref{position}.  
The field configurations of the hypermultiplet scalars 
$H^1$ are conveniently summarized in the following 
$N_{\rm F}\times N_{\rm F}$ matrix 
\begin{eqnarray}
 H^1{}^\dagger H^1
 =e^{My}H_0{}^\dagger \Omega ^{-1}H_0e^{My}.
\end{eqnarray}

The informations of the gauge field configurations 
can also be obtained by using the gauge invariant 
quantities. 
In the case of $N_{\rm C}=2$, for instance, it is useful 
to consider the following gauge invariant quantity 
\begin{eqnarray}
I_{\rm gauge}
={1\over |\Sigma |^2}
\left({|{\cal D}_y\Sigma |^2
-(\partial _y|\Sigma |)^2}\right),
\label{Ig}
\end{eqnarray} 
where the quantities in the right-hand side of 
the above formula are given by 
\begin{eqnarray}
|\Sigma |^2&\equiv& \sum_{I=1}^3(\Sigma ^I)^2
=2{\rm Tr}((\Sigma )^2)-({\rm Tr}(\Sigma ))^2\nn\\
&=&({\rm Tr(\Sigma )})^2-4\det\Sigma 
=\left({\partial _y\det\Omega \over 2\det\Omega }\right)^2
-{\det\partial _y\Omega \over \det\Omega }.\qquad 
\label{eq:mag-sigma}
\end{eqnarray}
Here we used a property of a $2\times2$ matrix $X$: 
${\rm Tr}(X^2)=({\rm Tr}(X))^2-2\det(X)$, and,
\begin{eqnarray}
&&|{\cal D}_y \Sigma |^2
\equiv \sum_{I=1}^3({\cal D}_y\Sigma ^I)^2
\nn\\
&&=\left(\partial _y\left({\partial _y
\det\Omega \over 2\det\Omega }\right)\right)^2
-{\det(\partial _y^2\Omega 
-\partial _y\Omega \Omega ^{-1}\partial _y\Omega )
\over \det\Omega }.\qquad 
\end{eqnarray}
The gauge invariant (\ref{Ig}) is ill-defined at $|\Sigma|=0$.  
If we choose a gauge fixing of vanishing fifth component 
gauge field  $W_y=0$, 
the quantity (\ref{Ig}) measures a twist of the 
trajectory for the wall solution in 
the space of the adjoint scalar 
$(\Sigma ^1,\Sigma ^2,\Sigma ^3)$ 
without changing the singlet scalar $\Sigma ^0$ 
of the $U(1)$ vector multiplet. 
If we choose another gauge of $\Sigma ^1=\Sigma ^2=0$ 
and $W_y^3=0$ instead, 
we obtain $I_{\rm gauge}$ as the sum of squares of the 
gauge fields 
\begin{eqnarray}
  I_{\rm gauge}=(W_y^1)^2+(W_y^2)^2.
\label{eq:sum-gauge-field}
\end{eqnarray}
Note that $I_{\rm gauge}$ is generically nontrivial 
around the regions where the walls have nontrivial 
profile as we will explain later. 
Regions far away from the walls are essentially close to 
vacua. 
In these regions, 
all fields $W_y$ and $\Sigma$ vanish or approach to a 
constant, resulting in 
$I_{\rm gauge}\simeq 0$.  
In Sec.~\ref{sc:finite-coupl}, we will define the notion of 
a `factorizable moduli' for models with the 
infinite gauge coupling. 
We will find that the above gauge invariant 
quantity $I_{\rm gauge}$ for walls 
with the factorizable moduli vanishes except at 
$\Sigma=0$ where 
$I_{\rm gauge}$ becomes ill-defined. 

\subsection{General Properties of the Moduli Matrix $H_0$}
\label{sc:moduli-matrix}
From the arguments of previous section, we 
should consider only the moduli contained in 
the moduli matrix $H_0$. 
Therefore the number of complex 
moduli parameters is given by 
\begin{eqnarray}
 \dim_{\bf C} {\cal M}_{N_{\rm F},N_{\rm C}} \equiv 
N_{\rm C} N_{\rm F} -N_{\rm C}^2 
=N_{\rm C} \tilde{N}_{\rm C},\label{DoF-moduli}
\end{eqnarray}
where we have denoted the moduli space by 
${\cal M}_{N_{\rm F},N_{\rm C}}$ 
and have defined 
\begin{eqnarray}
 \tilde N_{\rm C}\equiv N_{\rm F}-N_{\rm C} .
\end{eqnarray}

We now examine walls or vacua implied by the moduli matrices. 
Let us begin with the 
simplest case of the moduli matrix $H_0$ given by 
\begin{eqnarray}
(H_0)^{rA}=\sqrt{c}\delta ^{A_r}{}_A,\label{vacuum-moduli}
\end{eqnarray} 
where the flavor locked with the color $r$ is denoted as 
$A_r$ and is chosen as 
\begin{eqnarray}
 1\leq A_r\not=A_s\leq N_{\rm F}, \qquad {\rm for~}r\not=s .
\end{eqnarray} 
Let us define a matrix 
$\sigma (M):\sigma (M)^r{}_s =m_{A_r}\delta ^r{}_s$, which 
satisfies the relation, $H_0e^{My}=e^{\sigma (M)y}H_0$,
with the moduli matrix (\ref{vacuum-moduli}). 
By using this relation, we find that 
\begin{eqnarray}
 \Omega =e^{2\sigma (M)y},\quad (S=e^{\sigma (M)y})
\end{eqnarray}
gives a solution of the 
BPS equation 
(\ref{diff-eq-S}) with the moduli matrix (\ref{vacuum-moduli}). 
This solution corresponds to a vacuum 
\begin{eqnarray}
\quad H^1=H_0,\quad  \Sigma =\sigma (M),\quad W_y=0, 
\end{eqnarray}
apart from the freedom of gauge transformations. 
Since 
there is a one-to-one 
correspondence between the BPS solution
and 
the moduli matrix $H_0$ after fixing the world-volume 
symmetry, this moduli matrix 
describes the vacuum. 
We denote the moduli matrix corresponding to the vacuum 
$\langle A\rangle\equiv 
\langle A_1 \cdots A_{N_{\rm C}}\rangle $ 
as $H_{0\langle A\rangle}$.

The redundancy of moduli matrix $H_0$ due to the 
world-volume symmetry (\ref{art-sym}) can be fixed 
in several ways. 
The first possibility to fix the 
world-volume symmetry is to choose 
$H_0$ in the following form 
\begin{eqnarray}
&&\hspace{6em}A_1\hspace{3.7em}A_2\ \ 
\cdots \ \ \ A_{N_{\rm C}}\nn\\
 H_0&\!=\!&\sqrt{c}\left(
\begin{array}{ccccccc}
0\cdots 0&1&*\cdots *&0      &*\cdots *&0&*\cdots *    \\
0\cdots 0&0&0\cdots 0&1      &*\cdots *&0&*\cdots * \\
         & &         &\vdots &         & &             \\
0\cdots 0&0&0\cdots 0&0      &0\cdots 0&1&*\cdots* \\
\end{array}\right),\qquad \label{RREF}
\end{eqnarray}
which is useful for some purposes. 
This is the so-called row-reduced echelon form. 
It is known in the theory of the linear algebra 
that any $N_{\rm C} \times N_{\rm F}$ matrix can be 
transformed into 
this form uniquely by using $GL(N_{\rm C},{\bf C})$ 
in Eq.~(\ref{art-sym}).

We find, however, that the following form 
is more useful. 
Let us choose the form for the moduli 
matrix $H_0$ by using the transformation (\ref{art-sym}) as 
\begin{eqnarray}
&&\hspace{5.2em}A_1\hspace{3.1em}A_2
\hspace{2.7em}\stackrel{y\rightarrow \infty }{\longleftarrow }
\hspace{2.6em}B_1\hspace{4.2em}B_2\nn\\
 H_0&=&\sqrt{c}\left(
\begin{array}{cccccccccccc}
\cdots 0&1&*\cdots &*&\cdots  & &\cdots *&e^{v_1} &0\cdots     \\
 & &\cdots 0&1&*\cdots & &\cdots  &    &\cdots*&e^{v_2}&0\cdots\\
 & &\vdots  & &        & &\vdots  &      &                 &&& \\
 & &  & &\cdots 0&1&*\cdots &\cdots *&e^{v_{N_{\rm C}}}&0\cdots\\
\end{array}\right). 
\label{standard-form}\\
&&\hspace{13.8em}A_{N_{\rm C}}\hspace{5.2em} B_{N_{\rm C}} \nn
\end{eqnarray}
In the $r$-th row, all the elements 
before the $A_r$-th flavor are eliminated, the $A_r$-th 
flavor is normalized to be unity, 
and the last non-vanishing element $e^{v_r}$ 
($\in {\bf C}-\{ 0 \}$) occurs 
at the $B_r$-th flavor . 
We can choose these flavors $A_r, B_r$ to be
\begin{equation}
1\leq A_1<A_2<\cdots<A_{N_{\rm C}}\leq N_{\rm F},
\label{eq:echelon-ordering}
\end{equation}
\begin{equation}
 A_r\leq B_r,
\label{vacua-ord}
\end{equation}
\begin{equation}
 B_r\not=B_s,\quad  {\rm for~} r\not=s .
\end{equation} 
When the set of flavors 
$\{B_r\}$ are not ordered like $\{A_r\}$ in 
Eq.~(\ref{eq:echelon-ordering}), 
we must eliminate some more elements 
to remove the redundancy due to the world-volume 
symmetry. 
This procedure to eliminate these elements can be 
unambiguously defined 
as is described in Appendix~\ref{TSFH01}. 
We call this form the ``standard form''. 
We show in Appendix~\ref{TSFH01} 
that the general moduli matrix $H_0$ can be 
uniquely transformed to the standard form 
by means of the world-volume symmetry (\ref{art-sym}) 
and that the world-volume symmetry is completely 
fixed by transforming $H_0$ to the standard form. 

In the standard form 
it is easy to read vacua at the both 
boundaries $y=\pm \infty$ for walls (or vacua) 
corresponding to the moduli matrix $H_0$. 
To see this point, note that the form of solution for 
$H^1$ in Eq.~(\ref{sol-H}) implies the transformation of 
the moduli matrix 
\begin{eqnarray}
 H_0\rightarrow H_0e^{My_0}
\label{H0-translation}
\end{eqnarray}
under a translation $y\rightarrow y+y_0$.
Since the world-volume symmetry allows us to 
multiply the matrix $(V)^r{}_s=e^{-m_{A_r}y_0}\delta ^r{}_s$ 
from the left of $H_0$, the matrix $VH_0e^{My_0}$ 
remains finite when taking 
the limit $y_0\rightarrow \infty $ to give 
\begin{eqnarray}
H_0^{rA}\simeq 
\left\{\begin{array}{cc}
 0& ,A<A_r \\ 
\sqrt{c} & ,A=A_r \\ 
O(e^{-(m_{A_r}-m_A)y_0}) & ,A>A_r
       \end{array}\right.
{\rightarrow }
\sqrt{c}\delta ^{A_r}{}_A, \quad 
\end{eqnarray}
where we used the property of the standard 
form (\ref{standard-form}).   
The symbol $\simeq$ denotes the equivalence by using 
the world-volume symmetry. 
Similarly, we can choose another transformation 
$(V)^r{}_s=e^{-m_{B_r}y_0-v_r}$ with $v_r$ defined 
in Eq.~(\ref{standard-form}) in taking the limit 
$y_0\rightarrow -\infty $, to find that 
$H_0$ approaches to 
\begin{eqnarray}
  H_0^{rA}\rightarrow \sqrt{c}\delta ^{B_r}{}_A.
\end{eqnarray}
These observations mean that 
{\it a multi-wall solution corresponding to the moduli matrix 
(\ref{standard-form}) interpolates between a vacuum 
labeled by 
$\langle A_1A_2\cdots A_{N_{\rm C}}\rangle $ 
at $y\rightarrow \infty $ and 
a vacuum $\langle B_1B_2\cdots B_{N_{\rm C}}\rangle $ 
at $y\rightarrow -\infty $}. 
We will denote such a wall solution by 
$\langle  A_1A_2\cdots A_{N_{\rm C}} 
\leftarrow 
B_1B_2\cdots B_{N_{\rm C}} \rangle$. 
One should note that we enclosed 
both boundary conditions at $y=\pm \infty$ 
into a single bracket $\langle \; \rangle $, 
since we have used a gauge fixed 
representation for the multi-wall solution, 
as described in Sec.~\ref{sc:vacua-UN}. 
We denote the moduli matrix corresponding to the 
topological sector for a multi-wall interpolating 
between the vacuum 
$\langle A\rangle\equiv 
\langle A_1 \cdots A_{N_{\rm C}}\rangle $ 
at $y=\infty$ 
and the vacuum 
$\langle B\rangle\equiv 
\langle B_1 \cdots B_{N_{\rm C}}\rangle $ 
at $y=-\infty$ 
as $H_{0\langle A \leftarrow B\rangle}$.

For later convenience, we give some definitions for 
wall solutions associated with particular standard forms. 
We call a ``single wall'' if the solution is generated by 
$H_0$ in a particular standard form which contains 
only one non-vanishing element $e^{v_s}$ 
other than unit elements corresponding to the vacuum 
at $y=\infty$, 
namely if 
$B_r=A_r$ for $r \neq s$ and 
$B_s = A_s + l + 1$ with $l ( \geq 0)$ zero elements 
between $B_s$ and $A_s$, 
like 
\begin{eqnarray}
&&\hspace{5.2em}A_1\hspace{3em}A_s
\hspace{3.9em} B_s 
\hspace{8.0em} 
\nn\\
 H_0&=&\sqrt{c}\left(
\begin{array}{cccccccccccc}
\cdots 0&1&0\cdots &0&\cdots                  & \vdots&          \\ 
        & &\vdots  & &                        &  0    &  \\
        & &\cdots 0&1&\underbrace{0 \cdots 0}_l& e^{v_s}   &  {\bf 0} 
         \vspace{-2.5ex}\\
        & &\vdots  & &                        &   0   &    \\
        & &        & &\cdots 0                & \vdots&          \\
\end{array}\right) .
\label{single_walls} 
\end{eqnarray} 
This $H_0$ generates a wall labeled by 
$\langle 
A_1,A_2,\cdots ,A_s, \cdots, A_{N_{\rm C}}
\leftarrow
A_1,A_2,\cdots, A_s + l + 1, \cdots, A_{N_{\rm C}} \rangle$. 
We call $l$ the ``level" of the single wall. 
We call a single wall an ``elementary wall" or a ``compressed wall"
if its level $l$ is zero or non-zero, respectively.
\subsection{Topological Sectors in Moduli Space} 
\label{topo.sectors}
Any moduli matrix in the standard form has one-to-one 
correspondence with a point in the moduli space 
because of the uniqueness of the standard form 
as proved in Appendix \ref{TSFH01}.
The moduli manifold corresponding to 
a boundary condition 
$\langle A_1A_2\cdots A_{N_{\rm C}}\rangle $ 
at $y\rightarrow \infty $ and 
a boundary condition 
$\langle B_1B_2\cdots B_{N_{\rm C}}\rangle $ 
at $y\rightarrow -\infty $ defines a topological sector 
denoted by 
\begin{eqnarray} 
 {\cal M}_{N_{\rm F},N_{\rm C}}^{\langle A_1A_2
\cdots A_{N_{\rm C}}\rangle 
 \leftarrow \langle B_1B_2\cdots B_{N_{\rm C}}\rangle}. 
  \label{topo.-sect.}
\end{eqnarray} 
The standard form of the moduli matrix 
is quite useful to classify the moduli manifold 
into these topological sectors, since the boundary conditions 
can readily be read off as we have seen above. 
The boundary condition at $y\rightarrow \infty$ 
is uniquely specified by 
the standard form, whose label 
$\langle A_1A_2\cdots A_{N_{\rm C}}\rangle $ is 
ordered as in Eq.~(\ref{eq:echelon-ordering}). 
A given boundary condition at $y=-\infty$, however, 
corresponds to several different standard forms, 
since different labels 
$\langle B_1B_2\cdots B_{N_{\rm C}}\rangle $ 
and $\langle C_1C_2\cdots C_{N_{\rm C}}\rangle $ 
stand for the same boundary condition if they are just 
different orderings of the same set $\{B_r\}=\{C_r\}$. 
Therefore a single topological sector cannot be 
covered by a single standard form. 
Several patches of the coordinates 
corresponding to several different moduli matrices 
in the standard form are needed to cover 
the whole moduli space in that topological sector.

On the other hand, the row-reduced echelon form 
(\ref{RREF}) specifies only the vacuum 
$\langle A_1A_2\cdots A_{N_{\rm C}}\rangle $ 
at the boundary $y=\infty $. 
All possible BPS multi-wall solutions with 
that boundary condition at $y=\infty$ are 
covered by a single row-reduced echelon form, 
since the row-reduced echelon form 
does not distinguish 
the boundary condition at $y=-\infty $ at all. 
{\it One topological sector is covered by only 
one patch of the coordinates in the 
row-reduced echelon form}, 
which is not useful 
to classify topological sectors. 
Therefore the row-reduced echelon form (\ref{RREF}) 
is useful to 
discuss the relation between submanifolds 
covered by different patches of coordinates 
in the standard form. 

In this paper, we use the standard form, 
except otherwise stated.  
Once a topological sector is given, 
there exist $N_{\rm C}!$ 
moduli matrices 
$H_{0\langle A_1\cdots A_{N_{\rm C}}\leftarrow 
B_1\cdots B_{N_{\rm C}}\rangle}$ in the standard form 
(\ref{standard-form}) 
corresponding to the ordering of 
the label $\langle B_1B_2\cdots B_{N_{\rm C}}\rangle $ 
for the vacuum at $y=-\infty $. 
Components in each $H_0$ are coordinates  
in that topological sector, 
and every topological sector is completely 
covered by these sets of coordinate patches. 
Moreover every point in the topological sector 
is covered by only one of them without double counting, 
because the standard form is unique as shown in Appendix \ref{TSFH01}.

If the label 
$\langle B_1B_2\cdots B_{N_{\rm C}}\rangle $ 
happens to be ordered 
\begin{eqnarray}
 B_1\leq B_2\leq \cdots\leq  B_{N_{\rm C}}, \label{order-B}
\end{eqnarray}
then the submanifold represented by the moduli 
matrix in the standard form 
has the maximal dimension in that sector, 
since the world-volume symmetry (\ref{art-sym}) 
is fixed completely to determine $A_r$ and $B_r$  
and we have no more freedom to eliminate any 
elements between $A_r$ and $B_r$.
Its real dimension is calculated 
straightforwardly as 
\begin{eqnarray}
 \dim_{\bf C} {\cal M}^{\langle 
A \rangle \leftarrow \langle B
\rangle }_{N_{\rm F},N_{\rm C}
} 
 =  \left(\sum_{r=1}^{N_{\rm C}} B_r
-\sum_{r=1}^{N_{\rm C}} A_r  \right).
\label{dim-formula}
\end{eqnarray}
Thus we call such a moduli matrix in the 
standard form and the corresponding submanifold as 
the ``generic moduli matrix" and the ``generic submanifold" 
for each topological sector, respectively.  

On the other hand, if $B_r$ in $H_0$ in the standard form 
is not ordered as (\ref{order-B}) 
$H_0$ has smaller dimension than (\ref{dim-formula}) 
because we have to eliminate some elements between 
$A_r$ and $B_r$ to fix (\ref{art-sym}) completely. 
Its dimension can be counted by the method given 
in (\ref{ex-NC=6}) in Appendix~\ref{TSFH01}. 
Submanifolds represented by one coordinate patch
other than the generic submanifold 
are considered to be ``boundaries" of the generic submanifold. 
We will explain this in later sections.

The ``maximal topological sector" is defined by 
the sector that represents domain walls 
interpolating between vacua 
$
\langle 1,2,\cdots,N_{\rm C}\rangle 
\leftarrow 
\langle N_{\rm F}-N_{\rm C}+1,
\cdots,N_{\rm F}-1,N_{\rm F}
\rangle 
$. 
Its generic moduli matrix is given by 
\begin{eqnarray}
 H_0&=&\sqrt{c}\left(
\begin{array}{ccccccccc}
1     &*     &*     &\cdots&* &0     &\cdots&0      \\
0     &1     &*     &      &  &\ddots&\ddots&\vdots \\
\vdots&\ddots&\ddots&\ddots&  &      &*     &0      \\
0     &\cdots&0     &1     &* &\cdots&*     &*
\end{array}\right).\\
&&\hspace{9em} \vspace{-4em}\underbrace{{}\hspace{8em}}\vspace{4em} \nn\\
&&\hspace{9em} N_{\rm F}-N_{\rm C}+1\nn
 \label{maximal}
\end{eqnarray}
By using the formula (\ref{dim-formula}), 
we find that the number of the complex moduli 
parameters, $N_{\rm C} \tilde N_{\rm C}$ given 
in Eq.~(\ref{DoF-moduli}), is equal to 
the complex dimension of the maximal topological sector:
\begin{eqnarray}
&& \dim_{\bf C} {\cal M}^{
\langle 1,2,\cdots,N_{\rm C}\rangle 
\leftarrow 
\langle N_{\rm F}-N_{\rm C}+1,
\cdots,N_{\rm F}-1,N_{\rm F}
\rangle 
} 
\nn \\
&& = \dim_{\bf C} {\cal M}_{N_{\rm F},N_{\rm C}} 
(= N_{\rm C} \tilde N_{\rm C}).
\end{eqnarray}

Let us now count the number of topological sectors, 
which are defined by boundary conditions at $y=\pm \infty$. 
The restriction (\ref{vacua-ord}) of the 
labels in the standard form corresponds to the restriction 
for the boundary condition to allow BPS 
saturated domain walls.
\footnote{We have chosen one set of four supercharges to be 
conserved. Solutions conserving the other four supercharges 
are called anti-BPS walls and are not counted except for 
vacua which conserve all eight supercharges.} 
Due to this restriction, 
the number of different topological sectors 
in the moduli manifold which allow BPS domain walls is 
given by 
\begin{eqnarray}
N_{\rm BPS} \equiv 
{N_{\rm F}!\over N_{\rm C}!\tilde N_{\rm C}!}
{(N_{\rm F}+1)!\over (N_{\rm C}+1)!(\tilde N_{\rm C}+1)!}, 
\label{eq:num-top-sec-BPS}
\end{eqnarray}
where we identified the BPS and anti-BPS walls with the 
boundary conditions at $y=\pm \infty$ exchanged and 
counted only once. 
We have confirmed this formula for 
lower values of $N_{\rm C}$ by 
actually counting the number of different 
maximal moduli matrices. 
A proof for general $N_{\rm C}$ and $N_{\rm F}$ 
is given in Appendix~\ref{PONNBPS}. 

If we allow both boundary conditions for BPS 
and non-BPS walls, 
we can choose arbitrary two vacua at the boundaries. 
Consequently the number of topological sectors is 
larger, and is given by 
\begin{eqnarray}
N_{\rm top.sec.} \equiv {1\over 2}{}_{N_{\rm F}}C_{N_{\rm C}}
\left({}_{N_{\rm F}}C_{N_{\rm C}}+1\right),
\label{eq:tot-num-top-sect}
\end{eqnarray}
where we identified two boundary conditions at 
$y=\pm \infty$ exchanged. 
The difference between (\ref{eq:tot-num-top-sect}) and 
(\ref{eq:num-top-sec-BPS}) should be the number of 
topological sectors for non-BPS domain walls;
\begin{eqnarray}
&&  N_{\rm {non-BPS}} \equiv N_{\rm top.sec.} - N_{\rm BPS} \nn\\  
&& = {1\over 2}{N_{\rm F}!\over N_{\rm C}!\tilde N_{\rm C}!}
\left({N_{\rm F}!\over N_{\rm C}!\tilde N_{\rm C}!}- 
{2(N_{\rm F}+1)!\over (N_{\rm C}+1)!(\tilde N_{\rm C}+1)!}+1\right).
\qquad \label{non-BPS-number}
\end{eqnarray}

\subsection{Wall Positions and (Quasi-)Nambu-Goldstone 
Modes}\label{position-NGQNG}
Now, let us discuss how to extract informations on 
positions of walls from the moduli matrix $H_0$. 
As explained in Appendix~\ref{position}, 
positions of walls are best read off from the profile 
of the energy density 
${\cal E}=(c/2) \partial_y^2{\rm log }({\rm det}\Omega )$ 
given by Eqs.~(\ref{eq:bogomolnyi}) and (\ref{eq:sigm-Omega}). 
We can, however, guess positions of walls 
roughly from the moduli matrix $H_0$ 
without an explicit solution for $\Omega $. 
For simplicity, let us discuss the case of 
$N_{\rm C}=1$ with a generic moduli 
matrix for the maximal topological sector 
\begin{eqnarray}
 H_0&=&\sqrt{c}\left(e^{r_{1}},\,
e^{r_2},\,\cdots,\,e^{r_{N_{\rm F}}}\right) ,\quad r_1\equiv 0,
\end{eqnarray}
where, $r_A$ are the complex moduli parameters. 
Let us define new complex parameters $Y_A$ by 
\begin{eqnarray}
 Y_A\equiv -{r_A-r_{A+1}\over m_A-m_{A+1}}, 
\quad A=1, \cdots, N_{\rm F}
.
\end{eqnarray}
We denote ${\rm Re}(Y_A)=y_A$. By using a translation 
(\ref{H0-translation}) and the  
world-volume symmetry transformation 
(\ref{art-sym}) with $V=e^{-r_B -m_B y_0}$, 
the $B$-th flavor component of 
$H_0$ becomes unity 
\begin{eqnarray}
e^{-r_B -m_B y_0} H_0 e^{My_0} =\hspace{16em} \nn\\
\sqrt{c}
\left(\cdots,e^{(m_{B-1}-m_{B})(y_0-Y_{B-1})},
\,1,\,e^{-(m_{B}-m_{B+1})(y_0- Y_{B})},\cdots\right). \nn
\end{eqnarray}
If we assume 
$y_{1}\gg y_2\gg 
\cdots\gg y_{N_{\rm F}-1}$ 
for simplicity and consider the region of 
$y_{B-1}\gg y_0\gg y_B$, 
then the $B$-th flavor component is 
dominant whereas the other components become negligible 
\begin{eqnarray}
e^{-r_B -m_B y_0}\left(H_0 e^{My_0}\right)^A 
\sim 
\sqrt{c}\delta^{AB}, 
\end{eqnarray}
corresponding to the vacuum specified by that flavor 
$B$. 
As $y_0$ decreases, the dominant element shifts 
to the right 
gradually in the flavor space (to larger values of 
flavor index) as: 
$
\sqrt{c}\delta^{A B} \rightarrow 
\sqrt{c}\delta^{A (B+1)} \cdots 
$. 
This shift of the 
vacuum from $B$ to $B+1$ occurs around 
the transition point $y_B$. 
Therefore $y_B$ should approximately the position of the 
domain wall separating the vacuum $B$ and $B+1$. 
Thus we find that the number of moduli parameters 
for positions of walls is 
$N_{\rm F}-1$ for the maximal topological sector 
in this $N_{\rm C}=1$ case. 

We can repeat the same argument for each color component 
in the general $N_{\rm C}$ case. 
Therefore the number of moduli parameters for positions 
of walls in the maximal topological sector is given by 
\begin{eqnarray}
N_{\rm wall} = 
 N_{\rm C}(N_{\rm F}-N_{\rm C})
=N_{\rm C}\tilde N_{\rm C},  
\label{eq:number-of-wall}
\end{eqnarray}
which is nothing but the maximum number of distinct walls. 
One of them is 
the center of masses of a multi-wall configuration, 
which gives an exact Nambu-Goldstone mode 
corresponding to the broken translational symmetry. 
The others are approximate Nambu-Goldstone modes, 
since the position of each wall can be translated 
independently in the limit where the wall is 
infinitely separated from other walls. 

There also exist Nambu-Goldstone modes 
for internal symmetry. 
In our case of non-degenerate mass, 
the global flavor symmetry acting on the hypermultiplets
is $U(1)^{N_{\rm F}-1}$. 
It is spontaneously broken by wall configurations 
in the maximal topological sector (\ref{maximal}) completely. 
We have $N_{\rm F}-1$ moduli parameters which can be 
attributed to the Nambu-Goldstone theorem 
associated with the spontaneously broken 
flavor symmetry. 
The remaining moduli parameters cannot be explained 
by the spontaneously broken symmetry. 
They are called the quasi-Nambu-Goldstone modes, 
and are required by unbroken SUSY to make the moduli space 
a complex manifold~\cite{HNOO,Ni}.
The number of the quasi-Nambu-Goldstone modes 
is given by 
\begin{eqnarray}
 N_{\rm QNG} \equiv 
 2N_{\rm C}\tilde N_{\rm C}-N_{\rm C}\tilde N_{\rm C}
-(N_{\rm F}-1)=(N_{\rm C}-1)(\tilde N_{\rm C}-1).
\end{eqnarray}

When we construct effective field theories on walls, 
these (quasi-)Nambu-Goldstone modes 
are promoted to (quasi-)Nambu-Goldstone bosons.
Together with fermionic zero modes, 
they constitute chiral multiplets. 
The effective theory is a nonlinear sigma model on 
a K\"ahler manifold as a target space. 
Corresponding to Nambu-Goldstone bosons, 
this target K\"ahler manifold admits 
$U(1)^{N_{\rm F}-1}$ isometry. 
We return to effective theories in Sec.~\ref{eff_thoery}.

\subsection{Infinite Gauge Coupling and Nonlinear Sigma Models}
\label{BPSWSIGC}

SUSY gauge theories reduce to 
nonlinear sigma models in general 
in the strong gauge coupling limit 
$g_0 ,\ g \to \infty$.
In the case of theories with eight supercharges, 
they are hyper-K\"ahler (HK) nonlinear sigma models~\cite{Zu,AF2}
on the Higgs branch~\cite{APS,AP} of gauge theories as 
their target spaces.\footnote{
This construction of HK manifold is known as 
a HK quotient~\cite{LR,HKLR}.
} 
Since the BPS equations are drastically simplified  
to become solvable in some cases, we often consider this limit. 
In fact the BPS domain walls in theories with 
eight supercharges were first obtained in 
HK nonlinear sigma models~\cite{AT}. 
They have been the only known examples for models with eight 
supercharges~\cite{GTT2,To,ANNS,AIN} 
until exact wall solutions 
at finite gauge coupling were found 
recently~\cite{KS,IOS1,IOS2}.
When hypermultiplets in gauge theories are massless,  
HK nonlinear sigma models do not have potentials, 
whereas a nontrivial potential is needed to obtain 
domain wall solutions. 
If hypermultiplets have masses, 
the corresponding nonlinear sigma models 
have potentials, which can be written 
as the square of the tri-holomorphic 
Killing vector on the target manifold~\cite{AF2}. 
These models are called massive HK nonlinear sigma models. 
By this potential most vacua are lifted 
leaving some discrete degenerate points as vacua,
which are characterized by fixed points of the Killing vector. 
In these models interesting composite $1/4$ BPS solitons like 
intersecting walls~\cite{GTT1}, intersecting lumps~\cite{NNS1,PT}  
and composite of wall-lumps~\cite{GPTT,INOS2} 
were constructed.
 
The BPS equation (\ref{diff-eq-S}) for the gauge invariant $\Omega$ 
reduces to an algebraic equation 
in the strong gauge coupling limit, given by 
\begin{eqnarray}
 \Omega_{g \to \infty} 
 = (SS^\dagger)_{g \to \infty}  
 = c^{-1}H_0 e^{2My}H_0^{\dagger}. 
  \label{SS-H0}
\end{eqnarray}
Therefore in the infinite gauge coupling 
we do not have to solve the second order differential 
equation for $\Omega$ 
and can explicitly construct wall solutions 
once the the moduli matrix $H_0$ is given. 
We will work out the cases of $\NC=2$ and 
$\NF=3,4$ in detail as illustrative examples 
in Sec.~\ref{CEWSIC}. 
Qualitative behavior of walls 
for finite gauge couplings 
is not so different from that
in  infinite gauge couplings. 
This is because the right hand side of 
Eq.~(\ref{diff-eq-S}) tend to zero at 
both spatial infinities even for finite $g$. 
Hence wall solutions for finite $g$ asymptotically 
coincides with those for infinite $g$, 
and they differ only at finite region.  
In fact in \cite{IOS1} 
we have constructed exact wall solutions 
for finite gauge couplings 
and found that their qualitative behavior 
is the same as the infinite gauge coupling cases 
found in the literature~\cite{AT,GTT2,To,ANNS}. 
Unfortunately we have also found that 
the $1/g^2$ expansion does not converge uniformly 
in extra-dimensional coordinate $y$~\cite{IOS1}.

Let us give the concrete action of nonlinear sigma models 
in the rest of this subsection. 
Since the gauge kinetic terms for $W_M$ and $\Sigma$ 
(and their superpartners) 
disappear in the strong coupling limit, 
they become auxiliary fields whose 
equations of motion enable us to 
express them in terms of hypermultiplets as
\begin{eqnarray}
 && W^{I}_M= i (A^{-1})^{IJ} 
   {\rm Tr}
  [(H^i \overleftrightarrow{\partial}_M H^{i\dagger} )T_J ],  
   \nonumber \\
 && \Sigma^I = 2 (A^{-1})^{IJ} 
  {\rm Tr}(H^{i\dagger} T_J H^i M),
 \label{constraint1}
\end{eqnarray}
where $(A^{-1})^{IJ}$ is an inverse matrix of $A_{IJ}$ defined by
\begin{eqnarray}
&& A_{IJ} = {\rm Tr}(H^{i\dagger} \{ T_I,T_J \} H^i) .
\end{eqnarray}
The auxiliary fields $Y^a$ serve as Lagrange multiplier 
fields to give constraints as their equations of motion 
\begin{eqnarray}
 H^1H^{1\dagger} - H^2H^{2\dagger} = c {\bf 1}_{N_{\rm C}}, 
  \quad
 H^2 H^{1\dagger} = 0.
   \label{constraint2}
\end{eqnarray}
As a result, in the limit of infinite coupling, 
the Lagrangian reduces to 
\begin{eqnarray}
\mathcal{L}^{g\rightarrow\infty}&\!=\!&{\rm Tr}[({\cal D}_M H^i)^\dagger {\cal D}^M H^i]\nn \\
&&{} + {\rm Tr}[(H^{i\dagger} \Sigma - M H^{i\dagger})(\Sigma H^i - H^i M)],\label{reduced-L}
\end{eqnarray}
with the constraints (\ref{constraint1}) 
and (\ref{constraint2}).
This is the HK nonlinear sigma model 
on the cotangent bundle over 
the complex Grassmann manifold~\cite{LR,ANS} 
\begin{eqnarray}
 {\cal M}^{M=0}_{\rm vac} \simeq
 T^* G_{N_{\rm F},N_{\rm C}} \simeq 
 T^* \left[SU(N_{\rm F}) \over 
 SU(N_{\rm C}) \times SU(\tilde N_{\rm C}) \times U(1) 
 \right] \, . \label{T*Gr}
\end{eqnarray}
In our choice of the FI parameters, 
$H^1$ parametrize the base Grassmann manifold 
whereas $H^2$ its cotangent space as fiber.\footnote{
Setting $H^2 =0$ we obtain the K\"ahler nonlinear 
sigma model on 
the Grassmann manifold~\cite{HN}. 
We thus have found the bundle structure.
}
The isometry of the metric, which is the symmetry 
of the kinetic term, is $SU(\NF)$, although it 
is broken to its maximal Abelian subgroup $U(1)^{\NF-1}$ 
by the potential. 
In the massless limit $M=0$, the potential $V$ vanishes 
and the whole manifold become vacua, 
the Higgs branch of our gauge theory. 
So we have denoted the target manifold by 
${\cal M}^{M=0}_{\rm vac}$ in (\ref{T*Gr}).
Turning on the hypermultiplet masses, 
we obtain the potential allowing 
only discrete points as SUSY vacua~\cite{ANS}, 
which are fixed points 
of the invariant subgroup $U(1)^{\NF-1}$ 
of the potential.
The number of vacua is of course $\NF !/\NC ! \tilde\NC !$, 
which is the same as the case (\ref{vacnumber}) of the finite gauge coupling.

\begin{figure}
\begin{center}
\includegraphics[width=5cm,clip]{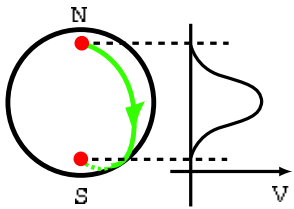}
\caption{ ${\bf C}P^1$ and the potential $V$.
The base space of $T^*{\bf C}P^1$, 
${\bf C}P^1 \simeq S^2$, is displayed.  
This model contains two discrete vacua denoted by 
$N$ and $S$.  
The potential $V$ is also displayed on 
the right of the ${\bf C}P^1$.
It admits a single wall solution 
connecting these two vacua expressed by a curve.
The $U(1)$ isometry around the axis connecting 
$N$ and $S$ is spontaneously broken by the wall configuration.
}
\label{nit-fig1}
\end{center}
\end{figure}

In the case of $\NC =1$ the target space 
reduces to the cotangent bundle over 
the complex projective space 
$T^* {\bf C}P^{\NF-1}
=T^* [SU(\NF)/SU(\NF -1) \times U(1)]$~\cite{T*CPN}
endowed with the Calabi metric~\cite{Ca}. 
Since the metric is invariant under the $SU(\NF)$ isometry,
whose maximal Abelian subgroup is $U(1)^{\NF-1}$, 
it is a toric HK manifold. 
This model has discrete $\NF$ vacua and 
admits $\NF-1$ parallel domain walls~\cite{GTT2,To}.  
Moreover if $\NF=2$ the target space 
$T^*{\bf C}P^1$ is the simplest HK manifold, 
the Eguchi-Hanson space~\cite{EH} 
(see Fig.~\ref{nit-fig1}). 
This model contains two vacua and a single BPS wall 
solution~\cite{AT,ANNS}.

\medskip
From the target manifold (\ref{T*Gr}) one can easily 
see that there exists a duality 
between theories with the same flavor and 
two different gauge groups 
in the case of the infinite gauge coupling~\cite{AP,ANS}:
\begin{eqnarray}
 U(N_{\rm C}) \leftrightarrow 
 U(\tilde N_{\rm C}) = U(N_{\rm F}-N_{\rm C}) \,.
\end{eqnarray}
This duality holds for the Lagrangian of the 
nonlinear sigma models, and 
leads to the duality of the BPS equations 
between these two theories. 
The BPS equation for a dual theory is discussed in 
Appendix~\ref{BPSDBPSIC}. 
This duality holds also for the moduli space 
of domain wall configurations.

\subsection{Factorizable Moduli and 
Solutions with Finite 
Coupling} 
\label{sc:finite-coupl}
If the moduli matrix $H_0$ takes a certain restricted form 
which will be defined below as the  
`$U(1)$-factorizable moduli', 
the BPS equation (\ref{diff-eq-S}) for our non-Abelian case 
can be decomposed into a direct sum of BPS equations for 
the Abelian case. 
In such circumstances, we can construct exact solutions 
for finite, but special values of gauge coupling 
by using the solutions found in our previous 
paper~\cite{IOS2}. 

The BPS equation (\ref{diff-eq-S}) is covariant 
under the  world-volume transformation 
(\ref{art-sym}), where the matrix $H_0e^{2My}H_0{}^\dagger$ 
transforms with multiplication of 
constant matrices $V$ and  
$V^\dagger $ from both sides of this matrix. 
The  world-volume symmetry allows us to 
make this matrix $H_0e^{2My}H_0{}^\dagger$ diagonal 
at one point of the extra dimension, 
say, $y=y_0$. 
If the matrix $H_0e^{2My}H_0{}^\dagger$ with this gauge 
fixing remains diagonal at every other points 
in the extra dimension $y\not=y_0$, 
\begin{eqnarray}
 H_0e^{2My}H_0{}^\dagger =c\,{\rm diag.}
\left({\cal W}_1(y),{\cal W}_2(y),\cdots,
{\cal W}_{N_{\rm C}}(y)\right),
\label{factorizable-case}
\end{eqnarray}
then, we call that moduli matrix $H_0$ 
as `$U(1)$-factorizable'. 
Note that such a property is a characteristic 
inherent in each moduli matrix $H_0$, 
and is independent of the choice of the 
initial coordinate $y_0$. 
Thus the $U(1)$-factorizability 
is a property intrinsically attached to 
each point on the moduli manifold 
of the BPS solution. 
If the moduli matrix is $U(1)$-factorizable, 
off-diagonal components of the 
matrix $H_0e^{2My}H_0{}^\dagger $ vanishes 
at any point of the extra dimension $y$ by definition. 
This implies that each coefficient of 
$e^{2m_Ay}$ in the off-diagonal components 
must vanish. 
We consider in this 
paper the case of non-degenerate masses, 
unless otherwise stated. 
In the non-degenerate case, the condition for the 
$U(1)$-factorizability can be written for 
each flavor $A$ 
\begin{eqnarray}
 (H_0)^{r A}\left((H_0{})^{s A}\right)^*=0, 
\quad {\rm for~} r\not=s , 
\label{factorizable-moduli}
\end{eqnarray}
where we do not take sum over the flavor indices $A$. 
In other words, $(H_0)^{rA}$ can be non-vanishing 
in only one color component $r$ 
for each flavor $A$. 
For instance, we can choose $H_0$ as 
\begin{eqnarray}
 H_0=\sqrt{c}
\left(
\begin{array}{ccccc}
0      &0      &e^{r_3}&0      &\\
0      &0      &0      &0      &\cdots\\
0      &e^{r_2}&0      &e^{r_4}&\\
e^{r_1}&0      &0      &0      &\\
       &\vdots &       &       &\ddots
\end{array}\right).
\end{eqnarray}
We can rearrange 
these moduli matrix to a standard form 
(an echelon form) in Eq.~(\ref{standard-form}) 
with the 
 world-volume symmetry 
keeping the forms 
(\ref{factorizable-case}),(\ref{factorizable-moduli}). 
Moduli matrices representing points of $U(1)$-factorizable 
moduli do not always  satisfy the condition 
(\ref{factorizable-moduli}), 
because of the redundancy of the  world-
volume symmetry.
We can always establish the $U(1)$-factorizability of 
moduli matrices by checking the condition
(\ref{factorizable-moduli}) in the standard form.

For such a $U(1)$-factorizable moduli, it is 
sufficient to take an ansatz where only the 
diagonal components of the matrix $\Omega $ survive 
\begin{eqnarray}
 \Omega ={\rm diag.}\left(e^{2 \psi _1},\,e^{2\psi _2},
\,\cdots,e^{2\psi _{N_{\rm C}}}\right),
\end{eqnarray} 
where $\psi_r(y)$'s are real functions. 
With this ansatz, the BPS equations (\ref{diff-eq-S}) 
for the non-Abelian gauge theories 
with the $U(1)$-factorizable moduli with 
the condition (\ref{factorizable-case}) reduce to a set
of the BPS equations~\cite{To,IOS1} 
for the Abelian gauge theory 
\begin{eqnarray}
 \partial _y^2\psi _r
={g^2c\over 2}\left(1-e^{-2\psi _r}{\cal W}_r\right), 
\qquad 
{\rm for ~} r=1,2,\cdots N_{\rm C}, 
\label{U1BPS}
\end{eqnarray}
where the functions ${\cal W}_r$ defined in (\ref{factorizable-case}) are given by 
\begin{eqnarray}
 {\cal W}_r=\sum_{A\in {\cal A}_r}e^{2 m_Ay+2 r_A }. 
\end{eqnarray}
${\cal A}_r$ is a set of flavors of the 
hypermultiplet scalars whose $r$-th 
color component is non-vanishing. 
Note that the condition (\ref{factorizable-moduli}) 
of the $U(1)$-factorizability can be rewritten as 
${\cal A}_r\cap{\cal A}_s=\phi $ for $r\not=s$. 
In this case, the vector multiplet scalars $\Sigma $ 
and the hypermultiplet scalars $H^{1rA}$ 
are given by~\cite{IOS1}  
\begin{eqnarray}
 \Sigma 
={\rm diag.}\left(\partial _y\psi _1,\,
\partial _y\psi _2,\,\cdots,\,
\partial _y\psi _{N_{\rm C}}\,\right) ,
\label{eq:factor-sigma}
\end{eqnarray}
\begin{eqnarray}
 H^{1rA} 
=\sqrt{c} e^{-\psi _r(y) +m_A (y-y_0) + r_A} ,
\label{eq:factor-hyper}
\end{eqnarray}
with a gauge choice of $W_y=0$.
The energy density ${\cal E}$ of the BPS multi-walls in 
Eq.(\ref{eq:bogomolnyi}) are 
obtained by a summation of energy density for 
each individual wall 
$c\partial_y^2 \psi _r$ as 
\begin{eqnarray}
{\cal E}
=c\, \partial_y {\rm Tr}(\Sigma )
=c \sum_{r=1}^{N_{\rm C}}\partial^2_y\psi _r.
\end{eqnarray}
Therefore, the profile of the energy density 
for the BPS multi-walls are obtained by a 
simple summation of 
those of individual wall generated from different 
$\psi _r$. 
Since moduli parameters contained in the BPS equation 
(\ref{U1BPS}) for each $\psi _r$ are independent 
of each other, 
we find that the walls 
originated in different $\psi _r$ 
can have positive and negative relative positions 
other maintaining their identity. 
When two walls can go through each other like here, 
they are called penetrable each other. 
More generally, if we take up two sets of walls belonging to 
two diagonal entries of Eq.~(\ref{eq:factor-sigma}) 
of the $U(1)$-factorizable case, 
these two sets are mutually penetrable, in the sense that 
they can go through each other provided the relative 
distances between walls in the same diagonal entry 
are fixed. 

We have found previously that 
the $U(1)$ gauge theories allow exact BPS solutions for 
finite gauge couplings~\cite{IOS1}. 
These finite gauge couplings have been found to be restricted 
to specific values in relation to mass splittings: 
exact solutions for single-walls 
at 
$g^2c=4(m_1-m_2)^2/k^2$, for $k=2,3,4$, 
and double wall at 
$g^2c=(m_1-m_2)^2=(m_2-m_3)^2$. 
A number of exact solutions of the BPS multi-walls 
for our non-Abelian gauge theory 
can be obtained 
in the $U(1)$-factorizable cases 
by embedding these known 
solutions into the equations (\ref{U1BPS}) 
for the $U(1)$ factor groups. 
For example, in the case of 
$N_{\rm C}=2,\,N_{\rm F}=4$ with
\begin{eqnarray}
 g^2c=(\Delta m)^2 ,\quad m_1-m_2=m_3-m_4\equiv \Delta m, 
\end{eqnarray} 
and with a $U(1)$-factorizable moduli matrix 
\begin{eqnarray}
 H_0=\sqrt{c}\left(
\begin{array}{ccccc}
e^{-m_1y_1}  &e^{-m_2 y_1}&0 &0 \\
0  &0        &e^{-m_3y_2} &e^{-m_4 y_2}
\end{array}\right)
\end{eqnarray}
with real parameters $y_1, \ y_2$. Then an exact solution 
is given by two copies of the solution for $k=2$ as 
\begin{eqnarray}
 \psi _1&=&\log\left(e^{m_1(y-y_1)}+e^{m_2(y-y_1)}\right),
\quad \nn\\ 
\psi _2&=&\log\left(e^{m_3(y-y_2)}+e^{m_4(y-y_2)}\right).
\end{eqnarray}
This solution represents a double wall that 
are located at $y=y_1$ and $y=y_2$. 
More complicated exact solutions can be obtained if we 
take a larger number of flavor and color. 
For instance, in the case of $N_{\rm C}=3,\,N_{\rm F}=7$ 
with 
\begin{eqnarray}
g^2c=4m^2, \ M
={\rm diag.}\left(2m,\,{3\over 2}m,\,m,\,0,\,-m,\,
-{3\over 2}m,\,-2m\right)\quad  
\end{eqnarray}
and with a $U(1)$-factorizable moduli matrix 
\begin{eqnarray}
 H_0=\sqrt{c}\left(
\begin{array}{ccccccc}
e^{-2my_1}\hspace{-1em}   &0& 0&e^{mR} & 0&0&e^{2my_1} \\
0   &0&e^{-my_2}\hspace{-1em} &0&e^{my_2}\hspace{-1em}&0&0\\
0&e^{-{3\over 2}my_3}\hspace{-1em} &0&0&0&e^{-{3\over 2}my_3}\hspace{-1em}&0
\end{array}\right), \quad 
\end{eqnarray}
we obtain an exact solution for a BPS four-walls 
\begin{eqnarray}
\psi _1
&=&\log\left(e^{2m(y-y_1)}+e^{-2m(y-y_1)}
+\sqrt{6+e^{2mR}}\right),\nn\\ 
\psi _2
&=&\log\left(e^{m(y-y_2)}+e^{-m(y-y_2)}\right),\quad \nn\\
\psi _3
&=&{3\over 2}\log\left(e^{m(y-y_3)}+e^{-m(y-y_3)}\right).
\end{eqnarray}
Although we have given only the solution for $\psi$, 
the vector multiplet scalar $\Sigma$ and the hypermultiplet 
scalar $H^1$ can be obtained readily from $\psi$ 
by using Eqs.~(\ref{eq:factor-sigma}) and 
(\ref{eq:factor-hyper}). 

\section{Constructing Explicit 
Solutions at Infinite Coupling}
\label{CEWSIC}
In this section, as explicit examples, 
we construct BPS wall solutions and 
investigate their properties  
in the $N_{\rm C}=2$, $N_{\rm F}=3,4$ cases. 
General $N_{\rm C}$ and/or $N_{\rm F}$ cases are similar.
In the first subsection, 
we work with the simplest case of 
$N_{\rm C}=2,\ N_{\rm F}=3$ 
to illustrate methods to construct the solutions and 
relations between the moduli matrices $H_0$ and 
profiles of solutions for domain walls. 
This case is, however, equivalent to 
the case of $N_{\rm C}=1,\,N_{\rm F}=3$ by duality 
$N_{\rm C} \leftrightarrow \tilde N_{\rm C}$, 
and thus the properties of 
walls are also equivalent to the Abelian case.   
In the second subsection, 
we consider the case of $N_{\rm C}=2,\,N_{\rm F}=4$, 
which is the simplest case that possesses characteristic 
properties of genuine non-Abelian walls.
We will define matrix operators acting on the moduli space 
to create multi-wall solutions from a solution with walls 
less by one.

\subsection{$N_{\rm C}=2, \ N_{\rm F}=3$ Case}
\label{NC2NF3Case}
\begin{figure} 
\begin{center}
\includegraphics[width=5cm,clip]{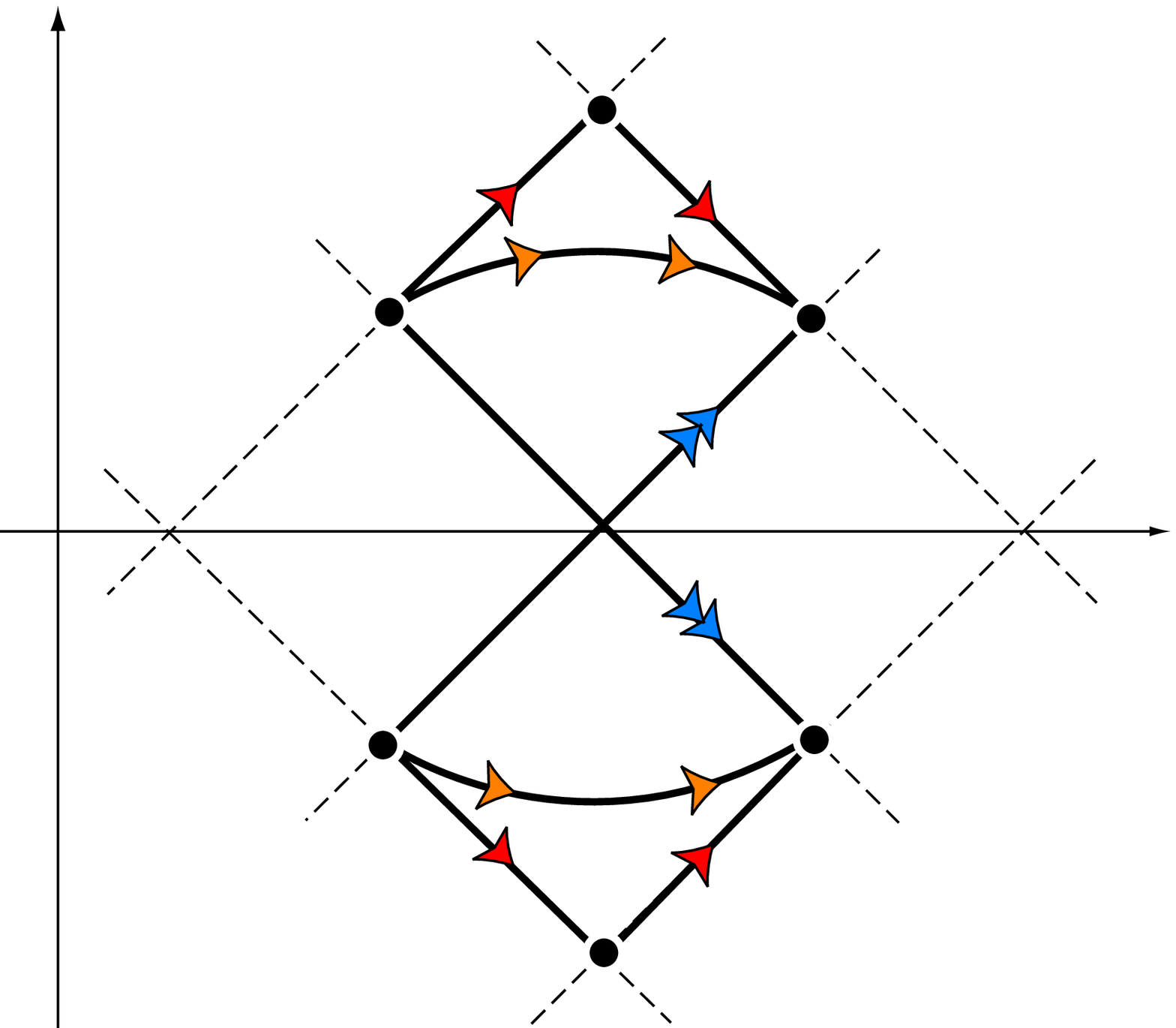}
\put(-77,-8){\footnotesize $\langle 31 \rangle$}
\put(-35,30){\footnotesize $\langle 21 \rangle$}
\put(-120,30){\footnotesize$\langle 32 \rangle$}
\put(-77,125){\footnotesize $\langle 13 \rangle$}
\put(-35,85){\footnotesize $\langle 12 \rangle$}
\put(-120,85){\footnotesize $\langle 23 \rangle$}
\put(-20,45){\footnotesize $2m_1$}
\put(-78,45){\footnotesize $2m_2$}
\put(-128,45){\footnotesize $2m_3$}
\put(0,55){$\Sigma^0$}
\put(-150,120){$\Sigma^3$}
\caption{Walls for $N_{\rm C}=2$ and $N_{\rm F}=3$. 
This model admits three single walls. 
Two of them are elementary walls and the other 
is a compressed wall.
The latter is obtained in a particular limit 
of the double wall configuration. 
Meaning of arrows is explained in Fig.~\ref{yajirusi}. 
}
\label{3f-all-wall}
\end{center}
\end{figure}

\begin{figure} 
\begin{center}
\includegraphics[width=1.8cm,clip]{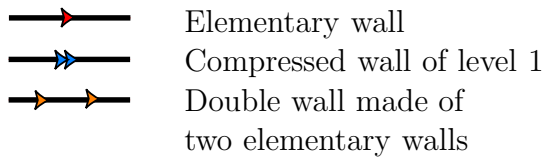}
\put(20,35){Elementary wall}
\put(20,20){Compressed wall of level $1$}
\put(20,5){Double wall made of} 
\put(20,-10){two elementary walls}
\caption{An arrow with a single 
arrowhead denotes an elementary wall. 
An arrow with duplicate ($l$-uninterrupted) 
arrowhead denotes a compressed wall 
of level $1$ ($l-1$). 
An arrow with two separate arrowheads denotes 
a double wall consisting of two single walls with 
the relative distance as a moduli. 
}
\label{yajirusi}
\end{center}
\end{figure}

In this case, there exist $3 (={}_3C_2)$ vacua and 
maximally $2 (=N_{\rm C}\tilde N_{\rm C})$ walls 
interpolating between these vacua. 
Fig.~\ref{3f-all-wall}  illustrates 
the diagram of SUSY vacua and walls 
in the space of the scalars of 
vector multiplets $\Sigma $. 
Let us construct explicit expressions of 
the exact solutions for the BPS equations. 
First of all, it is important to  classify 
arbitrary $2\times 3$ moduli matrices $H_0$ 
in the standard form (\ref{standard-form}) 
into several types of matrices. 
The standard form matrices
\begin{eqnarray}
H_{0\langle 12\rangle}=
\sqrt{c}\left( 
\begin{array}{ccc}
1&0 &0\\
0&1 &0  
\end{array}
\right),
\ \ 
H_{0\langle 13\rangle}=
\sqrt{c}\left( 
\begin{array}{ccc}
1&0 &0\\
0&0 &1  
\end{array}
\right),
\ \ 
H_{0\langle 23\rangle}=
\sqrt{c}\left( 
\begin{array}{ccc}
0&1 &0\\
0&0 &1  
\end{array}
\right)
\end{eqnarray}
correspond to the three vacua 
$\langle 12\rangle ,\langle 13\rangle $ 
and $\langle 23\rangle $, 
respectively as 
illustrated in Fig.~\ref{3f-all-wall}. 
The three matrices 
with complex parameters $r_1,\,r_2$ and $r_3$,
\begin{eqnarray}
H_0{}_{\langle 12\leftarrow 13\rangle }&=&
\sqrt{c}\left(
\begin{array}{ccc}
  1& 0 &0 \\0& 1 & e^{r_1}
\end{array}
\right),\quad  -\infty < {\rm Re}(r_1)< \infty ,
\nonumber\\
H_0{}_{\langle 13\leftarrow 23\rangle }&=&
\sqrt{c}\left(
\begin{array}{ccc}
  1& e^{r_2}&0 \\0& 0& 1
\end{array}
\right),\quad -\infty <{\rm Re} (r_2)<\infty ,\nonumber\\
H_0{}_{\langle 12\leftarrow 32\rangle }&=&\sqrt{c}\left(
\begin{array}{ccc}
1& 0&e^{r_3}\\
 0&1&0 
\end{array}
\right),\quad -\infty < {\rm Re}(r_3)<\infty  
\label{N3-single-moduli}
\end{eqnarray}
describe single-wall configurations, where the suffix  
$\langle A_1A_2\leftarrow B_1B_2\rangle $ denotes  
a moduli matrix describing a BPS state interpolating 
from the vacuum 
$\langle B_1B_2\rangle $ at $y=-\infty$ 
to the vacuum $\langle A_1A_2\rangle $ at $y=+\infty$ .
By these labels, 
we recognize the first two of the matrices 
(\ref{N3-single-moduli}) describe
elementary walls and the last one  
a compressed wall of level one 
as defined in (\ref{single_walls}).
As explained in Sec.~\ref{position-NGQNG}, 
positions $y_1, y_2$ and $y_3$ of the single-walls 
labeled by $\langle 12\leftarrow 13\rangle ,
\langle 13\leftarrow 23\rangle $ and 
$\langle 12\leftarrow 32\rangle $ can be guessed
roughly as 
\begin{eqnarray}
y_1&\equiv & {{\rm Re}(r_1)\over m_2-m_3},
\quad y_2\equiv   {{\rm Re}(r_2)\over m_1-m_2},
\quad y_3\equiv {{\rm Re}(r_3)\over m_1-m_3}, 
\label{N3-positions}
\end{eqnarray}
respectively.
Finally, the moduli matrix  
\begin{eqnarray}
H_0{}_{\langle 12\leftarrow 23\rangle }&=&
\sqrt{c}\left(
\begin{array}{ccc}
  1& e^{r_2}&0 \\0& 1& e^{r_1}
\end{array}
\right),\quad  -\infty < {\rm Re}(r_1)< \infty ,
\quad -\infty <{\rm Re} (r_2)<\infty  
\label{N3-double-moduli}
\end{eqnarray} 
corresponds to a double wall 
interpolating from the vacuum $\langle 23\rangle $ 
to the vacuum $\langle 12\rangle $ through the 
vicinity of 
the vacuum $\langle 13\rangle $. 
Note that we have distinguished 
the moduli matrix (\ref{N3-double-moduli}) 
from the third one in Eq.~(\ref{N3-single-moduli}) 
by their orders of flavors. 
As described in Sec.~\ref{sc:vacua-UN}, 
it is convenient to 
distinguish the vacua with different order of labels.
This is because there are no freedom of local gauge 
transformation, if we fix the gauge by eliminating 
the off-diagonal components of the scalar $\Sigma $.  
In that gauge, the labels for moduli matrices 
reflect trajectories 
in the space of the diagonal components 
$(\Sigma ^0,\Sigma ^3)$ as illustrated in 
Fig.~\ref{3f-all-wall}. 
This is the most convenient gauge to represent 
the solutions, which we usually use. 
Various types of walls are distinguished by arrows as explained in Fig.~\ref{yajirusi}.

The above identification between moduli matrices 
and BPS objects are performed without constructing 
the exact solutions. 
It is also easy to investigate relations between 
these moduli matrices. 
For instance, 
the moduli matrix 
$H_0{}_{\langle 12\leftarrow 23\rangle }$ 
approaches $H_0{}_{\langle 12\leftarrow 13\rangle }$ 
in the limit of $r_2\rightarrow -\infty$. 
Since $y=y_2$ corresponds to the position of the wall 
interpolating between 
$\langle 23\rangle $ and $\langle 13\rangle $, 
$r_2\rightarrow -\infty$ implies expelling the wall 
to negative infinity $y_2 \rightarrow -\infty$. 
This explains $H_0{}_{\langle 12\leftarrow 23\rangle }
\rightarrow H_0{}_{\langle 12\leftarrow 13\rangle }$ 
in the limit of $r_2\rightarrow -\infty$. 
The moduli matrices 
$H_0{}_{\langle 12\leftarrow 32\rangle }$ 
for single-wall and 
$H_0{}_{\langle 12\leftarrow 23\rangle }$ for
double wall 
describe BPS states interpolating between the same pair 
of the vacua at the 
boundaries, in other words these moduli 
matrices describes different 
submanifolds of the same topological sector. 
To understand how these 
submanifold connect with each others, 
it is convenient to consider 
these moduli matrices in the row-reduced 
echelon form (\ref{RREF}).
Let us perform the world-volume symmetry transformation 
(\ref{art-sym}) 
on the $H_{0\langle 12\leftarrow 23\rangle }$ 
so that the moduli matrix becomes a 
row-reduced echelon form.  
\begin{eqnarray}
 H_{0\langle 12\leftarrow 23\rangle }&\rightarrow &
\left(
\begin{array}{cc}
  1& -e^{r_2}\\0& 1 
\end{array}\right)
H_{0\langle 12\leftarrow 23\rangle }=
\sqrt{c}\left(
\begin{array}{ccc}
  1& 0&-e^{r_1+r_2} \\0& 1& e^{r_1}
\end{array}\right).
\end{eqnarray}
Here, if we take the limit of $r_1\rightarrow -\infty$, 
keeping the parameter  
\begin{eqnarray}
 r_3=r_1+r_2+\pi i \label{r3}
\end{eqnarray}
finite ($r_2\rightarrow \infty$), 
we find that $H_{0\langle 12\leftarrow 23\rangle }$ 
in the row-reduced echelon form 
becomes $H_0{}_{\langle 12\leftarrow 32\rangle }$. 
This relation between 
$H_{0\langle 12\leftarrow 23\rangle }$ 
and $H_0{}_{\langle 12\leftarrow 32\rangle }$
is quite different from that between 
$H_{0\langle 12\leftarrow 23\rangle }$ 
and $H_0{}_{\langle 12\leftarrow 13\rangle }$, 
while one describes a double wall and the other 
describes a single wall in the both case.  
Since the boundary conditions 
are not changed by transition form 
$H_{0\langle 12\leftarrow 23\rangle }$ 
to $H_0{}_{\langle 12\leftarrow 32\rangle }$ 
in this case, the transition means that  
the two walls approach each other 
and are compressed to a single wall.   
In the region of $y_1\leq y_2$, 
a profile of the 
energy density of two constituent walls are 
compressed into a profile of a single peak. 
The parameters $y_1$ and $y_2$ do not represent 
positions of the walls. Instead, 
the parameter $y_1-y_2$ represents the extent 
of compression of the walls. 
The relation (\ref{r3}) implies
the parameter $y_3$, which denotes 
the position of the single wall labeled by 
$\langle 12\leftarrow 32\rangle $, is related to 
the center of mass of the double wall formally 
\begin{eqnarray}
 y_3={T_{\langle 2\leftarrow 3\rangle }y_1
+T_{\langle 1\leftarrow 2\rangle }y_2\over 
T_{\langle 2\leftarrow 3\rangle }
+T_{\langle 1\leftarrow 2\rangle }}.
\end{eqnarray}
We will discuss this compression of walls more using 
an exact
solution in the latter part of this subsection. 
This phenomenon of compressed wall 
has also occurred in the Abelian case~\cite{GTT2,To,IOS1}. 
Actually, we find that this $N_{\rm C}=2, N_{\rm F}=3$ 
case is dual to the $N_{\rm C}=1, N_{\rm F}=3$ case, 
which is the case of the Abelian gauge theory.  

\begin{figure}[htb]
\begin{center}
\includegraphics[width=7cm,clip]{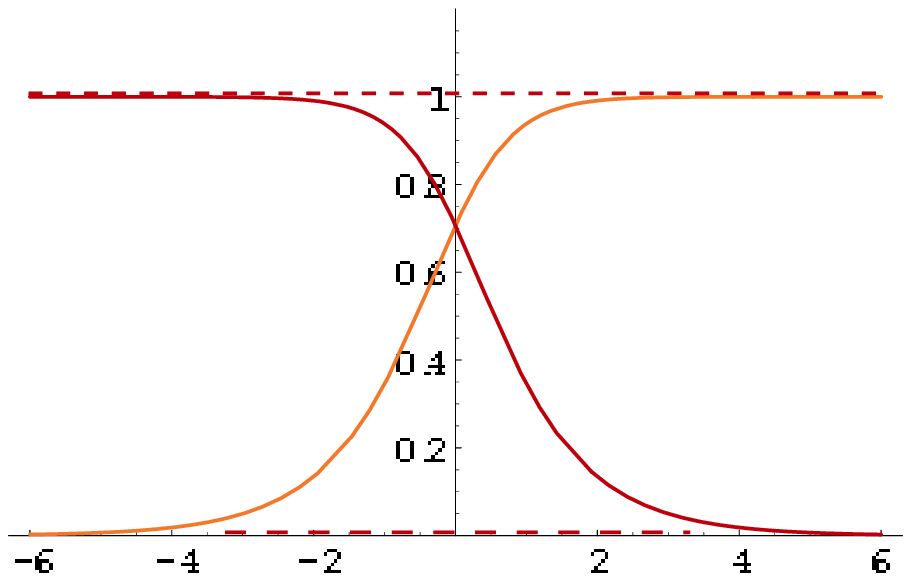}
\put(0,10){$y$}
\put(-170,80){$(H^1)^{22}$}
\put(-50,80){$(H^1)^{23}$}
\put(-100, -10){a)}
\hspace{0.3cm}
\includegraphics[width=6cm,clip]{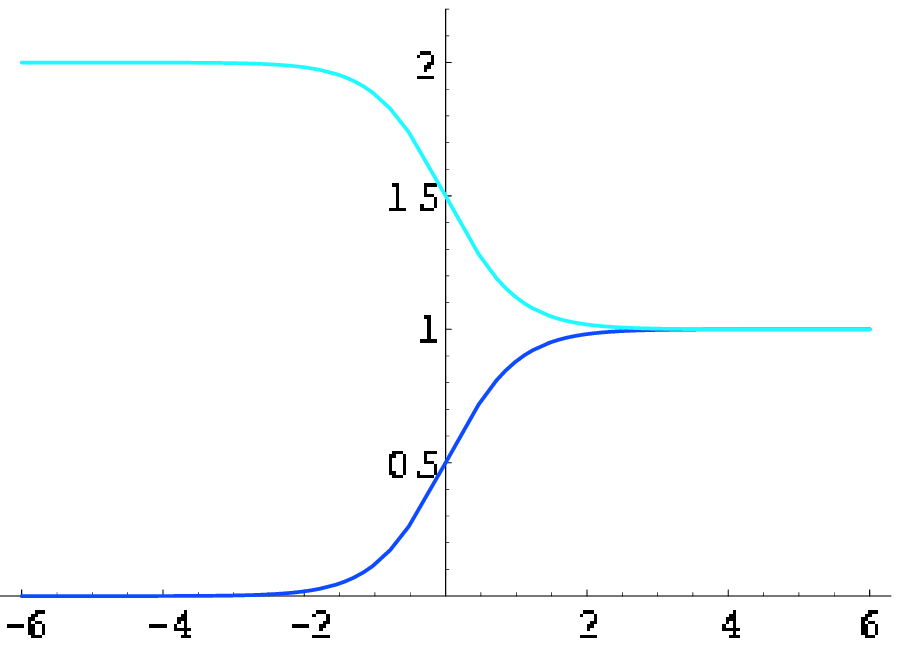}
\put(10,10){$y$}
\put(-150,95){$\Sigma ^3$}
\put(-150,20){$\Sigma ^0$}
\put(-90,-10){b)}
\end{center}
\caption{ Configurations for $\Sigma ^0,\Sigma ^3$ and $(H^1)^{22},(H^1)^{23}$, 
in the case of $(m_1,\,m_2,\,m_3)=(1,0,-1)$ and $r_1=0$.
}
\label{single-wall}
\end{figure}

Now let us construct exact solutions explicitly 
with the infinite gauge 
coupling by the formula for solutions (\ref{SS-H0}) 
and discuss the behavior of solutions. 
As our first example, 
let us start with solutions for the moduli matrices 
$H_{0 \langle 12\leftarrow 13\rangle }$ to confirm 
that $H_{0\langle 12\leftarrow 13\rangle }$ in fact 
gives a domain wall interpolating between 
SUSY vacua ${\langle 13\rangle}$ and ${\langle 12\rangle}$.  
Note that the moduli matrix
$H_0{}_{\langle 12\leftarrow 13\rangle }$ are
$U(1)$-factorizable and the $\Omega $ for 
above $H_0{}_{\langle 12\rangle}$
calculated by (\ref{SS-H0}) forms a diagonal matrix, 
thus we can easily find $S$ as
\begin{eqnarray}
S_{\langle 12\leftarrow 13 \rangle}&=&
\left( 
\begin{array}{cc}
e^{m_1y} & 0 \\
       0 & \sqrt{ e^{2m_2y} + e^{2m_3y+2{\rm Re}(r_1)}}
\end{array}
\right),
\end{eqnarray}
where we take the easiest gauge choice.
Therefore, from (\ref{sol-H}), we obtain the following single wall solution
\begin{eqnarray}
H^1=
\sqrt{c}
\left( 
\begin{array}{ccc}
1&0 &0\\
0&\frac{e^{m_2(y-y_1)}}
{\sqrt{e^{2m_2(y-y_1)}+e^{2m_3(y-y_1)}}} 
& \frac{e^{m_3(y-y_1)+i{\rm Im}(r_1)}}{\sqrt{e^{2m_2(y-y_1)}+e^{2m_3(y-y_1)}}} 
\end{array}
\right),\label{H0-13-12}
\end{eqnarray}
where ${\rm Im}(r_1)$ is a moduli with respect to broken $U(1)_{\rm F}$-phase.  
$\Sigma$ and $W_y$ can also be calculated from $S$ as
\begin{eqnarray}
\Sigma +iW_y=
\left( 
\begin{array}{cc}
m_1 & 0\\
0 & 
\frac{m_2e^{2m_2(y-y_1)}+m_3e^{2m_3(y-y_1)}}{e^{2m_2(y-y_1)}+e^{2m_3(y-y_1)}}
\end{array}
\right),
\end{eqnarray}
The components $\Sigma^0, \ \Sigma^3$ read 
\begin{eqnarray}
\Sigma^0&=&m_1+ \frac{m_2e^{2m_2(y-y_1)}+m_3e^{2m_3(y-y_1)}}{e^{2m_2(y-y_1)}+e^{2m_3(y-y_1)}},\nonumber\\
 \Sigma^3&=&m_1-\frac{m_2e^{2m_2(y-y_1)}+m_3e^{2m_3(y-y_1)}}{e^{2m_2(y-y_1)}+e^{2m_3(y-y_1)}},
\end{eqnarray}
while $\Sigma ^1,\,\Sigma ^2$ and $W_y$ vanish due to the $U(1)$-factorizability
of the moduli matrix $H_0{}_{\langle 13\rightarrow 12\rangle }$.
These wall solution for $\Sigma^0,\,\Sigma ^3$ and $H_0$ 
are illustrated in Fig.~\ref{single-wall}. 
From these solutions, we confirm that the parameter $y_1$ 
defined by (\ref{N3-positions}) is really the position 
of the wall in this 
case. 
The configuration approaches to the vacuum 
$\langle 13\rangle $ in the limit $y\rightarrow -\infty $ 
\begin{eqnarray}
H^1=
\sqrt{c}\left( 
\begin{array}{ccc}
1&0 &0\\
0&0 &1  
\end{array}
\right),\quad 
\Sigma =\left(
\begin{array}{cc}
   m_1 & 0\\ 0 &m_3
\end{array}
\right),
\end{eqnarray}
and to the vacuum $\langle 12\rangle $ 
in the limit $y\rightarrow \infty $ 
\begin{eqnarray}
 H^1=
\sqrt{c}\left( 
\begin{array}{ccc}
1&0 &0\\
0&1 &0  
\end{array}
\right),\quad 
\Sigma =\left(
\begin{array}{cc}
   m_1 & 0\\ 0 &m_2
\end{array}
\right).
\end{eqnarray} 
Moreover $\Sigma ^0+\Sigma ^3=2m_1$ implies that this one 
wall solution follows a straight line from 
${\langle 13\rangle}$ to ${\langle 12\rangle}$ 
in the $(\Sigma^0, \Sigma^3)$-plane when $y$ varies 
from $-\infty$ to $+\infty$, as shown in 
Fig.~\ref{3f-all-wall}.
Generally, we find that the configuration of 
the solution for single wall is a straight line 
segment linking two vacua in the 
$(\Sigma ^0,\,\Sigma ^3)$-plane with the gauge choice 
of $\Sigma ^1=\Sigma ^2=0$.

\begin{figure}[htb]
\includegraphics[width=6cm,clip]{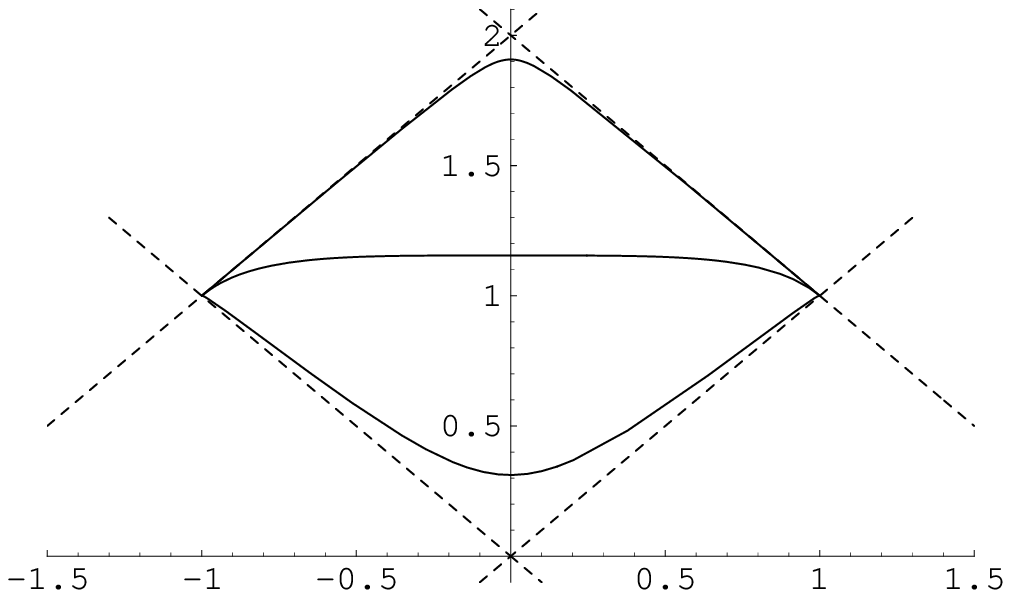}
\put(-90,-10){a)}
\put(-90,120){$\Sigma ^3$}
\put(5,5){$\Sigma ^0$}
\put(-65,85){$R=3$}
\put(-80,50){$R=0$}
\put(-60,20){$R=-3$}
\hspace{1cm}
\includegraphics[width=6cm,clip]{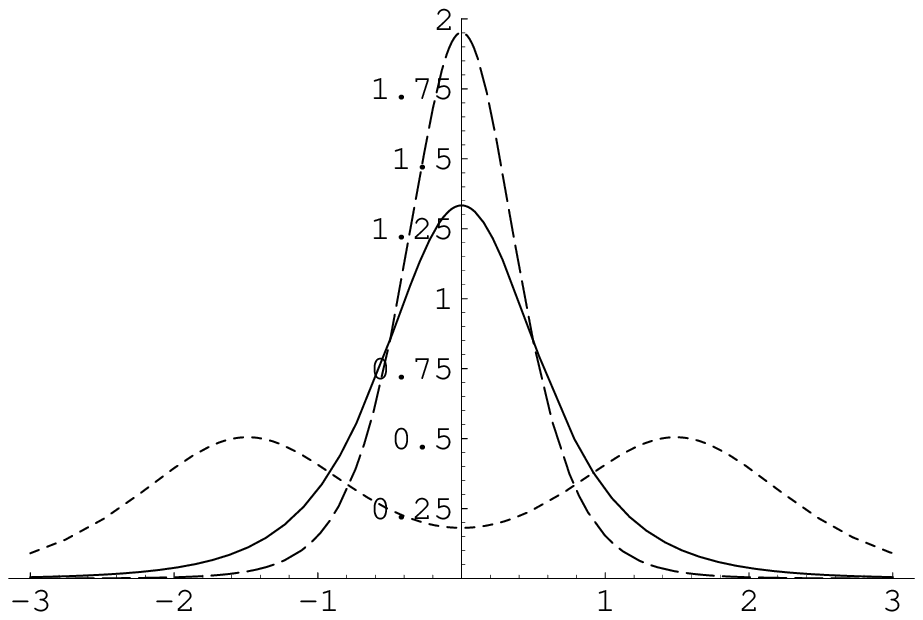}
\put(-90,-10){b)}
\put(-90,120){${\cal E}$}
\put(5,5){$y$}
\put(-40,90){:$R=3$ }\multiput(-70,93)(3,0){7}{\line(1,0){2}}
\put(-40,78){:$R=0$ }\multiput(-70,81)(1,0){1}{\line(1,0){20}}
\put(-40,66){:$R=-3$ }\multiput(-70,69)(6,0){4}{\line(1,0){5}}
\caption{Configuration for the double wall  with 
$m=1,$ $R=3,0,-3$ and $c=1$. 
Fig.~a) are paths of the solutions in the 
$(\Sigma ^0,\Sigma ^3)$-plane and Fig.~b)
are energy densities, 
${\cal E}=(c/2) \partial_y^2{\rm log }({\rm det}\Omega )$. 
Note that the trajectory for $R<0$ passes through 
the vicinity of the 
point $(\Sigma ^0,\Sigma ^3)=(0,0)$, 
which is not vacuum, while 
the trajectory for $R>0$ passes through 
the vicinity of the 
vacuum $\langle 13\rangle $.    
}
\label{double-sigma}
\end{figure}

The solution for the 
$H_0{}_{\langle 12\leftarrow 23\rangle}$ describing two 
walls can be obtained similarly.  
We are, however, faced with a little technical problem 
in this case.
 Substituting the explicit form of 
the moduli matrix (\ref{N3-double-moduli}) 
into the formula (\ref{SS-H0}),
 we obtain 
\begin{eqnarray}
\Omega =
\left( 
\begin{array}{cc}
 e^{2m_1y}+ e^{2m_2y+2{\rm Re}(r_2)} 
& e^{2m_2 y+r_2}  \\
  e^{2m_2 y+r_2^*} 
& e^{2my_2y} + e^{2m_3y+2{\rm Re}(r_1)} 
\end{array}
\right),
\end{eqnarray}
and hence, off-diagonal components appear. 
Therefore, we should consider the general case 
where $\Omega = SS^\dagger$ is given by
\begin{eqnarray}
\Omega = SS^\dagger =
\left( 
\begin{array}{cc}
\omega_+ & \omega_0^* \\
\omega_0 & \omega_-
\end{array}
\right),
\end{eqnarray}
with $\omega_\pm \in \mathbf{R}, \ \omega_0 \in \mathbf{C}$. 
Since the $U(2)$ gauge symmetry can be fixed by choosing 
$S$ to be a lower triangular matrix, 
we obtain an explicit form of the matrix $S$ 
\begin{eqnarray}
S&=&
\left( 
\begin{array}{cc}
\sqrt{\omega_+} & 0 \\
\frac{\omega_0}{\sqrt{\omega_+}} 
& \frac{\sqrt{\omega_+\omega_- - |\omega_0|^2} }{\sqrt{\omega_+}}
\end{array}
\right).
\end{eqnarray}
Although this gauge choice is appropriate to obtain an 
explicit form of the matrix 
$S$ from $\Omega $, it is not convenient to understand 
physics of walls since the off-diagonal 
part of both the vector multiplet scalars and the gauge fields 
are non-vanishing, $\Sigma ^{I}\not=0,\,W_y\not=0$ 
in this gauge.  
We can, however, calculate the gauge-invariant 
quantities without the above explicit form of 
the matrix $S$ as we explained in Sec.~\ref{TBPSEABC}. 
For simplicity, we set $M={\rm diag.}(m,\,0,\,-m),$ and 
$(r_1,r_2)=(mR/2+i\theta /2,-mR/2-i\theta /2)$, then 
the solutions for the scalar $\Sigma $ 
are obtained by the formula (\ref{eq:sigm-Omega}) and 
(\ref{eq:mag-sigma}) 
\begin{eqnarray}
\Sigma ^0&=&m{e^{2my}-e^{-2my}\over e^{2my}+e^{-2my}+e^{m R}}
\sim \left\{\begin{array}{cc}
   \pm m, & 2|y|\gg R, \quad y=\pm |y|\\
   0,& 2|y|\ll R
	  \end{array}\right.\nonumber\\
|\Sigma |&=&2m{\sqrt{(\cosh({2my})+e^{m R})^2-1}\over 
e^{2my}+e^{-2my}+e^{m R}}
\sim \left\{\begin{array}{cc}
   m, & 2|y|\gg R\\
   2m,& 2|y|\ll R
	  \end{array}\right.,
\end{eqnarray}
where $R$ represents the distance between the two walls 
if $R>0$. 
Configurations with several values of $R$ for this solution 
are illustrated in the $(\Sigma ^0,\Sigma ^3)$-plane 
with the gauge choice 
$\Sigma ^1=\Sigma ^2=0$ in Fig.~\ref{double-sigma}, 
whereas a configuration
corresponding to the 
$H_0{}_{\langle 12\leftarrow 32\rangle }$ 
is a straight line segment 
through the origin of the coordinate axis  from 
$\langle 32\rangle (\not=\langle 23\rangle )$ 
to $\langle 12\rangle $ in the same gauge $\Sigma ^1=\Sigma ^2=0$.  
The difference between profiles of these solutions 
can be understood as follows. 
If we allow local gauge transformations to eliminate 
the off-diagonal components $\Sigma ^1,\Sigma ^2$, 
we can rotate a vector $\vec \Sigma $ 
by $\pi $ around the $T^1$ axis 
only in a region of $y\ll  0$ 
so that the sign of $\Sigma ^3$ is flipped. 
This interpretation can be strengthened by examining 
the gauge invariant quantity $I_{\rm gauge}$ in Eq.~(\ref{Ig}).  
While $I_{\rm gauge}$ for single wall vanishes, it is nontrivial 
for the double wall 
\begin{eqnarray}
 I_{\rm gauge}&=&
4m^2e^{mR}{e^{2my}+e^{-2my}+e^{m R}\over 
((\cosh({2my})+e^{m R})^2-1)^2}
\sim \left\{\begin{array}{cc}
   16m^2e^{-m(6|y|-R)}, & 2|y|\gg R\\
   4m^2e^{-2m R},& 2|y|\ll R
	  \end{array}\right.
. 
\end{eqnarray}
\begin{figure}[htb]
\begin{center}
\includegraphics[width=6cm,clip]{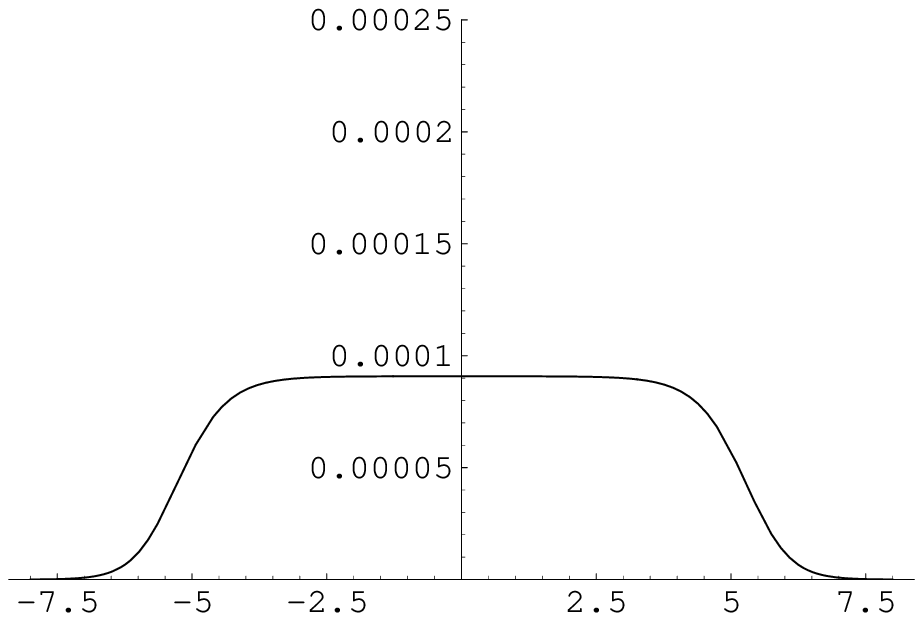}
\put(-110,-15){a)$R=10$}
\put(-90,110){$W_y^1$}
\put(0,5){$y$}
\hspace{1cm}
\includegraphics[width=6cm,clip]{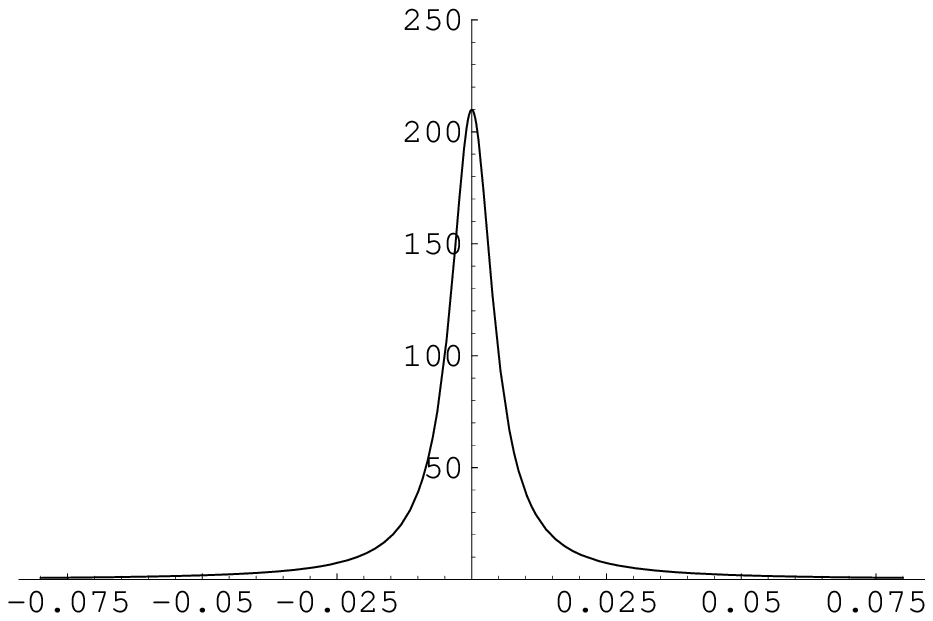}
\put(-110,-15){b)$R=-10$}
\put(-90,110){$W_y^1$}
\put(0,5){$y$}
\end{center}
\caption{Configuration of $\sqrt{I_{\rm gauge}}$ with $m=1$
$R=10,\,R=-10$. 
 If we take the gauge $\Sigma ^1=\Sigma ^2=0$ 
and set $\theta =0$, then 
$\sqrt{I_{\rm gauge}}=W_y^1$ and $W_y^2=0$.
Note that scales of the two figures are quite 
different. 
We find that a profile for $R\gg 0$ is a quite low 
plateau which
has edges on the double wall, and a profile for 
$R\ll 0$ approaches 
to a delta function on the compressed single wall.
 }  
\label{fig.Ig}
\end{figure}
In the limit of $R\rightarrow -\infty $, 
a profile of $\sqrt{I_{\rm gauge}}$ approaches to a 
delta function as illustrated in Fig.\ref{fig.Ig}
\begin{eqnarray}
\lim_{R\rightarrow -\infty }\sqrt{I_{\rm gauge}}
=\pi \delta (y), 
\label{eq:I-gauge-delta}
\end{eqnarray}
where the factor $\pi $ is obtained by integrating over 
the whole 
region of the coordinate $y$. 
Usually we use the gauge where $\Sigma^1=\Sigma^2=0, 
W_y^3=0$ 
unless otherwise stated. 
In that gauge the gauge invariant quantity is expressed 
in terms of gauge fields as in 
Eq.(\ref{eq:sum-gauge-field}). 
Since the gauge invariant quantity $\sqrt{I_{\rm gauge}}$ 
can then be interpreted as $W_y^1=\pi \delta (y)$, 
we can devise a local gauge transformation to eliminate 
the $W_y^1$. 
The gauge transformation which fixes the boundary condition 
at $y=\infty$ is given by a step function 
$\Lambda(y) = \pi \Theta (-y)$
\begin{eqnarray}
W_y^1 \rightarrow W_y^{1'}= W_y^1 + \partial_y \Lambda (y) =0. 
\end{eqnarray}
The resulting configuration turns out to be 
\begin{eqnarray}
\Sigma^{1'}=\Sigma^{2'}=0, \quad 
\Sigma^{3'}=\Sigma^3\epsilon(y) 
\quad W_y^{1'}=W_y^{2'}=W_y^{3'}=0
\end{eqnarray}
where $\epsilon(y)$ is a sign function. 
By this singular gauge transformation, 
a wall solution represented by a segment 
broken at $\Sigma^0 = |\Sigma| =0$ 
is transformed to a straight line segment 
which is generated by the third matrix in 
(\ref{N3-single-moduli}). 
The result (\ref{eq:I-gauge-delta}) 
appears to differ from the result $I_{\rm gauge}=0$ 
calculated from the moduli matrix 
$H_0{}_{\langle 12\leftarrow 32\rangle }$, in spite of 
the gauge-invariance of $I_{\rm gauge}$. 
This apparent discrepancy is due 
to the fact that $I_{\rm gauge }$ is ill-defined 
just at the point $|\Sigma |=0$. 

We summarize all the topological sectors and the associated 
moduli matrices in the case of $N_{\rm C}=2,
\,N_{\rm F}=3$ in Table \ref{N2N3-table}.

\begin{table}
\begin{center}
\caption{Six topological sectors in the case of $N_{\rm C}=2,
\,N_{\rm F}=3$. 
$a_1$ and $a_2$ are matrices 
which generate wall configuration as will be explained 
in the next subsection.
}
\label{N2N3-table} 
\begin{tabular}{|c|c|c|c|} \hline
top. sector&moduli matrix & dim.&objects\\ \hline\hline
$\langle 12\rangle \leftarrow \langle 12\rangle $
&$H_0{}_{\langle 12\rangle } $&0&vacuum $\langle 12\rangle $\\
$\langle 13\rangle \leftarrow \langle 13\rangle $
&$H_0{}_{\langle 13\rangle } $&0&vacuum $\langle 13\rangle $\\ 
$\langle 23\rangle \leftarrow \langle 23\rangle $
&$H_0{}_{\langle 23\rangle }$&0&vacuum $\langle 23\rangle $\\ 
$\langle 12\rangle \leftarrow \langle 13\rangle $
&$H_0{}_{\langle 12\leftarrow 13\rangle }$&2
& elementary wall $a_2$\\
$\langle 13\rangle \leftarrow \langle 23\rangle $
&$H_0{}_{\langle 13\rightarrow 23\rangle }$&2
& elementary wall $a_1$\\
$\langle 12\rangle \leftarrow \langle 23\rangle $
&$H_0{}_{\langle 12\rightarrow 23\rangle }
\oplus H_0{}_{\langle 12\rightarrow 32\rangle }$
&4& double wall $(a_1,a_2)\oplus $ 
compressed wall $a_1 a_2$  \\\hline
\end{tabular}
\end{center}
\end{table}

\subsection{$N_{\rm C}=2, \ N_{\rm F}=4$ Case}
\label{NC2NF4Case}
The $N_{\rm C}=2, \ N_{\rm F}=4$ case 
is the simplest example 
containing characteristic properties 
originated from a non-Abelian gauge group. 
In this case, 
there are six SUSY vacua, and
six elementary walls interpolating between these vacua. 
There exist 20 BPS topological sectors described by 
25 kinds of moduli matrices in the standard form, 
which we show 
explicitly in Appendix~\ref{STF}. 
Note that if we choose an arbitrary set
of vacua at both boundaries, we find 
that there are 21 topological
sectors, that is, there exists one non-BPS 
topological sector, which 
interpolates between vacua $\langle 14\rangle $ and 
$\langle 23\rangle $. 
If we consider 
the maximal topological sector interpolating between vacua 
$\langle 12\rangle $ and $\langle 34\rangle $, 
the moduli space is described by four complex
moduli parameters, of which four real parameters 
represent positions of four 
walls, and other four real parameters represent 
the orientation of walls in the target space. 
Among them, relative phase of vacua separated by the 
wall can be understood as the Nambu-Goldstone mode. 
We will also obtain 
one moduli parameter 
which cannot be attributed to the spontaneously broken 
symmetry, namely the quasi-Nambu-Goldstone mode.

\begin{figure} 
\begin{center}
\includegraphics[width=9cm,clip]{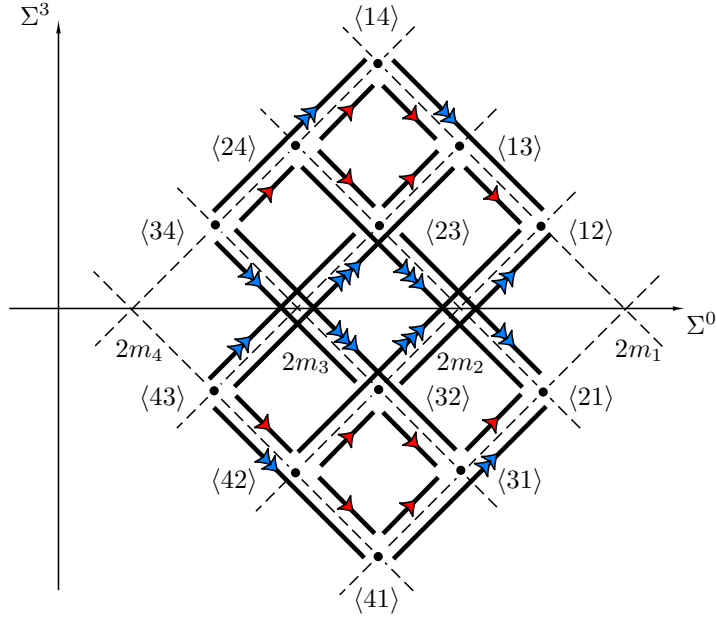}
\footnotesize
\put(-140,240){$\langle 14 \rangle$}
\put(-80,185){$\langle 13 \rangle$}
\put(-200,185){$\langle 24 \rangle$}
\put(-230,150){$\langle 34 \rangle$}
\put(-110,150){$\langle 23 \rangle$}
\put(-50,150){$\langle 12 \rangle$}
\put(-240,100){$ 2m_4 $}
\put(-170,95){$ 2m_3 $}
\put(-105,95){$ 2m_2 $}
\put(-30,100){$ 2m_1 $}
\put(-230,80){$\langle 43 \rangle$}
\put(-110,80){$\langle 32 \rangle$}
\put(-50,80){$\langle 21 \rangle$}
\put(-80,45){$\langle 31 \rangle$}
\put(-200,45){$\langle 42 \rangle$}
\put(-140,-5){$\langle 41 \rangle$}
\put(-0,110){$\Sigma^0$}
\put(-280,240){$\Sigma^3$}
\normalsize
\vspace{1cm}
\caption{All single walls for $N_{\rm C}=2$ and $N_{\rm F}=4$.
Lines with a single arrow are 
elementary walls, whereas lines with a double  (triple) 
arrow are compressed single walls of level one (two). 
}
\label{f4-single}
\end{center}
\end{figure}

\begin{figure}[htb]
\begin{center}
\includegraphics[width=7cm,clip]{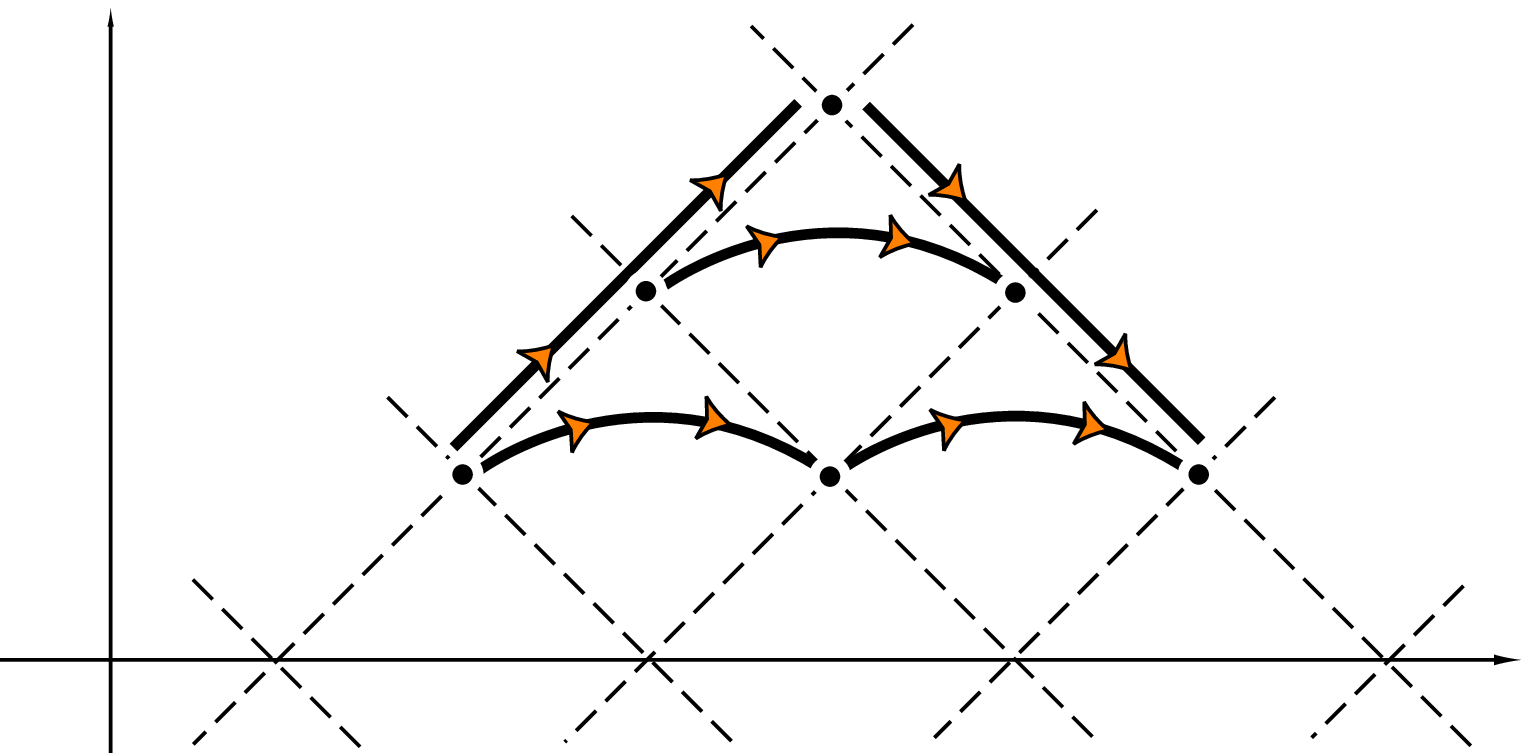}
\put(-100,100){$\langle 14 \rangle$}
\put(-30,35){$\langle 12 \rangle$}
\put(-148,60){$\langle 24 \rangle$}
\put(-55,60){$\langle 13 \rangle$}
\put(-170,35){$\langle 34 \rangle$}
\put(-100,20){$\langle 23 \rangle$}
\put(-170,-10){$ 2m_4 $}
\put(-120,-10){$ 2m_3 $}
\put(-73,-10){$ 2m_2 $}
\put(-25,-10){$ 2m_1 $}
\put(0,0){$\Sigma^0$}
\put(-205,90){$\Sigma^3$}
\put(-90, -30){a)}
\hspace{1cm}
\includegraphics[width=7cm,clip]{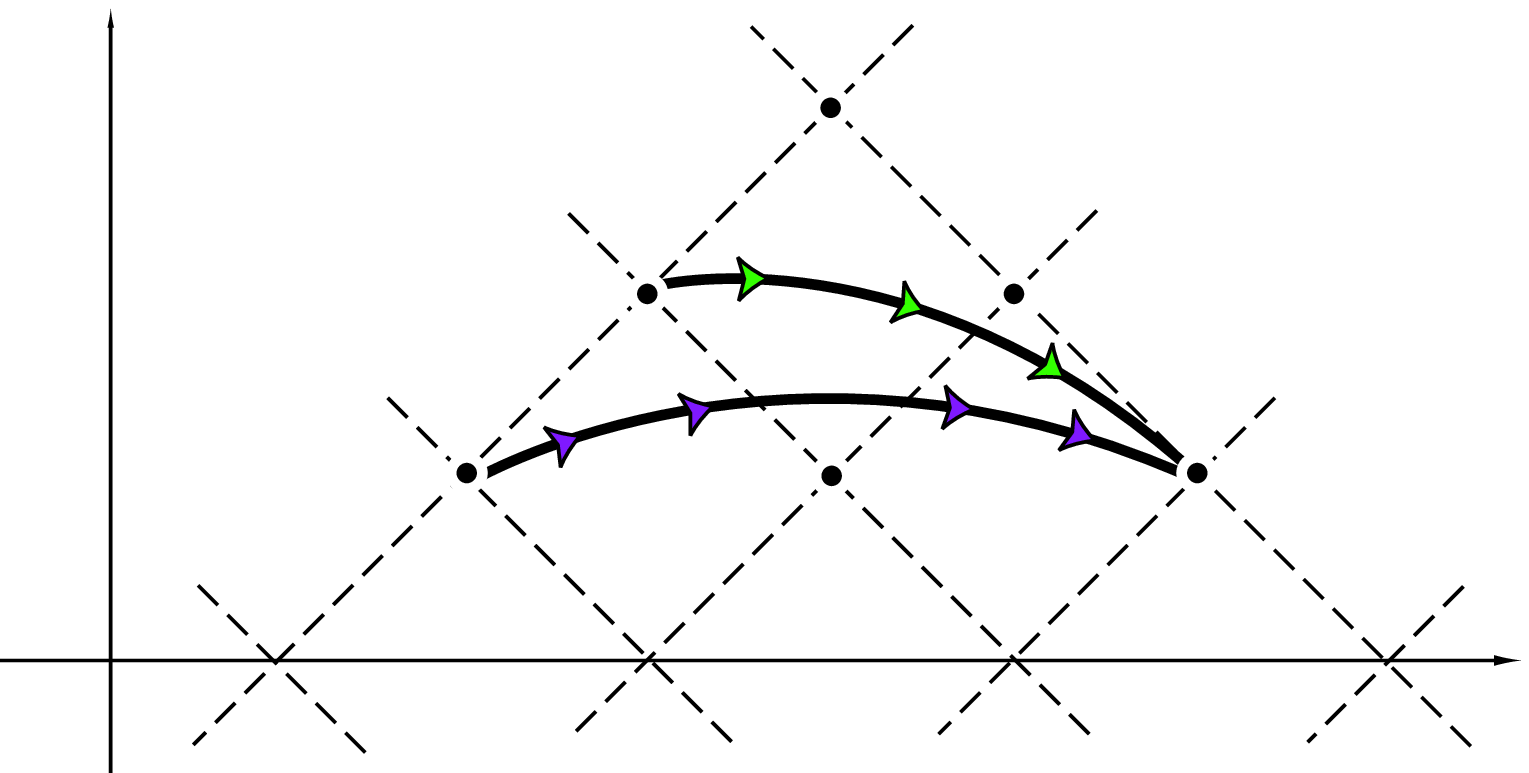}
\put(-100,100){$\langle 14 \rangle$}
\put(-30,35){$\langle 12 \rangle$}
\put(-148,60){$\langle 24 \rangle$}
\put(-55,60){$\langle 13 \rangle$}
\put(-170,35){$\langle 34 \rangle$}
\put(-100,20){$\langle 23 \rangle$}
\put(-170,-10){$ 2m_4 $}
\put(-120,-10){$ 2m_3 $}
\put(-73,-10){$ 2m_2 $}
\put(-25,-10){$ 2m_1 $}
\put(0,0){$\Sigma^0$}
\put(-205,90){$\Sigma^3$}
\put(-90,-30){b)}
\end{center}
\caption{Multi-walls. 
a) All double wall configurations composed of two elementary walls
are displayed.
b) Examples of triple-wall and the four-wall configurations 
in the maximal topological sector are displayed.
}
\label{multi-walls}
\end{figure}

All single walls including elementary walls and compressed walls 
are displayed in Fig.~\ref{f4-single}.
\unitlength 0.9pt
\unitlength 1pt
Multi-wall solutions are displayed 
in Fig.~\ref{multi-walls}.\footnote{
In Fig.~\ref{multi-walls}-a), 
five double-wall configurations are drawn. 
However two of them $\langle 14 \leftarrow 34\rangle$ 
and  $\langle 12 \leftarrow 14\rangle$ are 
straight lines whose position moduli parameters 
are not visible in this figure. 
This is because we have displayed only 
the configuration projected to 
the $\Sigma$-space, while 
the full configuration space is larger. 
The wall $\langle 14 \leftarrow 34\rangle$ 
($\langle 12 \leftarrow 14\rangle$) does not go 
through the $\langle 24 \rangle$ ($\langle 13 \rangle$) 
vacuum, as can be seen by the full configuration 
besides the  $\Sigma$-space. 
}

\unitlength 1pt

\begin{figure}[htb]
\begin{center}
\includegraphics[width=3.5cm,clip]{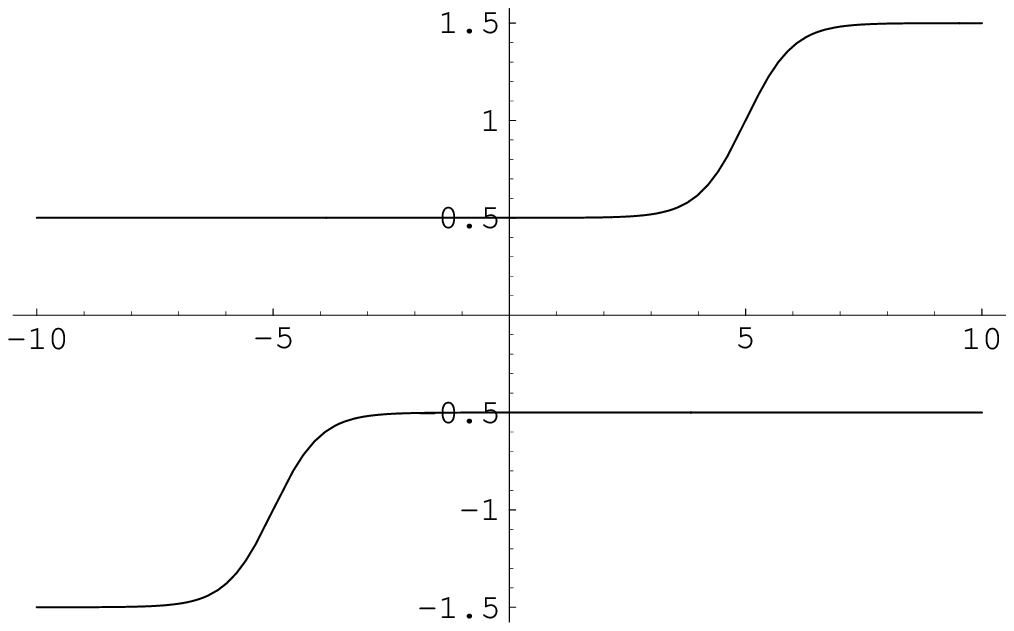}
\put(5,60){$y$}
\put(-200,90){\footnotesize  $\Sigma ^0+\Sigma ^3$}
\put(-60,20){\footnotesize $\Sigma ^0-\Sigma ^3$}
\put(-90, -30){a)}
\hspace{2em}
\includegraphics[width=3.5cm,clip]{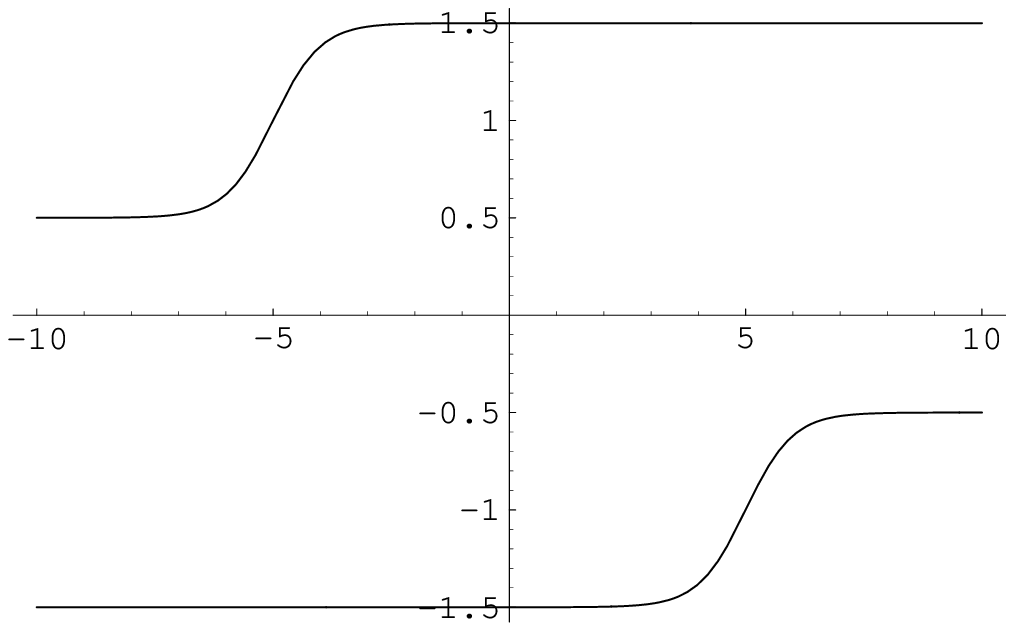}
\put(5,60){$y$}
\put(-60,100){\footnotesize$\Sigma ^0+\Sigma ^3$}
\put(-200,10){\footnotesize$\Sigma ^0-\Sigma ^3$}
\put(-90,-30){b)}\\
\vspace{2em}
\includegraphics[width=3.5cm,clip]{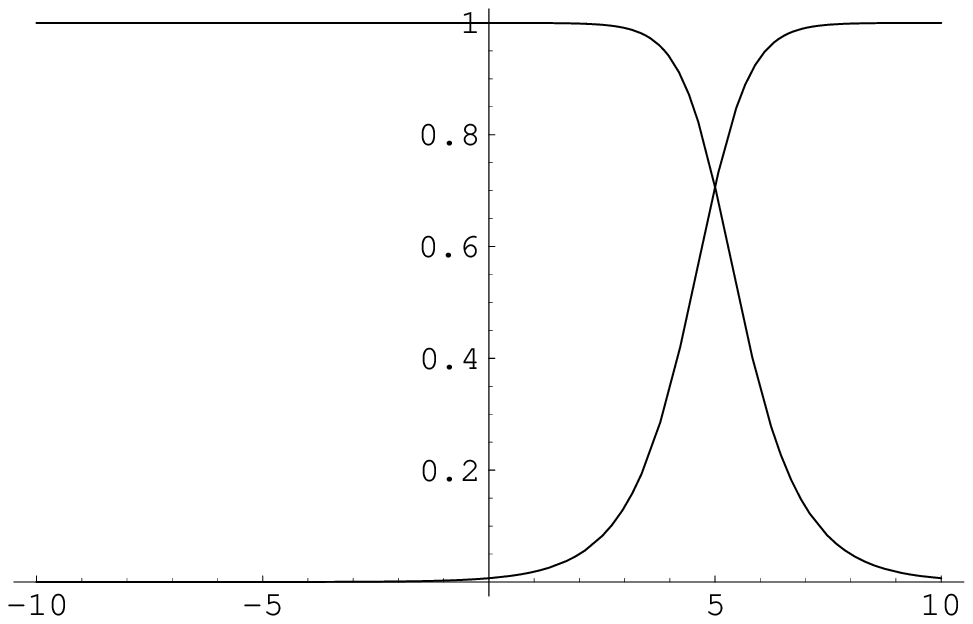}
\put(-90,-30){a)}
\put(-0,10){$y$}
\put(-200,95){\footnotesize$(H^1)^{12}$}
\put(-40,95){\footnotesize$(H^1)^{11}$}
\hspace{2em}
\includegraphics[width=3.5cm,clip]{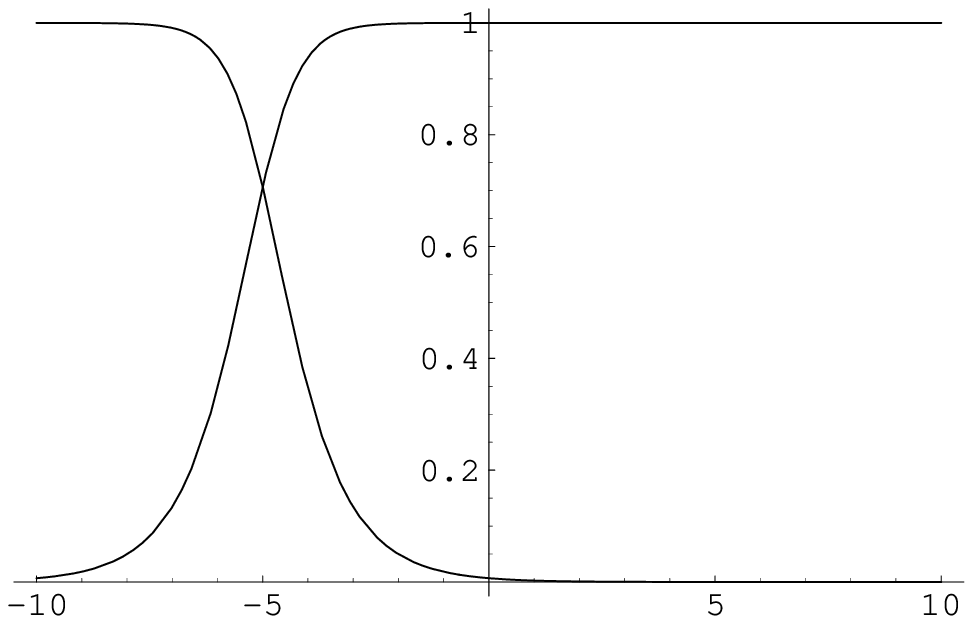}
\put(-90,-30){b)}
\put(0,10){$y$}
\put(-215,95){\footnotesize$(H^1)^{12}$}
\put(-40,95){\footnotesize$(H^1)^{11}$}\\
\vspace{2em}
\includegraphics[width=3.5cm,clip]{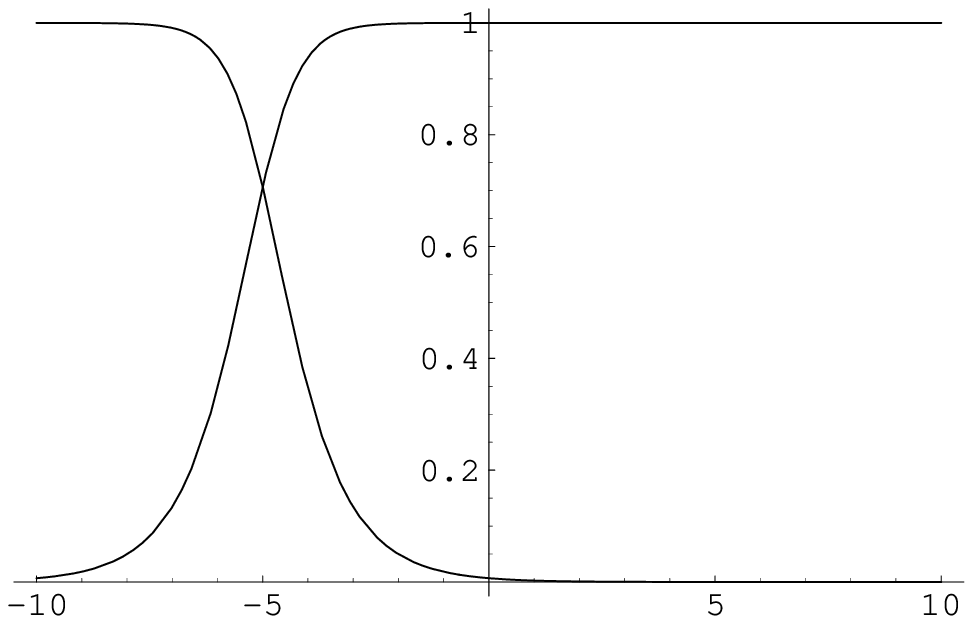}
\put(-90,-30){a)}
\put(-0,10){$y$}
\put(-215,95){\footnotesize$(H^1)^{24}$}
\put(-40,95){\footnotesize$(H^1)^{23}$}
\hspace{2em}
\includegraphics[width=3.5cm,clip]{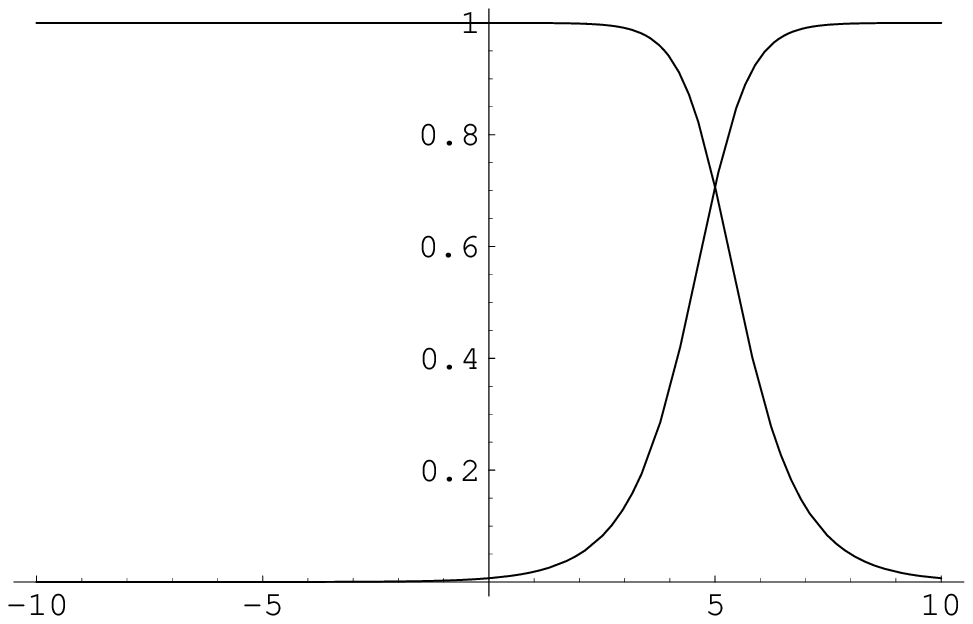}
\put(-90,-30){b)}
\put(0,10){$y$}
\put(-200,95){\footnotesize$(H^1)^{24}$}
\put(-40,95){\footnotesize$(H^1)^{23}$}
\end{center}
\caption{
Penetrable walls of $\langle 13 \leftarrow 24 \rangle$ in $N_{\rm C}=2$, 
$N_{\rm F}=4$ case. 
$M={\rm diag.}({3\over 2},{1\over 2},-{1\over 2},-{3\over 2})$. 
a)
 $r_3=5,\,r_2=-5$: 
the intermediate region shows $\langle 23 \rangle$ vacuum. 
b) $r_3=-5,\,r_2=5$, $c=1$: 
the intermediate region shows $\langle 14 \rangle$ vacuum. 
}
\label{commute-walls}
\end{figure}

A remarkable phenomenon in this case is that there exists 
a pair of walls whose positions can commute with each other. 
Let us show this property, considering the 
topological sector labeled by 
$\langle 13\rangle \leftarrow \langle 24\rangle $. 
A moduli matrix for this sector 
$H_{0\langle 13\leftarrow 24\rangle}$ 
is given by two complex moduli 
parameters $r_2,r_3$ by 
\begin{eqnarray}
H_{0\langle 13\leftarrow 24\rangle}=
\sqrt{c}\left( 
\begin{array}{cccc}
1&e^{r_3} &0 &0\\
0&0 & 1& e^{r_2} 
\end{array}
\right).
\end{eqnarray} 
We notice that this moduli matrix is 
$U(1)$-factorizable, and 
a solution obtained with this moduli matrix 
is given by 
\begin{eqnarray}
 \Sigma ^0+\Sigma ^3&=&{m_1e^{2m_1y}+m_2e^{2m_2y+2{\rm Re}(r_3)}
\over e^{2m_1y}+e^{2m_2y+2{\rm Re}(r_3)}},\nonumber\\
 \Sigma ^0-\Sigma ^3&=&{m_3e^{2m_3y}+m_4e^{2m_4y+2{\rm Re}(r_2)}
\over e^{2m_3y}+e^{2m_4y+2{\rm Re}(r_2)}}
\end{eqnarray}
and $\Sigma ^1=\Sigma ^2=W_y=0$, and
 \begin{eqnarray}
 H^1=
 \sqrt{c}\left( 
 \begin{array}{cccc}
 {e^{m_1y}\over \sqrt{e^{2m_1y}+e^{2m_2y+2{\rm Re}(r_3)}}} & 
 {e^{m_2y+r_3}\over \sqrt{e^{2m_1y}+e^{2m_2y+2{\rm Re}(r_3)}}} &0 &0\\
 0&0 & {e^{m_3y}\over \sqrt{e^{2m_3y}+e^{2m_4y+2{\rm Re}(r_2)}}} &{e^{m_4y+r_2}\over \sqrt{e^{2m_3y}+e^{2m_4y+2{\rm Re}(r_2)}}} 
 \end{array}
 \right).\nonumber \\
\end{eqnarray}
Profiles of $\Sigma ^0\pm \Sigma ^3$ and $H^1$ 
are illustrated in Fig.~\ref{commute-walls}. 
\unitlength 0.5pt
\unitlength 1pt
This solution describes a configuration of double wall. 
In this case, position of the walls are exactly expressed by 
\begin{eqnarray}
y_3= {{\rm Re}(r_3)\over m_1-m_2},\quad y_2={{\rm Re}(r_2)\over m_3-m_4}.
\end{eqnarray} 
In a region $y_2>y_3$, the configuration 
interpolates between vacua $\langle 13\rangle $
and $\langle 24\rangle $ 
through the vicinity of the vacuum $\langle 14\rangle $. 
The wall at $y_2$ ($y_3$) interpolates between 
vacua $\langle 13\rangle $ and $\langle 14\rangle $ 
($\langle 24\rangle $ and $\langle 14\rangle $). 
On the other hand, in a region $y_3>y_2$ 
the wall at $y_3$ ($y_2$) interpolates 
as $\langle 13\rangle \leftarrow \langle 23\rangle $ 
($\langle 23\rangle \leftarrow \langle 24\rangle $). 
Note that  while 
intermediate vacua are different from one region $(y_3 > y_2)$
 to the other $(y_3 < y_2)$, 
the two parameters $y_2,y_3$ retain the physical meaning as 
positions of the walls, in contrast to the example of 
the compressed wall in the last subsection.  
The wall represented 
by the position $y_2$ changes 
flavors of non-vanishing hypermultiplet scalar 
from 4 to 3 and changes the value of $\Sigma ^0-\Sigma ^3$, 
while the wall at $y_3$ changes flavors from 2 to 1 and 
changes $\Sigma ^0+\Sigma ^3$.  
Thus, it is quite natural to identify the wall 
represented by the same parameter,  
although interpolated vacua are
different. 
With this identification, we interpret the above
configuration as two walls commuting with each other.

In this and more generic cases with larger flavors and colors,
we find many pairs of walls which commute with 
each other. 
On the other hand, there are 
also pairs of walls which do not commute with each other. 
These pairs are compressed to a single wall, if their relative 
distance go to the infinity to the direction 
where they do not commute. 
Here we propose an algebraic method to distinguish 
whether two walls commute with each other or are compressed. 
Our goal is to express the single walls as creation 
operators on the (Fock) space of vacua.  
To this end it is useful to define nilpotent matrices in the Lie algebra 
${\cal SU}(N_{\rm F})$ by 
$(E_{i,j})_{kl} = \delta_{ik}\delta_{jl}$ ($1<i<j<N_{\rm F}$). 
The matrices $a_i \equiv E_{i,i+1}$ generate elementary wall as seen below.
In our $N_{\rm F}=4$ case, they are
$a_1,a_2,a_3$ given by 
\begin{eqnarray}
 a_1 = E_{12} = \left(\begin{array}{cccc}
  0&1&0&0 \\0&0&0&0\\0&0&0&0\\0&0&0&0
	    \end{array}\right),\ 
 a _2=E_{23} =\left(\begin{array}{cccc}
  0&0&0&0 \\0&0&1&0\\0&0&0&0\\0&0&0&0
	    \end{array}\right), \ 
 a_3 = E_{34} = \left(\begin{array}{cccc}
  0&0&0&0 \\0&0&0&0\\0&0&0&1\\0&0&0&0
	    \end{array}\right).
\end{eqnarray}
We also define $a_i (r), E_{i,j}(r)$ by
\begin{equation}
 a _i(r)\equiv e^r a _i ,\quad E_{i,j}(r)\equiv e^rE_{i,j}.
\label{eq:operator-Heisenberg-rep}
\end{equation}
Then they act on the moduli matrices from the right like 
\begin{eqnarray}
  H_0{}_{\langle 12\rangle }e^{a _2(r_1)}
&=&H_0{}_{\langle 12\rangle }({\rm 1}_3+a _2(r_1))
=\sqrt{c}\left(
\begin{array}{cccc}
	1  &0&0&0 \\0&1&0&0
 \end{array}\right)
\left(\begin{array}{cccc}
  1&0&0&0 \\0&1&e^{r_1}&0\\0&0&1&0\\0&0&0&1
	    \end{array}\right)\nonumber\\
&=&\sqrt{c}\left(\begin{array}{cccc}
  1&0&0&0 \\0&1&e^{r_1}&0
	    \end{array}\right)
=H_0{}_{\langle 12\leftarrow 13\rangle }.
\end{eqnarray}
Similarly we find
\begin{eqnarray}
 H_0{}_{\langle 13\rangle }e^{a _3(r_2)}
&=&H_0{}_{\langle 13\leftarrow 14\rangle },\quad
 H_0{}_{\langle 13\rangle }e^{a _1(r_3)}
=H_0{}_{\langle 13\leftarrow 23\rangle },\nonumber\\ 
 H_0{}_{\langle 14\rangle }e^{a _1(r_3)}
&=&H_0{}_{\langle 14\leftarrow 24\rangle }, \quad 
 H_0{}_{\langle 23\rangle }e^{a _3(r_2)}
=H_0{}_{\langle 23\leftarrow 24\rangle },\nonumber\\
 H_0{}_{\langle 24\rangle }e^{a _2(r_4)}
&=&H_0{}_{\langle 24\leftarrow 34\rangle }.
\end{eqnarray}
In other cases, $e^{a _i(r)}$ acts on moduli 
matrices for vacua as the identity operator 
up to the world-volume symmetry. 
Thus 
these matrices $e^{a _i(r)}$ 
can be interpreted as {\it operators 
generating elementary walls}. 
The elementary wall defined in Sec.~\ref{sc:moduli-matrix} 
changes the flavor by one unit $i\leftarrow i+1 $ 
in the same color component and carries the tension 
$T_{\langle i\leftarrow i+1\rangle }$. 
This elementary wall is 
realized by the matrix $e^{a _i(r)}$ 
which we call as ``elementary-wall operator''.
Interestingly, the mass matrix $c M$ can be 
interpreted as Hamiltonian 
for elementary walls 
\begin{eqnarray}
 [c\,M,a _i]
&=&c(m_i-m_{i+1})a _i
=T_{\langle i\leftarrow i+1\rangle }a _i.
\end{eqnarray}
Only the following matrices are generated from commutators 
of the matrices $a_1,a_2,a_3$ 
\begin{eqnarray}
 E_{13}=a _1a _2=[E_{12},E_{23}],
\quad E_{24}= a _2a _3=[E_{23},E_{34}],
\end{eqnarray}
which generate level-$1$ compressed single walls made
 by compressing two elementary walls, for instance 
\begin{eqnarray}
  H_0{}_{\langle 12\rangle }e^{E_{13}(r_5)}
&=&\sqrt{c}\left(\begin{array}{cccc}
  1&0&e^{r_5}&0 \\0&1&0&0
	    \end{array}\right)
=H_0{}_{\langle 12\leftarrow 32\rangle }.
\end{eqnarray} 
By further taking commutators including these 
new operators corresponding to compressed walls, 
we obtain a new operator 
\begin{eqnarray}
\quad E_{14}= a _1a _2a _3
=[E_{12},E_{24}]=[E_{13},E_{34}], 
\end{eqnarray}
which generates a level-$2$ compressed single wall made by compressing 
an elementary wall with a level-$1$ compressed wall. 

Let us work out how to 
express arbitrary multi-wall states 
by these operators.
An arbitrary complex upper triangular 
matrix ${\cal V}$ can be written by means of 
the matrices $E_{i,j}$ 
\begin{eqnarray}
 {\cal V}\equiv {\bf 1}_{N_{\rm F}}+\sum_{i>j}f^{i,j}E_{i,j}
=\prod_\alpha \exp\left({E_{i_\alpha,j_\alpha}(r_\alpha)}\right) 
\end{eqnarray} 
with complex parameters $f^{i,j}$ and $r_\alpha$. 
We thus find that an arbitrary  
moduli matrix in the row-reduced echelon form (\ref{RREF}) 
can be constructed by multiplying this upper triangular 
matrix to the moduli matrices representing the vacua. 
However the moduli matrix in the row-reduced echelon form 
do not distinguish vacua at $y=-\infty$ 
and include several topological sectors in one matrix 
as was explained in Sec.~\ref{topo.sectors}. 
Therefore we have to throw away some parameters to obtain 
matrices describing a single topological sector.
We propose an alternative method, which we call the operator method, 
to construct moduli matrices for 
a multi-wall by multiplying $e^{a_i(r)}$ 
from the right of moduli matrices for less walls by one. 
We will see that this operator method provides 
efficiently multi-wall solutions, 
but unfortunately they do not in general coincide with 
the moduli matrices in 
the standard form as seen below.\footnote{ 
For our $N_{\rm F}=4$ case, the difference of the 
parametrizations for four walls 
in the operator method and in 
the standard form can be recognized by 
comparing Eq.~(\ref{4-wall-matrix}) and 
the first matrix in Eq.~(\ref{4-wall-st}).
}

Moduli matrices for a double wall 
can be constructed by multiplying the 
operator for a single wall to the moduli matrix for a single 
wall, for instance 
\begin{eqnarray}
H_0{}_{\langle 13\leftarrow 14\rangle }e^{a _1(r_3)}
&=&\sqrt{c}\left(\begin{array}{cccc}
  1&e^{r_3}&0&0 \\0&0&1&e^{r_2}
	    \end{array}\right)
=H_0{}_{\langle 13\leftarrow 24\rangle },\nonumber\\
H_0{}_{\langle 12\leftarrow 13\rangle }e^{a _1(r_3)}
&=&\sqrt{c}\left(\begin{array}{cccc}
  1&e^{r_3}&0&0 \\0&1&e^{r_1}&0
	    \end{array}\right)
=H_0{}_{\langle 12\leftarrow 23\rangle }.
\end{eqnarray}
As we observed, a pair of walls is either 
penetrable or impenetrable each other. 
The double wall constructed by these wall operators 
reproduce this distinction and facilitate 
its understanding as follows. 
Noting that $[a _1,\,a _3]=0$, we obtain 
\begin{eqnarray}
 H_0{}_{\langle 13\leftarrow 14\rangle }e^{a _1(r_3)}&=&
H_0{}_{\langle 13\rangle }e^{a _3(r_2)}e^{a _1(r_3)}=
H_0{}_{\langle 13\rangle }e^{a _1(r_3)}e^{a _3(r_2)}
=H_0{}_{\langle 13\leftarrow 23\rangle }e^{a _3(r_2)}.
\end{eqnarray}
On the other hand, since 
$[a _1,\,a _2]=E_{13}\not =0$, we obtain 
\begin{eqnarray}
&&H_0{}_{\langle 12\rangle }e^{a _2(r_1)}e^{a _1(r_3)}
=H_0{}_{\langle 12\rangle }e^{a _1(r_3)}
e^{a _2(r_1)-[a _1(r_3),\,a _2(r_1)]}
\nonumber\\
&&\simeq H_0{}_{\langle 12\rangle }
e^{a _2(r_1)-[a _1(r_3),\,a _2(r_1)]}
= H_0{}_{\langle 12\rangle }
e^{a _2(r_1)-E_{13}(r_1+r_3)},
 \label{mat.}
\end{eqnarray}
where we used the world-volume symmetry as
\begin{eqnarray}
 H_0{}_{\langle 12\rangle }e^{a _1(r)}=\left(
\begin{array}{cc}
1&e^{r} \\0&1\end{array}\right)H_0{}_{\langle 12\rangle }
\simeq H_0{}_{\langle 12\rangle }.
\end{eqnarray}
In the limit $r_1\rightarrow -\infty$ with 
$r' \equiv r_1+r_3+\pi i$ fixed, 
the matrix in Eq.~(\ref{mat.}) approaches to a limit 
\begin{eqnarray}
H_0{}_{\langle 12\rangle }e^{a _2(r_1)}e^{a _1(r_3)}
\rightarrow  H_0{}_{\langle 12\rangle }
e^{E_{13} (r')}=H_0{}_{\langle 12\leftarrow 32\rangle }.
\end{eqnarray}
Therefore, we find the remarkable fact: 

{\it Two walls are penetrable 
each other, if they are generated by 
operators $E_{i,j},\,E_{k,l}$ which are commutative 
$[E_{i,j},E_{k,l}]=0$. 
} 
{\it If two operators are non-commutative each other, 
two walls are impenetrable, and are compressed 
to a single wall generated by the single operator 
$E_{i,l}=[E_{i,j},\,E_{k,l}]$, with $j=k$ not summed.
} 

\medskip
In the end, we obtain a moduli matrix for four walls 
interpolating between the
vacuum $\langle 12\rangle $ and the 
vacuum $\langle 24\rangle $ 
in the maximal topological sector 
by the operator formalism as
\begin{eqnarray}
 H_0{}_{\langle 12\leftarrow 34\rangle }
&=&H_0{}_{\langle 12\rangle }e^{a _2(r_1)}
e^{a _3(r_2)
+a _1(r_3)}e^{a _2(r_4)}=\sqrt{c}\left(
\begin{array}{cccc}
1  & e^{r_3}&e^{r_3+r_4} &0 \\0&1&e^{r_1}+e^{r_4}&e^{r_1+r_2} 
\end{array}\right). 
 \label{4-wall-matrix}
\end{eqnarray}
This is a generic moduli matrix in 
the maximal topological sector. 
Note that the (2,3) element differs from that 
of the corresponding 
matrix in the standard form given in Appendix~\ref{STF}.

\begin{figure}[htb]
\begin{center}
\includegraphics[width=7cm,clip]{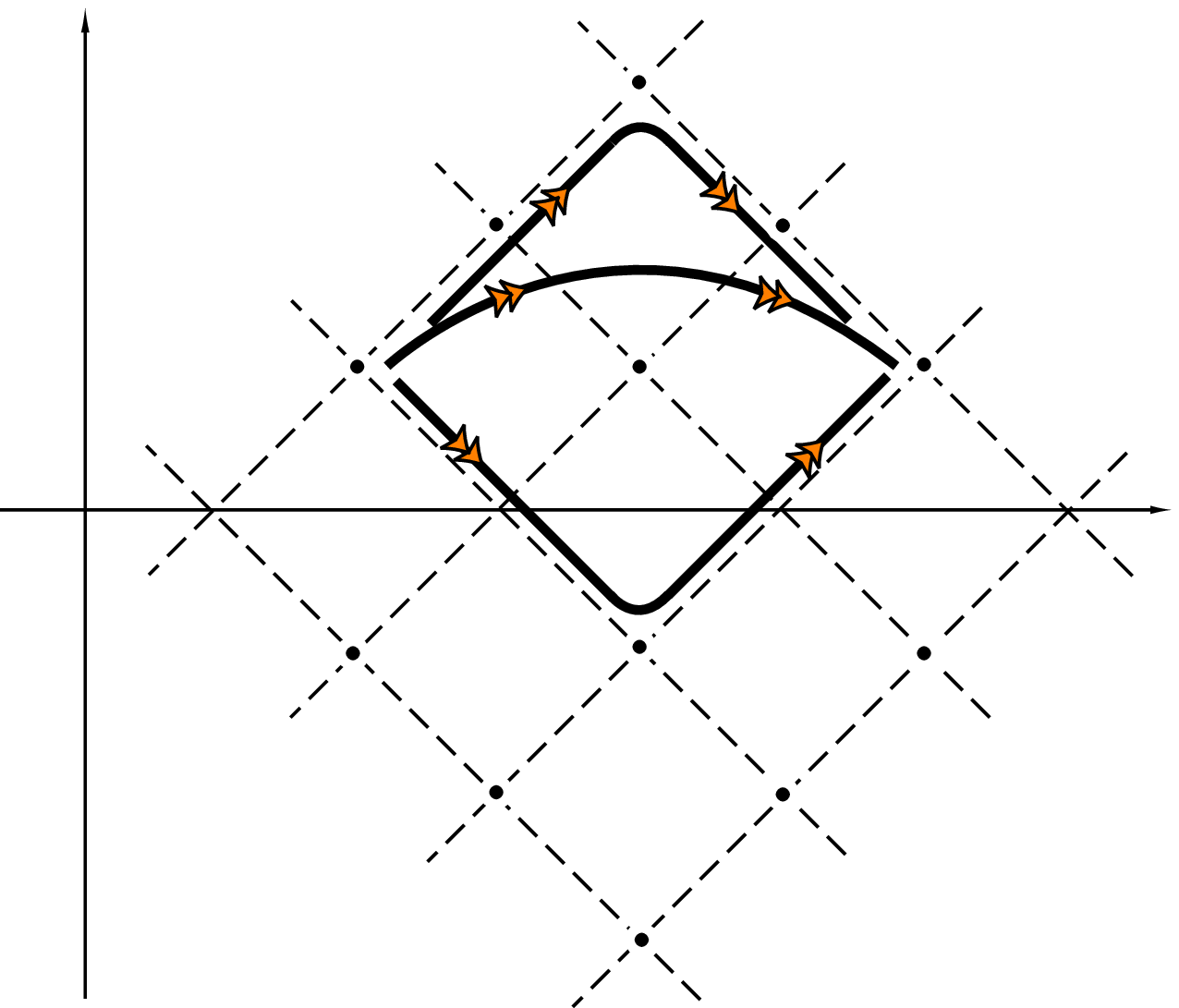}
\footnotesize
\put(-100,170){$\langle 14 \rangle$}
\put(-57,132){$\langle 13 \rangle$}
\put(-142,132){$\langle 24 \rangle$}
\put(-164,107){$\langle 34 \rangle$}
\put(-82,107){$\langle 23 \rangle$}
\put(-36,107){$\langle 12 \rangle$}
\put(-171,69){$ 2m_4 $}
\put(-123,69){$ 2m_3 $}
\put(-73,69){$ 2m_2 $}
\put(-82,57){$\langle 32 \rangle$}
\normalsize
\put(-0,78){$\Sigma^0$}
\put(-199,170){$\Sigma^3$}
\put(-90, -20){a)}
\hspace{1cm}
\includegraphics[width=7cm,clip]{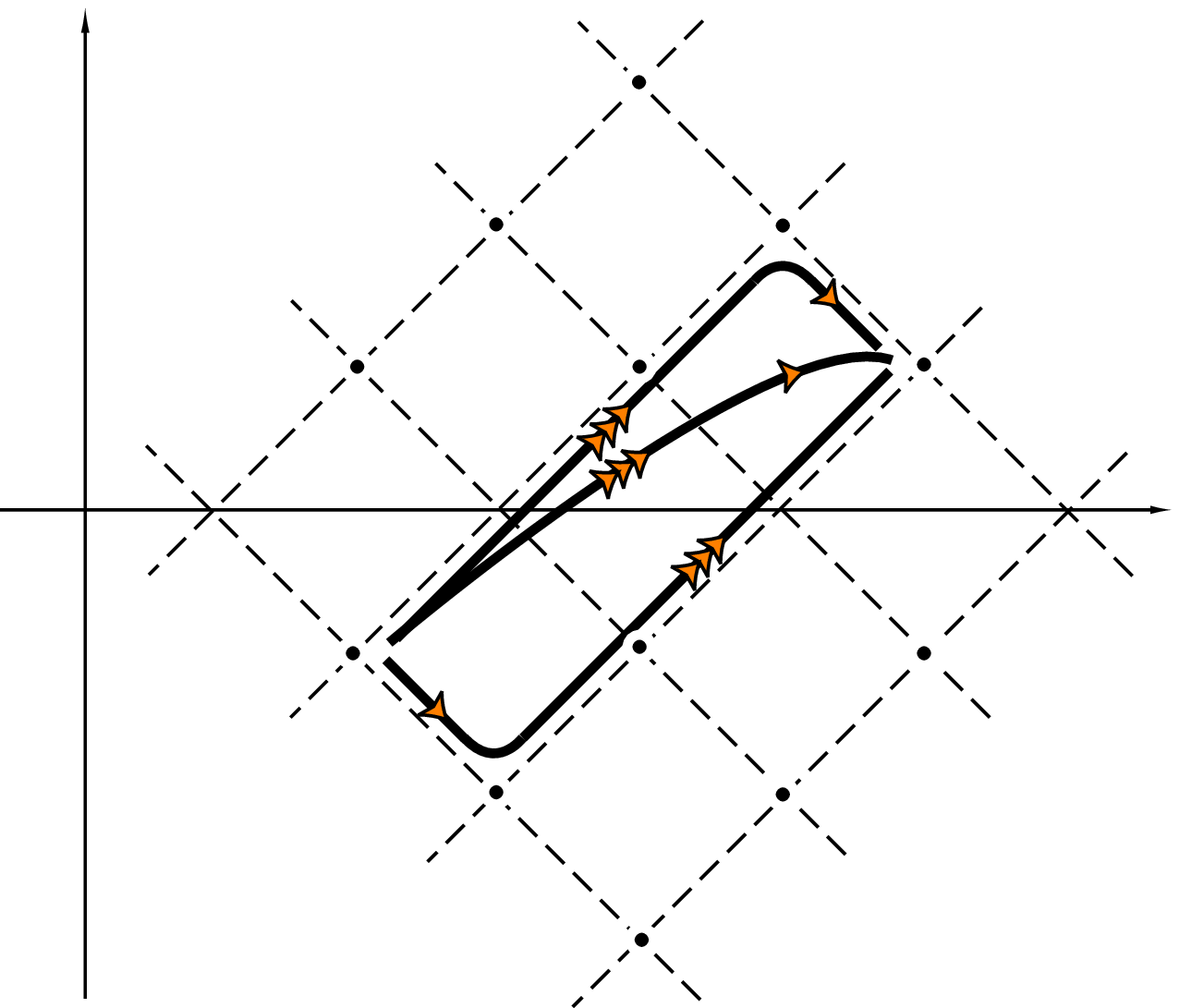}
\footnotesize
\put(-57,132){$\langle 13 \rangle$}
\put(-100,122){$\langle 23 \rangle$}
\put(-36,107){$\langle 12 \rangle$}
\put(-171,69){$ 2m_4 $}
\put(-123,97){$ 2m_3 $}
\put(-73,69){$ 2m_2 $}
\put(-22,69){$ 2m_1 $}
\put(-164,57){$\langle 43 \rangle$}
\put(-82,57){$\langle 32 \rangle$}
\put(-142,32){$\langle 42 \rangle$}
\normalsize
\put(-0,78){$\Sigma^0$}
\put(-199,170){$\Sigma^3$}
\put(-90, -20){b)}
\end{center}
\caption{Compressed multi-walls in the maximal topological sector. 
a) Two level-$1$ compressed walls forming a double wall 
are penetrable. 
b) A level-$2$ compressed wall and 
a single wall 
 forming a double wall 
are penetrable. 
}
\label{commuting-pairs}
\end{figure}

The four walls sector exhibits several interesting phenomena.
From the algebraic 
structure of the wall operators, 
we find a number of the corresponding pairs of walls 
which are penetrable 
each other. 
First, 
two elementary walls generated by the commuting 
operators $a_1$ and $a_3$ 
are penetrable in the same way 
with the two-wall sector 
explained at the beginning of this subsection. 
Second, two level-$1$ compressed walls generated by 
$a_1 a_2$ and $a_2 a_3$ 
are penetrable 
each other due to the relation 
$[a_1 a_2,a_2 a_3]=0$ 
as shown in Fig.~\ref{commuting-pairs}-a). 
Third, the elementary wall generated by $a_2$ and 
the level-$2$ compressed wall generated by 
$a_1a_2a_3$ are 
penetrable each other because of the relation  
$[a_2,a_1a_2a_3]=0$.
The third one is realized in two ways. 
One is shown in Fig.~\ref{commuting-pairs}-b) 
and the other is realized as a mirror ($\Sigma^3 \to - \Sigma^3$) 
of this figure.
We summarize moduli matrices for all topological sectors in this model 
in Table~\ref{N2N4-table}.
\begin{table}
 \begin{center}
 \caption{Topological sectors in the case of 
 $N_{\rm C}=2,\,N_{\rm F}=4$}\label{N2N4-table} 
 \begin{tabular}{|c|c|c|c|} \hline
 top. sector&moduli matrix & dim.&objects\\ \hline\hline
 $\langle 12\rangle \leftarrow \langle 12\rangle $
 &$H_0{}_{\langle 12\rangle } $&0&vacuum $\langle 12\rangle $\\
 $\langle 12\rangle \leftarrow \langle 13\rangle $
 &$H_0{}_{\langle 12\leftarrow 13\rangle }$
 &2& elementary wall $a _2$\\
 $\langle 12\rangle \leftarrow \langle 14\rangle $
 &$H_0{}_{\langle 12\leftarrow 14\rangle }$
 &4& double wall $(a _2,a _3)\oplus $ 
 compressed wall $a _2a _3$\\
 $\langle 12\rangle \leftarrow \langle 23\rangle $
 &$H_0{}_{\langle 12\leftarrow 23\rangle }
 \oplus H_0{}_{\langle 12\leftarrow 32\rangle }$
 &4& double wall $(a _2,a _1)\oplus$ 
 compressed wall $ a _1a _2$  \\
 $\langle 12\rangle \leftarrow \langle 24\rangle $
 &$H_0{}_{\langle 12\leftarrow 24\rangle }
 \oplus H_0{}_{\langle 12\leftarrow 42\rangle }$
 &6&$(a _2,a _3,a _1)\oplus 
 (a _2a _3,a _1)\oplus
 (a _1a _2,a _3)
 \oplus a _1a _2a _3$ \\
 $\langle 12\rangle \leftarrow \langle 34\rangle $&
 $H_0{}_{\langle 12\leftarrow 34\rangle }$
 &8& $(a _2,a _1,a _3,a _2)\oplus
 (a _2a _3,a _1,a _2)
 \oplus (a _2a _3,a _1a _2)$ \\
 &\qquad $\oplus H_0{}_{\langle 12\leftarrow 43\rangle }$&
 &\qquad $\oplus(a _1a _2,a _3,a _2)\oplus
 (a _1a _2a _3,a _2)$\\
 $\langle 13\rangle \leftarrow \langle 13\rangle $
 &$H_0{}_{\langle 13\rangle } $&0&vacuum $\langle 13\rangle $\\
 $\langle 13\rangle \leftarrow \langle 14\rangle $
 &$H_0{}_{\langle 13\leftarrow 14\rangle }$
 &2& elementary wall $a _3$\\
 $\langle 13\rangle \leftarrow \langle 23\rangle $
 &$H_0{}_{\langle 13\leftarrow 23\rangle }$&2
 &  elementary wall $a _1$\\
 $\langle 13\rangle \leftarrow \langle 24\rangle $
 &$H_0{}_{\langle 13\leftarrow 24\rangle }$
 &4&penetrable double wall $(a _1,a _3)$   \\
 $\langle 13\rangle \leftarrow \langle 34\rangle $
 &$H_0{}_{\langle 13\leftarrow 34\rangle }\oplus 
 H_0{}_{\langle 13\leftarrow 43\rangle }$
 &6&$(a _3,a _1,a _2)\oplus 
 (a _1,a _2a _3)\oplus(a _3,a _1a _2)
 \oplus a _1a _2a _3$  \\
 $\langle 14\rangle \leftarrow \langle 14\rangle $&
 $H_0{}_{\langle 14\rangle } $&0&vacuum $\langle 14\rangle $\\
 $\langle 14\rangle \leftarrow \langle 23\rangle $& -&-
 & (non-BPS state)\\
 $\langle 14\rangle \leftarrow \langle 24\rangle $&
 $H_0{}_{\langle 14\leftarrow 24\rangle }$&2
 & elementary wall $a _1$\\
 $\langle 14\rangle \leftarrow \langle 34\rangle $&
 $H_0{}_{\langle 14\leftarrow 34\rangle }$
 &4& double wall $(a _1,a _2)\oplus$ 
 compressed wall $ a _1a _2$ \\
 $\langle 23\rangle \leftarrow \langle 23\rangle $
 &$H_0{}_{\langle 23\rangle } $&0&vacuum $\langle 23\rangle $\\
 $\langle 23\rangle \leftarrow \langle 24\rangle $
 &$H_0{}_{\langle 23\leftarrow 24\rangle }$&2&
 elementary wall $a _3$\\
 $\langle 23\rangle \leftarrow \langle 34\rangle $&
 $H_0{}_{\langle 23\leftarrow 34\rangle }\oplus 
 H_0{}_{\langle 23\leftarrow 43\rangle }$
 &4& double wall $(a _3,a _2)\oplus$ 
 compressed wall $ a _2a _3$\\
 $\langle 24\rangle \leftarrow \langle 24\rangle $&
 $H_0{}_{\langle 24\rangle } $&0&vacuum $\langle 24\rangle $\\ 
 $\langle 24\rangle \leftarrow \langle 34\rangle $&
 $H_0{}_{\langle 24\leftarrow 34\rangle }$&2
 &elementary wall $a _2$\\ 
 $\langle 34\rangle \leftarrow \langle 34\rangle $
 &$H_0{}_{\langle 34\rangle } $&0&vacuum $\langle 34\rangle $\\ 
 \hline
 \end{tabular}
 \end{center}
\end{table}

We now explain that 
the most of moduli parameters in the matrix 
(\ref{4-wall-matrix}) can be understood 
as Nambu-Goldstone modes except one moduli which is 
a quasi-Nambu-Goldstone mode. 
Let us concentrate on imaginary parts 
of moduli parameters in this
moduli matrix.
As described in Sec.~\ref{position-NGQNG}, 
$(N_{\rm C}-1)(\tilde N_{\rm C}-1)$ 
quasi-Nambu-Goldstone parameters are contained 
in such parameters,
while others are the Nambu-Goldstone parameters 
corresponding to broken global symmetry $U(1)^{N_{\rm F}-1}$. 
The case of $N_{\rm C}=2,N_{\rm F}=4$ is the simplest 
case containing quasi-Nambu-Goldstone parameters. 
We have only one quasi-Nambu-Goldstone mode: 
$(N_{\rm C}-1)(\tilde N_{\rm C}-1)=1$. 
Let us consider the global transformation of 
$U(1)^{N_{\rm F}-1}$,
\begin{eqnarray}
 H^1&\!\rightarrow\! &(H^1)'=H^1e^{i\Lambda }, \ 
\Lambda 
={\rm diag.}
(\lambda _1,\lambda _2,\lambda _3,\lambda _4),
\quad  
\end{eqnarray}
with $\sum_{i=1}^4\lambda _i=0$.
The operators for the elementary walls transform under this
transformation as 
\begin{eqnarray}
 a _i(r)\rightarrow (a _i(r))'
=e^{-i\Lambda }a _i(r)e^{i\Lambda }
=e^{-i(\lambda _i-\lambda _{i+1})}a _i(r)
=a _i(r'),
\end{eqnarray}
and we find the complex moduli parameters $r_i$ to transform 
\begin{eqnarray}
 r_1&\rightarrow &r_1'=r_1-i(\lambda _2-\lambda _3),\nonumber\\
r_2&\rightarrow &r_2'=r_2-i(\lambda _3-\lambda _4),\nonumber\\
r_3&\rightarrow &r_3'=r_3-i(\lambda _1-\lambda _2),\nonumber\\
r_4&\rightarrow &r_4'=r_4-i(\lambda _2-\lambda _3).
\end{eqnarray}
Thus we find 
the solution in this $N_{\rm C}=2,\,N_{\rm F}=4$ case 
contains one quasi-Nambu-Goldstone parameter 
$ {\rm Im}(r_1-r_4)$.

\section{Moduli Space for Non-Abelian Walls} \label{MSFNAW}
In the first subsection, the moduli space 
for non-Abelian domain walls is shown 
to be homeomorphic to the complex Grassmann manifold. 
In the second subsection, we construct the moduli metric and 
show that it is a deformed Grassmann manifold.

\subsection{Topology of the Wall Moduli Space} 

In this subsection we discuss the moduli space 
for non-Abelian domain walls.
The $U(N_{\rm C})$ SUSY gauge theory with 
$N_{\rm F}$ hypermultiplets 
maximally admits $N_{\rm wall}$ parallel domain walls, 
with $N_{\rm wall}$ given in Eq.~(\ref{eq:number-of-wall}). 
All possible solutions can be constructed once 
the moduli matrix $H_0$ is given. 
The 
moduli matrix $H_0$ has a redundancy expressed as 
the world-volume symmetry (\ref{art-sym}) : 
$H_0 \sim V H_0$ with 
$V \in GL(N_{\rm C},{\bf C})$. 
We thus find that the moduli space 
denoted by ${\cal M}_{N_{\rm F},N_{\rm C}}$
is homeomorphic to 
the complex Grassmann manifold:
\footnote{
The last expression by a coset space is found as follows.
Using $V$, $H_0$ can be fixed as 
$H_0 = ({\bf 1}_{N_{\rm C}},h)$ 
with $h$ an $N_{\rm C}$ by $\tilde N_{\rm C}$ matrix. 
Consider a $G=SU(N_{\rm F})$ transformation from the right: 
$H_0 \to H_0 g$ with $g \in SU(N_{\rm F})$. 
Although it is not a symmetry 
it is transitive; it transforms 
any point to any point on the moduli space.  
The isotropy group $K$ can be found by putting $h=0$ as
$H_0 = ({\bf 1}_{N_{\rm C}},{\bf 0})$. 
It is $K=SU(N_{\rm C})\times SU(\tilde N_{\rm C}) \times U(1)$ 
with $g_1\in U(N_{\rm C})$, $g_2\in U(\tilde N_{\rm C})$, 
det$g_1$det$g_2$=1, acting as 
$H_{\rm 0} = 
({\bf 1}_{N_{\rm C}},{\bf 0}) \to g_1^{-1} 
({\bf 1}_{N_{\rm C}},{\bf 0}) 
\begin{pmatrix} 
 g_1 & 0 \cr 0 & g_2 
\end{pmatrix}$. 
Hence we obtain 
$G/K \simeq  {SU(N_{\rm F}) / SU(N_{\rm C}) 
\times SU(\tilde N_{\rm C}) \times U(1)}$.
However note that it does not imply that 
the moduli admits isometry $G$. 
This consideration deals merely with the topology.}
\begin{eqnarray}
 {\cal M}_{N_{\rm F},N_{\rm C}}
 \simeq \{H_0 | H_0 \sim V H_0, V \in GL(N_{\rm C},{\bf C})\} 
 \simeq G_{N_{\rm F},N_{\rm C}}
 \simeq {SU(N_{\rm F}) \over 
 SU(N_{\rm C}) \times SU(\tilde N_{\rm C}) \times U(1)}\,.
  \label{Gr}
\end{eqnarray}
This is a {\it compact} (closed) set. 
On the other hand, for instance,  
scattering of two Abelian walls is described by 
a nonlinear sigma model 
on a {\it non-compact} moduli space~\cite{To,To2,IOS1}. 
We also find similar non-compact moduli such as relative 
distances and quasi-Nambu-Goldstone modes of orientational moduli 
in Sec.~\ref{position-NGQNG}, and by an explicit analysis 
of multiple non-Abelian walls in Sec.~\ref{CEWSIC}. 
This fact of the compact 
moduli space consisting of non-compact moduli parameters 
can be consistently understood, if 
we note that the moduli space 
${\cal M}_{N_{\rm F},N_{\rm C}}$ 
includes {\it all BPS topological sectors} 
(\ref{topo.-sect.}) 
determined by the different boundary conditions. 
It is decomposed into 
\begin{eqnarray}
 {\cal M}_{N_{\rm F},N_{\rm C}} 
 = \sum_{\rm BPS} 
 {\cal M}^{\langle A_1A_2\cdots A_{N_{\rm C}} \rangle 
\leftarrow 
 \langle B_1B_2\cdots B_{N_{\rm C}} 
\rangle }_{N_{\rm F},N_{\rm C}}, \; 
\end{eqnarray}
where the sum is taken over the BPS sectors. 
Note that it also includes the vacuum states with no walls 
$\langle A_1A_2\cdots A_{N_{\rm C}}\rangle \leftarrow 
\langle A_1A_2 \cdots A_{N_{\rm C}} \rangle$ 
which correspond to 
$N_{\rm F}!/N_{\rm C}! \tilde N_{\rm C}!$ points 
on the moduli space.
Although each sector (except for vacuum states) 
is in general not a closed set,
the total space is compact. 
We call ${\cal M}_{N_{\rm F},N_{\rm C}}$ as 
the ``total moduli space". 
It is useful to rewrite this as a sum over 
the number of walls:
\begin{eqnarray}
 {\cal M}_{N_{\rm F},N_{\rm C}} 
 = \sum_{k=0}^{N_{\rm C}\tilde N_{\rm C}} 
{\cal M}_{N_{\rm F},N_{\rm C}}^k 
 = {\cal M}_{N_{\rm F},N_{\rm C}}^0 
  \oplus {\cal M}_{N_{\rm F},N_{\rm C}}^1 \oplus \cdots 
  \oplus {\cal M}_{N_{\rm F},
N_{\rm C}}^{N_{\rm C}\tilde N_{\rm C}} \,,
 \label{decom}
\end{eqnarray}
with ${\cal M}_{N_{\rm F},N_{\rm C}}^k$ 
the sum of the topological sectors with $k$-walls. 
Since the maximal number of walls is $N_{\rm C}\tilde N_{\rm C}$,
${\cal M}_{N_{\rm F},N_{\rm C}}^{N_{\rm C}\tilde N_{\rm C}}$ 
is identical to the maximal topological sector.

This decomposition can be understood as follows. 
Consider a $k$-wall solution  
and imagine a situation such 
that one of the outer-most walls goes to spatial infinity. 
We will obtain a ($k-1$)-wall 
configuration in this limit. 
This implies that the $k$-wall sector 
in the moduli space is an open set compactified 
by the moduli space of ($k-1$)-wall sectors on its boundary. 
Continuing this procedure 
we will obtain a single wall configuration. 
Pulling it out to infinity we obtain a vacuum state in the end. 
A vacuum corresponds to a point as 
a boundary of a single wall sector 
in the moduli space.
The $k=0$ sector comprises  
a set of $N_{\rm F}!/N_{\rm C}! \tilde N_{\rm C}!$ points
and does not have any boundary. 
Summing up all sectors,
we thus obtain the total moduli space 
${\cal M}_{N_{\rm F},N_{\rm C}}$ 
as a compact manifold.  
Note again that we include zero-wall sector, 
vacua without any walls.

These procedures are understood 
as a compactification in 
the mathematical theory of the moduli space.
In the following we will see these in some simple examples.

\begin{figure} 
\begin{center}
\includegraphics[width=4cm,clip]{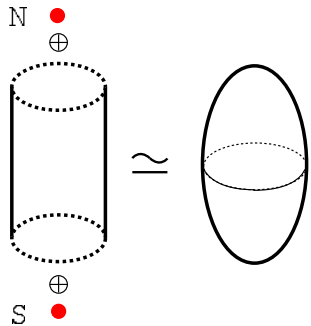}
\caption{The total moduli space in the case of 
$N_{\rm F}=2$, $N_{\rm C}=1$. 
 It is homeomorphic to a sphere, 
but its shape looks like a sausage when it is endowed with 
the metric.
}
\label{nit-fig2}
\end{center}
\end{figure}

\medskip
1) Let us discuss the simplest example of 
$N_{\rm C}=1$ and $N_{\rm F}=2$.
In the strong coupling limit $g \to \infty$
this model reduces to 
the massive HK nonlinear sigma model on
$T^* {\bf C}P^1$ or the Eguchi-Hanson space 
(see Fig.~\ref{nit-fig1}).
This model contains two vacua, $N$ and $S$ 
corresponding to the moduli matrices 
$H_0=(1,0)$ and $H_0=(0,1)$, respectively. 
Thus ${\cal M}_{2,1}^{k=0} \simeq N \oplus S$. 
It admits a single wall connecting them, 
corresponding to the moduli matrix 
$H_0 =(1,e^r)$ with $e^r \in {\bf C}-\{ 0 \}$. 
This single wall possesses two real moduli parameters 
corresponding to a translational zero mode 
${\rm Re}(r)$ 
and a zero mode ${\rm Im}(r) 
(\sim {\rm Im}(r) + 2\pi n,\, n \in {\bf Z})$ arising from 
spontaneously broken internal $U(1)$ symmetry. 
So the moduli space for the one wall solution 
is homeomorphic to a cylinder without boundary: 
${\cal M}_{2,1}^{k=1}  \simeq {\bf R} \times S^1$.
By taking a limit ${\rm Re}(r) \to - \infty$, 
the moduli matrix approaches to the vacuum $N$: $H_0 \to (1,0)$.
On the other hand, in the opposite limit 
${\rm Re}(r) \to + \infty$, 
it approaches to the other vacuum $S$ as 
$H_0 = (1,e^r) \sim (e^{-r},1) \to (0,1)$, 
where we have multiplied $e^{-r}$ using the 
world-volume symmetry  (\ref{art-sym}). 
Thus, adding two points $N$ and $S$ in ${\cal M}_{2,1}^{k=0}$ 
to two infinities of ${\cal M}_{2,1}^{k=1}$, 
we find that the total moduli space is homeomorphic to a sphere 
(see Fig.~\ref{nit-fig2})
\begin{eqnarray}
{\cal M}_{2,1} = 
{\cal M}_{2,1}^{k=0} \oplus {\cal M}_{2,1}^{k=1} 
\simeq \{N \oplus S\} \oplus \{ {\bf R} \times S^1 \} 
\simeq S^2 \simeq {\bf C}P^1 \, . \label{topo21}
\end{eqnarray}
This procedure of adding infinities is 
a two-point compactification 
in the mathematical moduli theory.
Physically this corresponds to the
following situation: 
Moving a wall to two spatial infinities,
the configuration approaches to two vacuum states 
of this theory.

The moduli metric 
on this ${\bf C}P^1$ manifold 
is of course not the round Fubini-Study metric. 
It is deformed to break two isometries in $SU(2)$
preserving one $U(1)$ isometry.  
Therefore it locally looks like a sausage 
but the distance to the infinities diverges.
We will return to this point in the end of this section but 
let us now make a comment on physical validity to consider 
the topology of the total moduli space 
instead of each topological sector. 
The authors in Ref.~\cite{GPTT} constructed a solution of 
1/4 BPS equation in a $D=4$, ${\cal N}=2$ SUSY theory.
It is a composite soliton made of a wall and vortices 
(strings) ending on it. 
Topological stability of this composite soliton can be understood 
by interpreting vortices as sigma model lumps 
in the effective theory on the wall  
if and only if we consider the total moduli space (\ref{topo21}) 
as the target space instead of the one-wall 
topological sector ${\cal M}_{2,1}^{k=1} \simeq {\bf R} \times S^1$.
This is because the second homotopy group $\pi_2$ 
ensuring the topological stability of lumps
is ${\bf Z}$ for ${\bf C}P^1$
but is trivial for ${\bf R} \times S^1$.
The total moduli space (\ref{Gr}) 
for general $N_{\rm F}$ and/or $N_{\rm C}$ 
provides the topological stability of 
more complicated composite solitons 
recently constructed by the present authors~\cite{INOS2}.

\begin{figure}[htb]
\begin{center}
\includegraphics[width=4cm,clip]{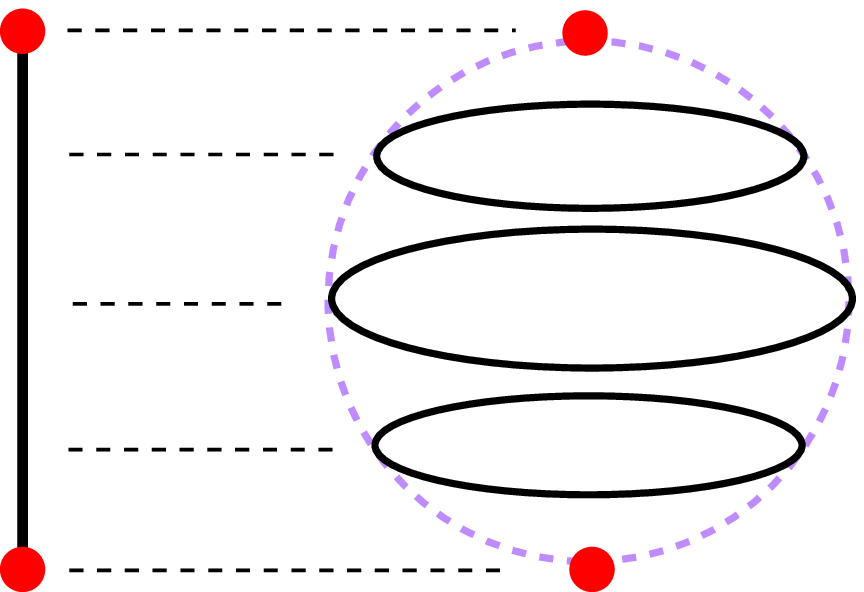}
\put(-70, -10){a)}
\hspace{0.3cm}
\includegraphics[width=4cm,clip]{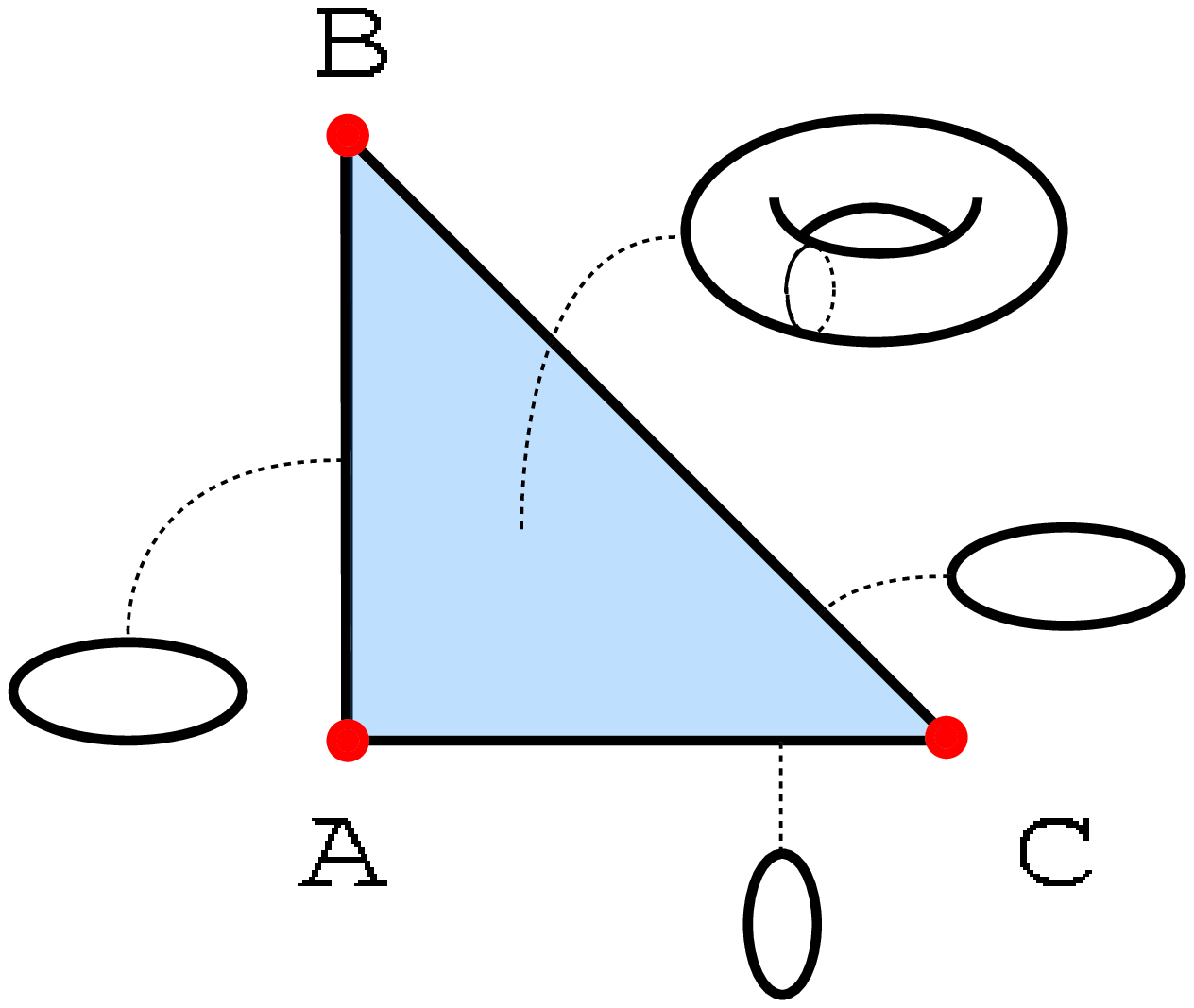}
\put(-70,-10){b)}
\vspace{0.3cm}
\caption{ Toric diagrams. 
a) The toric diagram for ${\bf C}P^1$. 
${\bf C}P^1$ admits one holomorphic $U(1)$ isometry as 
commuting $U(1)$ isometries. 
Dividing ${\bf C}P^1$ by this $U(1)$, 
we obtain the toric diagram for ${\bf C}P^1$ as 
a segment (1-simplex). 
It is parametrized by one $U(1)$ invariant 
which is the momentum map for the $U(1)$ action.
${\bf C}P^1$ is parametrized as 
a circle fibration over it. 
The circle shrinks at the endpoints of the segment. 
b) The toric diagram for ${\bf C}P^2$. 
The ${\bf C}P^2$ admits two commuting holomorphic 
$U(1)$ isometries. 
Dividing ${\bf C}P^2$ by $U(1)^2$ 
we obtain a toric diagram for ${\bf C}P^2$. 
It is parametrized by two $U(1)^2$ invariants 
which are the momentum maps for the $U(1)^2$ action. 
It consists of one triangle (2-simplex) without boundary, 
three open lines 
$A$-$B$, $B$-$C$ and $C$-$A$ on its edge
and three points denoted by
$A$, $B$ and $C$. 
${\bf C}P^2$ is parametrized as a torus  
fibration over this toric diagram, 
and one cycle shrinks on three edge lines 
and both cycles shrink at three points $A$, $B$ and $C$.
Three edges with circle fibers are ${\bf C}P^1$ submanifolds 
of ${\bf C}P^2$.
}
\label{toricdiag}
\end{center}
\end{figure}

\medskip
2) Next let us consider the case of 
$N_{\rm C}=1$ and $N_{\rm F}=3$. 
In the strong coupling limit $g \to \infty$
the model becomes the nonlinear sigma model on 
$T^* {\bf C}P^2$ homeomorphic to $T^* G_{3,2}$, 
the strong coupling limit of 
$N_{\rm C}=2$ and $N_{\rm F}=3$ discussed in 
Sec.~\ref{NC2NF3Case}. 
This model admits a double wall solution as the 
 maximal topological sector. 
We will show that moduli space is 
homeomorphic to ${\bf C}P^2$. 
To this end it is useful to introduce 
a toric diagram. 
The toric diagrams for ${\bf C}P^1$ and ${\bf C}P^2$ 
are displayed in Fig.~\ref{toricdiag} .

\begin{enumerate}
\item
This model contains three discrete vacua 
at $A$, $B$ and $C$ corresponding 
to the moduli matrices 
$H_0 = (0,0,1)$, $(0,1,0)$ and $(1,0,0)$, 
respectively. 
Therefore we have 
${\cal M}_{3,1}^{k=0} = A \oplus B \oplus C$.

\item
There exist two elementary walls 
interpolating between $A$ and $B$ and between 
$B$ and $C$ corresponding to the moduli matrices 
$H_0 = (0,1,e^s)$ and $(1,e^r,0)$ 
($e^r,e^s \in {\bf C} - \{0\} \simeq {\bf R} \times S^1$), 
respectively. 
They approach to vacua as was seen above.
A wall connecting $A$ and $C$ given by 
$H_0 = (1,0,e^s)$ is a compressed wall 
obtained 
from a double wall solution in a particular limit 
by compressing two walls without changing 
the topological sector or the wall tension, 
and so it 
is not included in ${\cal M}_{3,1}^{k=1}$.
Therefore we have 
${\cal M}_{3,1}^{k=1} \simeq {\bf R} \times S^1 \oplus 
{\bf R} \times S^1$.

\item
The moduli subspace for the double wall solution 
is generated by $H_0=(1,e^r,e^s)$ 
($e^r \in {\bf C}$, $e^s \in {\bf C}- \{0\}$). 
It has a topology of 
${\bf R} \times S^1 \times {\bf C}$. 
It sweeps inside the triangle in Fig.~\ref{toricdiag}-b) 
with the boundary 
line connecting $A$ and $C$  
and without the both lines $A$-$B$ and $B$-$C$ 
(because they are elementary wall solutions
\footnote{ 
The lines $A$-$B$, $B$-$C$ obtained by 
taking limits as 
$\lim_{{\rm Re}(s) \to -\infty} H_0 = (1,e^r,e^s)$ 
and $H_0 \sim (e^{-s},e^{r-s},1) \to (0,e^t,1)$ 
(${\rm Re}(s) \to + \infty$ with keeping 
$r-s \equiv t$ finite), respectively, 
have different boundary conditions and so are not included.
The open line $A$-$C$ corresponding to the compressed wall 
given by $H_0 \to (1,0,e^s)$ 
(${\rm Re}\, r \to - \infty$ in the above $H_0$) 
has the same boundary condition and therefore is included.
}) 
and without three points $A$, $B$ and $C$. 
The torus fiber $({\rm Im} (r), {\rm Im} (s)) \in T^2$ shrinks 
to ${\rm Im} (s) \in S^1$ at 
the 
line $A$-$C$ (${\rm Re}\, (r) \to - \infty$).
Such a shrinking cycle ${\rm Im} (r) \in S^1$ 
with a moduli for distance ${\rm Re} (r)$
between two walls constitute $r \in {\bf C}$. 
The non-shrinking cycle ${\rm Im} (s)$ 
with center of mass position ${\rm Re} (s)$
comprise ${\bf R} \times S^1$. 
Therefore the moduli subspace for two walls 
is homeomorphic to 
${\cal M}_{3,1}^{k=2} \simeq {\bf R} 
\times S^1 \times {\bf C}$ 
(see Fig.~\ref{nit-fig5}).
\begin{figure} 
\begin{center}
\includegraphics[width=8cm,clip]{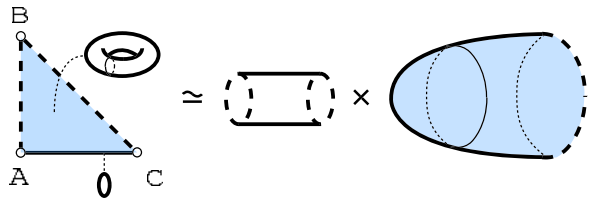}
\caption{The moduli subspace for the double wall 
configuration. 
It is homeomorphic to $ {\bf R} \times S^1 \times D$ with $D$ 
 a two-dimensional disc 
without a boundary. 
The disc $D$ is endowed with the metric looking like a cigar.
}
\label{nit-fig5}
\end{center}
\end{figure}

The factor ${\bf C}$ is homeomorphic to a disc $D$ without boundary.
As found in \cite{To,To2} $D$ is endowed with the metric 
looking like a cigar as shown in Fig.~\ref{nit-fig5}. 
It is also homeomorphic to 
the two dimensional Euclidean black hole metric.
In this case, the metric for finite gauge coupling was 
also calculated exactly in  \cite{IOS1} 
for particular values of $g$. 
The metric has the same topology, but 
is slightly deformed by $g$.

\end{enumerate}

Combining all of these, we obtain ${\bf C}P^2$. 
We thus have found that the total moduli space is 
homeomorphic to ${\bf C}P^2$ (see Fig.~\ref{nit-fig6}):
\begin{figure} 
\begin{center}
\includegraphics[width=8cm,clip]{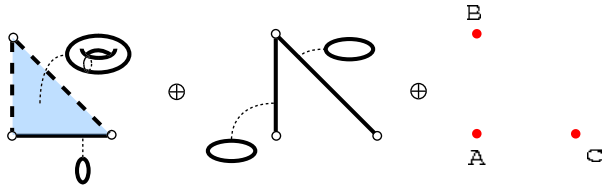}
\caption{The sum ${\cal M}_{3,1}^2 \oplus 
{\cal M}_{3,1}^1 \oplus {\cal M}_{3,1}^0$.
The double-wall and single-wall sectors are non-compact 
and the zero-wall (vacuum) sector is compact. 
The total space is ${\bf C}P^2$ in Fig.~\ref{toricdiag}-b).
}
\label{nit-fig6}
\end{center}
\end{figure}
\begin{eqnarray}
&&{\cal M}_{3,1} = 
{\cal M}_{3,1}^{k=0} \oplus {\cal M}_{3,1}^{k=1} 
 \oplus {\cal M}_{3,1}^{k=2}  \nonumber\\
&& \hs{10} \simeq 
 \{ A \oplus B \oplus C \} 
 \oplus \{ {\bf R} \times S^1 \oplus
           {\bf R} \times S^1 \} 
 \oplus \{{\bf R} \times S^1 \times D \}
 \simeq {\bf C}P^2 \, .
\end{eqnarray}
The topological sectors 
${\cal M}_{3,1}^{k=1}$ and ${\cal M}_{3,1}^{k=2}$ 
are open sets containing non-compact moduli parameters.

\medskip
3) 
In the case of $N_{\rm C} =1$ and an arbitrary $N_{\rm F}$, 
the model reduces in the strong coupling limit to
a nonlinear sigma model on $T^* {\bf C}P^{N_{\rm F}-1}$.
In the same way we can write down 
a toric diagram parametrized by 
$N_{\rm F}-1$ invariants. 
The maximal number of walls is $N_{\rm F}-1$ and 
the ($N_{\rm F}-1$)-wall solutions given by the moduli matrix
$H_0 = (1,e^{r_1},\cdots,e^{r_{N_{\rm F}-1}})$ 
($e^{r_1},\cdots,e^{r_{N_{\rm F}-2}} \in {\bf C}$, 
$e^{r_{N_{\rm F}-1}} \in {\bf C}-\{0\}$)
sweep a ($N_{\rm F}-1$)-simplex 
of the maximal topological sector  
${\cal M}_{N_{\rm F},1}^{k=N_{\rm F}-1}$ 
to which the $N_{\rm F}-1$ wall sector is homeomorphic. 
These parametrize almost all regions of 
${\bf C}P^{N_{\rm F}-1}$ except for infinities 
(and so it is open).
It is compactified by topological sectors with 
less numbers of walls by taking some limits. 

\begin{enumerate}
\item 
\label{same-top}
If we take a limit 
${\rm Re}\ (r_I) \to - \infty$ for $1\leq I \leq N_{\rm F}-2$, 
the configuration remains in the 
maximal topological sector  
${\cal M}_{N_{\rm F},1}^{N_{\rm F}-1}$, 
namely it does not give independent sectors.
\item 
\label{less-wall}
On the other hand, in the limit of 
${\rm Re}(r_{N_{\rm F}-1} ) \to - \infty$, 
we obtain $H_0 = (1,e^{r_1},\cdots,e^{r_{N_{\rm F}-2}},0)$ 
in a different topological sector with walls less by one, 
${\cal M}_{N_{\rm F},1}^{N_{\rm F}-2}$. 
Similarly in the limit of ${\rm Re}\, r_{N_{\rm F}-1} \to + \infty$ 
with keeping $r_I' \equiv r_{I+1} - r_1$ 
($1\leq I \leq N_{\rm F}-2$)
finite and ${\rm Re}\,r_1 \to \infty$, 
we obtain $H_0 = (0,1, e^{r_1'},\cdots,e^{r_{N_{\rm F}-2}'})$  
resulting again in ${\cal M}_{N_{\rm F},1}^{N_{\rm F}-2}$. 
Both sets of parameters parametrize almost all regions of 
${\bf C}P^{N_{\rm F}-2}$'s except for infinities again. 
\end{enumerate}

The procedure \ref{same-top}.~does not change topological 
sectors but 
\ref{less-wall}.~does.
Continuing these two procedures \ref{same-top}.~and 
\ref{less-wall}.~we reach at single walls. 
There exist ${}_{N_{\rm F}}C_2 
= \1{2} N_{\rm F}(N_{\rm F}-1)$ single wall solutions 
given by $H_0 = (0,\cdots,0,1,0,\cdots,0,e^r,0,\cdots,0)$ 
which correspond to one simplexes or 
${\bf C}P^1$'s without two points. 
Among them, elementary walls in 
${\cal M}_{N_{\rm F},1}^{1}$ are obtained as 
those given by the ($N_{\rm F}-1$) moduli matrices 
$H_0 = (0,\cdots, 0, 1,e^r, 0,\cdots,0)$
with a pair of adjacent non-vanishing elements. 
The rest are subspaces in ${\cal M}_{N_{\rm F},1}^{k \geq 2}$.  
By adding $N_{\rm F}$ vacua 
$H_0 = (0,\cdots,1,\cdots,0)$ as points 
in ${\cal M}_{N_{\rm F},1}^{0}$ further, 
we obtain the full moduli space as 
${\bf C}P^{N_{\rm F}-1}$.

In the above discussion, 
we did not use the fact that we took the $g  \to \infty$ 
limit, 
so this is true for finite $g$ (although the metric is 
deformed in general).

\medskip
4) In more general cases, the strong coupling limit is 
the sigma model on $T^* G_{N_{\rm F},N_{\rm C}}$, 
which is not toric anymore.
We do not have toric diagrams.
However it is now obvious that 
all informations for the topology of 
the total moduli space 
is encoded into $H_0$. 
We thus have found that the total moduli space 
is homeomorphic to the 
complex Grassmann manifold, 
${\cal M}_{N_{\rm F},N_{\rm C}} \simeq G_{N_{\rm F},N_{\rm C}}$, 
but the metric on it is not a homogeneous one. 
We construct the moduli metric in the next subsection.

\medskip
It is interesting to observe that 
in our decomposition of the moduli space (\ref{decom}),
the one wall sector ${\cal M}^1_{\NF,\NC}$ includes 
elementary walls only, i.e.
single walls that are not compressed walls.  
Compressed walls are contained in 
double- or multi-wall configurations 
as limits without changing boundary conditions. 
Therefore we can classify single walls into 
elementary or (non-elementary) compressed walls 
in terms of moduli space:
moduli subspace for non-elementary walls 
belong to boundaries of moduli space of 
double or multi-walls, 
$\partial {\cal M}^{k \geq 2}_{\NF,\NC}$, 
but elementary walls do not.

\subsection{Metric on the Wall Moduli Space 
and Effective Field Theory} \label{eff_thoery}

Effective theory on walls describes 
local fluctuations of moduli parameters depending on 
the world-volume coordinates $x^{\mu}$ ($\mu=0, \cdots, 3$) 
and the Grassmann coordinates $\theta, \bar \theta$ 
in the $D=4$, ${\cal N}=1$ superspace (with four supercharges). 

First we construct the effective action explicitly 
for the infinite gauge coupling limit $g\to \infty$.  
Following the Manton's method~\cite{Ma} 
we promote moduli parameters in the moduli matrix $H_0$ 
to weakly varying 
fluctuating fields depending on the world-volume 
coordinates of walls. 
We deal with bosonic fluctuation described by
complex scalars $\phi^X (x^{\mu})$  
with $X=1,\cdots, \NC \tilde \NC$ for a while 
and then will promote them to 
chiral superfields later. 
Then $H^1$ and $H^2$ are given by 
\begin{eqnarray}
 H^1(\phi ,\phi ^*,y) 
 =S^{-1}(\phi ,\phi ^*,y)H_0(\phi )e^{My},
 \quad H^2 = 0 \,. \label{fluc}
\end{eqnarray}
Using the constraints (\ref{constraint1}) and 
(\ref{constraint2}) with $H^2=0$,
\begin{eqnarray}
 H^1H^1{}^\dagger =c{\bf 1}_{N_{\rm C}}, \quad 
 W_\mu =ic^{-1}(\partial _\mu H^1)H^1{}^\dagger ,
\end{eqnarray}
the kinetic term in (\ref{reduced-L}) reduces to 
\begin{eqnarray}
 {\cal L}_{\rm kin} 
 = {\rm Tr}
   \left({\cal D}_\mu H^1{}^\dagger {\cal D}^\mu H^1\right)
 =
 {\rm Tr}
 \left(\partial _\mu H^1{}^\dagger \partial ^\mu H^1
  ({\bf 1}_{N_{\rm F}}-c^{-1}H^1{}^\dagger H^1)\right).
\end{eqnarray}
Substituting (\ref{fluc}) into this
and using (\ref{SS-H0}), 
we obtain the $y$-integrand of the effective Lagrangian 
\begin{eqnarray}
 {\cal L}_{\rm kin} = 
 c \, \partial_X \partial _X^*
  \left({\rm Tr}\left[\log\Omega \right]\right)
    \partial_\mu \phi ^{X}\partial^\mu \phi^{*X},
\end{eqnarray}
which is a K\"ahler nonlinear sigma model 
with the K\"ahler potential
\begin{eqnarray}
 K (\phi ,\phi^*,y) 
 \equiv c {\rm Tr}
    \left[\log\Omega (\phi ,\phi ^*,y)\right]
 = c \log \det 
    \left[\Omega (\phi ,\phi ^*,y)\right] \,.
\end{eqnarray}
Promoting bosonic fields $\phi^X$ to $D=4$, ${\cal N}=1$ 
chiral superfields $\phi^X (x,\theta,\bar\theta)$ 
we obtain the effective Lagrangian for 
(multi) domain walls describing dynamics of moduli 
including fermionic superpartners as
\begin{eqnarray}
 {\cal L}_{\rm walls}^{g \to \infty} 
  = c \int d^4 \theta \int dy \ \log \det \Omega_{g \to \infty} 
  = c \int d^4 \theta \int dy \ \log \det (H_0 e^{2 M y} H_0{}^\dagger) \;. 
 \label{moduli_metric}
\end{eqnarray}
Note that this Lagrangian is invariant under 
a transformation 
of $H_0$
\begin{eqnarray}
 H_0 \to H_0{}' = e^{\Lambda} H_0 \;
 \label{world-volume-gauge-symmetry2}
\end{eqnarray}
with an arbitrary $N_{\rm C} \times N_{\rm C}$ matrix 
of $D=4$, ${\cal N}=1$ chiral superfield 
$\Lambda(x,\theta,\bar\theta)$.  
The K\"ahler potential receives a K\"ahler transformation 
\begin{eqnarray}
 \log \det \Omega \to  \log \det \Omega 
+  \log \det \Lambda +  \log \det \Lambda^\dagger 
\label{Kahler-tr}
\end{eqnarray}
where the last two terms $\log \det \Lambda$ 
and $\log \det \Lambda^\dagger$ disappear 
under the $\theta$ integration in the Lagrangian  
(\ref{moduli_metric}). 
The transformation 
(\ref{world-volume-gauge-symmetry2}) is a local 
$U(N_{\rm C})$ gauge symmetry 
on the world volume of walls, 
although corresponding vector multiplet 
does not appear explicitly. 
It is actually enhanced to its complexification 
$U(N_{\rm C})^{\bf C} = GL(N_{\rm C},{\bf C})$ 
because the lowest component of $\Lambda$ are complex,  
and can be understood as 
a local symmetry originated from 
the global world-volume symmetry (\ref{art-sym}).

If the hypermultiplet masses are non-degenerate $M=0$, 
the integrand in the Lagrangian (\ref{moduli_metric}) 
is the K\"ahler potential for the Grassmann 
manifold with the maximal isometry $SU(N_{\rm F})$ 
acting on $H_0$ as $H_0 \to H_0 U$ with 
$U \in SU(N_{\rm F})$~\cite{HN}. 
In contrast, 
$SU(N_{\rm F})$ is explicitly broken down to $U(1)^{N_{\rm F}-1}$ 
by hypermultiplet masses $M$ for our non-degenerate 
mass case. 
Therefore the moduli space metric for 
non-Abelian domain walls 
is the {\it deformed Grassmann manifold}.

We have constructed the effective action on walls by 
the Manton's method. 
Here we discuss a symmetry approach to 
the low-energy effective action in the spirit of 
the chiral Lagrangian in QCD. 
Since we use only symmetry to construct the action in 
this method, it 
determines the effective action only up to an 
ambiguity in the K\"ahler potential 
with an arbitrary function of 
invariants.\footnote{
This kind of effective Lagrangian consistent with symmetry 
has been constructed for non-Abelian vortices in 
\cite{Eto:2004ii}. 
It was motivated to resolve discrepancy between 
the actions constructed by Manton's method and 
by the brane configuration~\cite{HT} 
in string theory. 
} 
The symmetry approach has an advantage of wide applicability, 
including the case of quantum corrections preserving 
the symmetry. 
In order to extract the true collective coordinates from the 
moduli matrix $H_0$ promoted to a matrix chiral superfield 
$H_0(x,\theta,\bar \theta)$, 
we have to take account of the redundancy 
expressed by the world-volume symmetry (\ref{art-sym}). 
Since the moduli are promoted to fields on the world volume, 
it is most convenient to promote the world-volume symmetry 
$GL(N_{\rm C},{\bf C})$ into a local gauge symmetry on 
the world volume. 
One should note that this local gauge symmetry is described by 
an $N_{\rm C} \times N_{\rm C}$ matrix vector superfield 
$V(x,\theta,\bar \theta)$ without kinetic terms.
Hence $V(x,\theta,\bar \theta)$ is a Lagrange multiplier field 
just to express the redundancy of 
the matrix chiral superfield 
$H_0(x,\theta,\bar \theta)$  
at least in the classical level. 
Then the world-volume gauge symmetry acts as
\begin{eqnarray}
 H_0 \to e^{\Lambda} H_0 \;, \hs{5}
 e^V \to e^{-\Lambda^\dagger} e^V e^{-\Lambda},
 \label{world-volume-gauge-symmetry}
\end{eqnarray}
with $\Lambda(x,\theta,\bar{\theta})$ 
defined in (\ref{world-volume-gauge-symmetry2}). 
This gauge symmetry is complexified to 
$GL(N_{\rm C},{\bf C})$ 
but we call this gauge symmetry $U(N_{\rm C})$ as usual.
Since the expression 
in Eq.~(\ref{Gr}) merely means the topology, 
the world-volume theory does not possess 
$SU(N_{\rm F})$ global symmetry. 
Instead $SU(N_{\rm F})$ was explicitly 
broken down to $U(1)^{N_{\rm F}-1}$ by 
the hypermultiplet masses. 
In this respect we decompose $H_0$ as
\begin{eqnarray}
  H_0 = (h_1 ,h_2, \cdots, h_{N_{\rm F}}), 
\end{eqnarray}
with $h_A (x,\theta,\bar{\theta})$ ($A=1,\cdots,N_{\rm F}$)
an $N_{\rm C}$-vector of chiral superfields. 
The world-volume theory is invariant under 
$U(N_{\rm C})_{\rm local} \times U(1)^{N_{\rm F}-1}$ 
in which $U(1)^{N_{\rm F}-1}$ acts on $h_A$ 
as $h_A \to e^{i \alpha_A }h_A$ 
with a constraint $\sum_A \alpha_A =0$. 
We define invariants under these symmetries\footnote{
These invariants parametrize the moduli space 
divided by the global symmetry 
$U(1)^{N_{\rm F}-1}$: 
$X_A \in {\cal M}_{N_{\rm F},N_{\rm C}} 
/ U(1)^{N_{\rm F}-1}$.
Since Nambu-Goldstone bosons parametrize 
the $U(1)^{N_{\rm F}-1}$-orbit, 
which was divided out, 
this space is parametrized by quasi-Nambu-Goldstone 
bosons~\cite{Ni}.
We thus find that $X_A$ correspond to quasi-Nambu-Goldstone 
bosons. 
} by
\begin{eqnarray}
 X_A \equiv h_A{}^\dagger e^V h_A \;.
\end{eqnarray}
When we construct the invariant K\"ahler potential for 
the low-energy effective Lagrangian  
these invariants bring in an ambiguity 
which cannot be determined by symmetry alone and 
depends on the detail of the theory. 
Thus the effective Lagrangian on the wall 
can be written by using an undetermined 
function $f_g$ of invariants $X_A$ which depends on 
the gauge coupling $g$ in 
the original gauge theory:
\begin{eqnarray}
 {\cal L}_{\rm wall}^g = \int d^4 \theta 
  [f_g (X_1,\cdots,X_{N_{\rm F}}) - C\, {\rm tr}\, V ],
 \label{wveff}
\end{eqnarray}
where $C$ is the FI parameter for the world-volume gauge theory.
Eliminating $V$ by its algebraic equation of motion 
we obtain the effective theory in terms of independent fields. 
This procedure is the K\"ahler quotient. We conjecture that Manton's
effective Lagrangian (\ref{moduli_metric}) 
may be contained in (\ref{wveff}) as a particular form of $f_g$, 
although we have not yet succeeded to demonstrate it explicitly.

\medskip

The effective theory (\ref{moduli_metric}) 
or (\ref{wveff}) 
on the world volume of walls includes all topological sectors. 
Which sector it describes depends completely on 
which point on moduli space one chooses as the background. 
As far as one analyzes the effective theory perturbatively,  
it describes the topological sector to which the background 
belongs. 
This is consistent with the fact that 
the geodesic distance diverges 
if one wants to reach other topological sectors.

Here we make a comment on 
the difference between effective 
Lagrangians for walls and other solitons. 
Other solitons admit D-brane pictures. 
Instantons can be interpreted for instance by 
the D0-D4 system. 
The instanton moduli space is obtained 
as an effective field theory on the D0-branes as found in \cite{Wi}. 
Taking its T-dual along the D4-brane world-volume 
it becomes D1-branes ending on D3-branes. 
The effective theory on D1-branes gives 
the monopole moduli space~\cite{brane-monopole}. 
Effective Lagrangian for non-Abelian vortices 
has recently been 
derived by Hanany and Tong~\cite{HT} 
in the Hanany-Witten brane configuration  
where vortices are identified with D1-branes. 
In all the cases, 
the gauge group $U(k)$ of these effective theories 
come from gauge invariance on $k$ D-branes.
In contrast to these cases, in our wall case,  
the gauge group $U(N_{\rm C})$ on the wall effective 
Lagrangian is identical to that of the original gauge theory 
not related with the number of walls. 
We have not yet found brane constructions 
 appropriate for the 
non-Abelian walls.

Before closing this section, we make a comment on 
a possible interpretation of our moduli space as 
a {\it special Lagrangian submanifold}. 
In the massless limit, 
the moduli space of vacua (the Higgs branch of the gauge theory) 
denoted as ${\cal M}_{\rm vac}^{M=0}$ was 
the cotangent bundle over the complex Grassmann manifold, 
${\cal M}_{\rm vac}^{M=0} \simeq T^* G_{N_{\rm F},N_{\rm C}}$ 
(\ref{T*Gr}). 
By introducing hypermultiplet masses, 
most points on ${\cal M}_{\rm vac}^{M=0}$ 
are lifted leaving some discrete points as vacua.
The moduli space ${\cal M}_{N_{\rm F},N_{\rm C}}$ for 
walls is shown to be homeomorphic to $G_{N_{\rm F},N_{\rm C}}$ 
which is a base manifold of ${\cal M}_{\rm vac}^{M=0}$. 
It is a special Lagrangian submanifold of 
$T^* G_{N_{\rm F},N_{\rm C}}$.
We expect that moduli space for non-Abelian 
domain walls in SUSY gauge theory 
with general gauge group and general matter contents is 
homeomorphic to a special Lagrangian submanifold in 
the massless Higgs branch of vacua ${\cal M}_{\rm vac}^{M=0}$
of that gauge theory.

\section{
Discussion}
\label{sc:discussion}

Let us list some of the interesting topics for 
future researches. 

As was shown in (\ref{moduli_metric}), 
the moduli space metric is deformed by hypermultiplet masses. 
We assumed that all masses are non-degenerate, 
so the global symmetry was $U(1)^{N_{\rm F}-1}$.
We expect that this expression is 
valid even in the case that 
there exist mass degeneracy. 
However Eq.~(\ref{moduli_metric}) can 
contain 
non-normalizable modes also. 
This is most easily exhibited for the case of $M=0$. 
In that case, the K\"ahler potential inside 
the $y$-integration is the standard one for 
the complex Grassmann manifold with the 
$SU(N_{\rm F})$ isometry. 
Is is a base manifold of the massless 
moduli space of vacua 
$T^* G_{N_{\rm F},N_{\rm C}}$. 
The integrand is independent of $y$ and 
hence the $y$-integration diverges. 
For partially degenerate masses 
we will obtain some non-Abelian localized zero modes also. 
To obtain effective theory on the world volume of walls, 
we have to throw away non-normalizable modes. 
Degenerate mass should also provides richer global 
symmetry~\cite{SY2}, 
which is likely to have implications to localization of 
non-Abelian gauge bosons on walls. 
The case of degenerate masses is one of the most 
interesting and immediate future problem. 

We have found previously for Abelian gauge theories 
that coupling the tensor multiplet provides the localized 
massless Abelian gauge boson on the wall~\cite{IOS1}. 
We wish to come back to the problem of coupling tensor 
multiplets to the non-Abelian gauge theories admitting 
non-Abelian walls. 
We hope to localize non-Abelian gauge bosons 
on the non-Abelian walls with four-dimensional world 
volume by coupling the tensor multiplets appropriately.

\begin{figure}[htb]
\begin{center}
\includegraphics[width=7cm,clip]{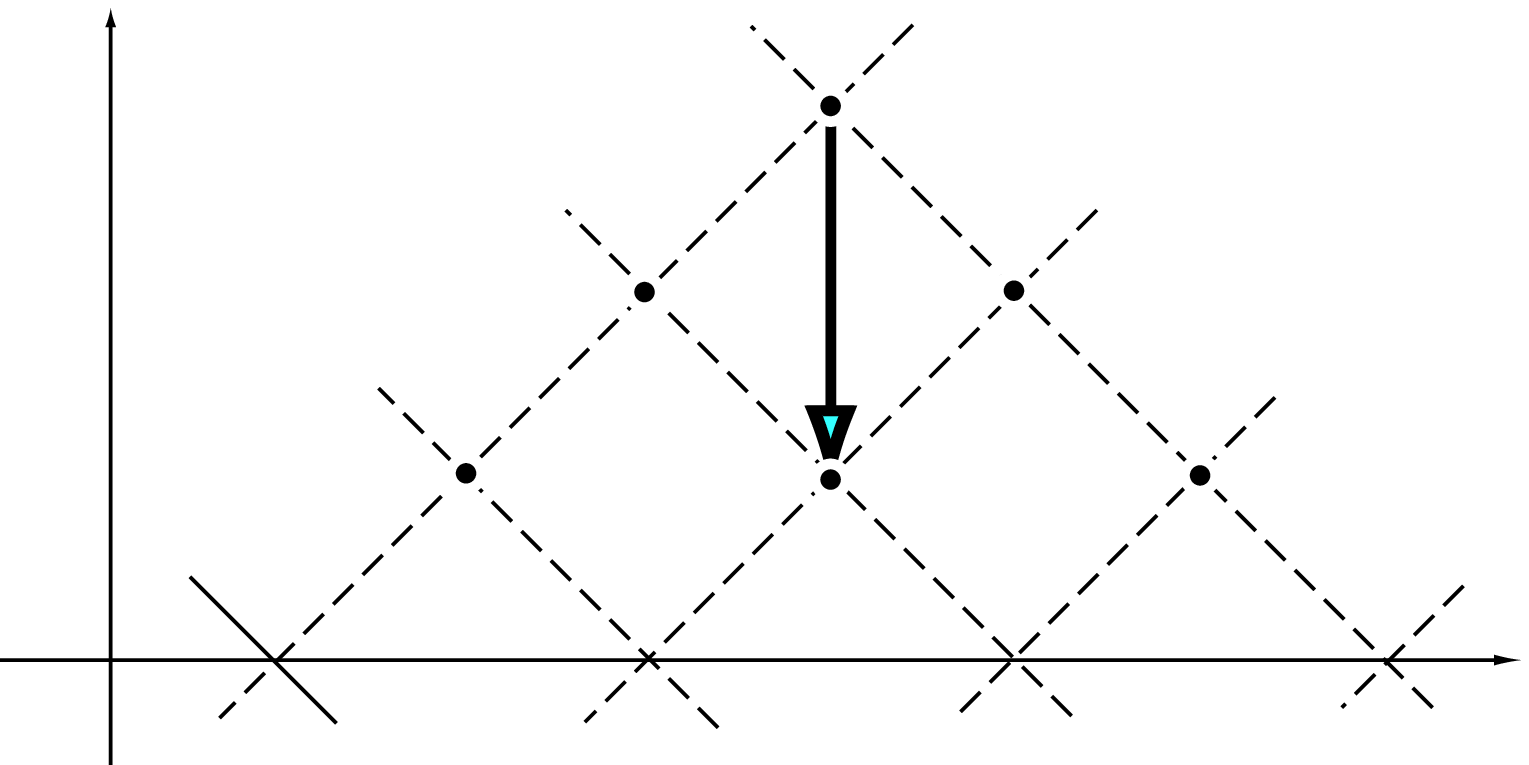}
\footnotesize
\put(-100,100){$\langle 14 \rangle$}
\put(-57,60){$\langle 13 \rangle$}
\put(-142,60){$\langle 24 \rangle$}
\put(-82,35){$\langle 23 \rangle$}
\put(-171,-0){$ 2m_4 $}
\put(-121,-0){$ 2m_3 $}
\put(-71,-0){$ 2m_2 $}
\put(-22,-0){$ 2m_1 $}
\normalsize
\put(-0,0){$\Sigma^0$}
\put(-199,100){$\Sigma^3$}
\put(-90, -20){a)}
\,\hspace{1cm}
\includegraphics[width=7cm,clip]{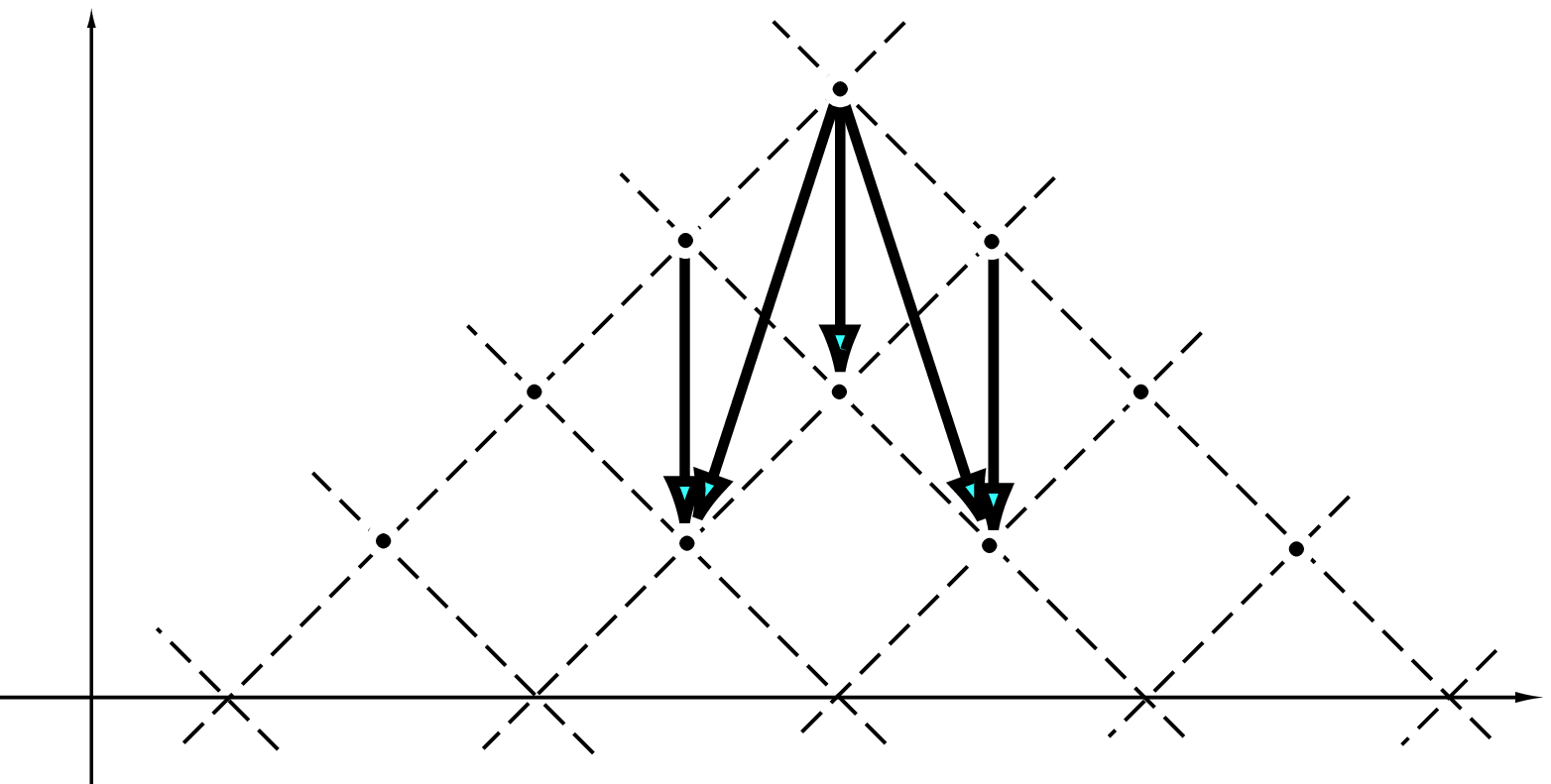}
\footnotesize
\put(-100,100){$\langle 15 \rangle$}
\put(-139,70){$\langle 25 \rangle$}
\put(-60,70){$\langle 14 \rangle$}
\put(-100,33){$\langle 24 \rangle$}
\put(-138,27){$\langle 34 \rangle$}
\put(-65,27){$\langle 23 \rangle$}
\put(-176,-3){$ 2m_5 $}
\put(-136,-3){$ 2m_4 $}
\put(-98,-3){$ 2m_3 $}
\put(-54,-3){$ 2m_2 $}
\put(-18,-3){$ 2m_1 $}
\normalsize
\put(-0,0){$\Sigma^0$}
\put(-199,100){$\Sigma^3$}
\put(-90, -20){b)}
\end{center}
\caption{Non-BPS walls. 
a) $N_{\rm C}=2$ and $N_{\rm F}=4$. 
b) $N_{\rm C}=2$ and $N_{\rm F}=5$. 
}
\label{non-BPS-fig}
\end{figure}

We have discussed only the topological sector containing 
BPS multi-walls. 
However, 
our model contains the topological sectors admitting no BPS 
walls also. 
The number of the topological sectors admitting 
no BPS walls has been given in 
(\ref{non-BPS-number}), which is one for 
$N_{\rm C}=2$ and $N_{\rm F}=4$ and 
five for $N_{\rm C}=2$ and $N_{\rm F}=5$ 
as seen in Fig.~\ref{non-BPS-fig}. 
For larger $N_{\rm C}$ or $N_{\rm F}$, 
our model contains many more topological sectors admitting 
no BPS walls. 
For instance in the case of  $N_{\rm C}=2$ and $N_{\rm F}=4$, 
a wall configuration $\langle 23 \rangle \leftarrow \langle 14 \rangle$ 
is made by a BPS wall $\langle 13 \rangle \leftarrow \langle 14 \rangle$
and an anti-BPS wall $\langle 23 \rangle \leftarrow \langle 13 \rangle$, 
or by a BPS wall $\langle 13 \rangle \leftarrow \langle 14 \rangle$
and an anti-BPS wall $\langle 23 \rangle \leftarrow \langle 24 \rangle$. 
The BPS bound for this topological sector 
is never saturated by these configurations, 
irrespective of the bound taking 
positive, zero, or negative values. 
We do not know if the walls composing these 
configurations are interacting or not. 
If they are interacting 
it is very interesting to investigate 
whether they form a stable bound state of walls 
or not, as was found in a simpler model in 
four dimensions~\cite{Sakai:2002wc}. 
The fate of $U(1)$ zero modes is also interesting. 
\unitlength 1pt
Even if non-BPS wall solutions are difficult to construct, 
we can construct the effective action by using the nonlinear realization method or the Green-Schwarz method, which has been worked out in four space-time 
dimensions~\cite{CNV}.

Coupling with gravity has been discussed 
for single wall and multi-wall solutions 
in the case of Abelian gauge theories~\cite{gravity1,gravity2}. 
In the limit of zero width of the wall, the BPS 
wall solution reduces to the  Randall-Sundrum model. 
Therefore the Abelian wall embedded into supergravity 
can be regarded as a smooth regularization of the AdS space 
in the Randall-Sundrum model 
by finite wall width constructed by physical scalar fields. 
It has been found that the necessary gravitational 
deformations can have another parameter giving asymmetry 
 of the bulk cosmological constant between left and right 
of the wall. 
To embed the globally SUSY nonlinear sigma model into supergravity, 
it has been difficult to obtain appropriate gravitational 
deformations of the target manifold. 
However, treating the model as a hyper-K\"ahler quotient 
using the gauge theory, it has been found to obtain the 
appropriate quaternionic manifold. 
It was essential to use the off-shell 
formalism~\cite{KugoOhashi} to 
couple the system to five-dimensional supergravity. 
Our formulation is a natural setting to embed 
the non-Abelian walls into the supergravity. 
We hope to complete the task in near future.

Moduli spaces for walls and vortices are K\"ahler 
whereas those for monopoles and instantons are hyper-K\"ahler.
In \cite{HT} the vortex moduli space was shown 
to be a special Lagrangian submanifold of 
the ADHM instanton moduli space.  
Investigating our wall moduli space as 
some middle dimensional manifold of the monopole moduli space 
is an interesting future problem.\footnote{
We would like to thank David Tong for a useful 
discussion.}

The moduli space for non-Abelian walls has turned out 
in this paper to be the complex Grassmann manifold. 
Taking a large $N_{\rm F}$ limit, 
$N_{\rm F} \to \infty$, 
it becomes the infinite dimensional Grassmann manifold.
It is a famous fact that the moduli space 
of the KdV equation is the infinite dimensional 
Grassmann manifold. 
Therefore we suspect that there may be some deep connection 
between our model and integrable systems.

\section*{Acknowledgement}
The authors thank David Tong for useful discussion 
and Masato Arai for a collaboration 
at an early stage. 
This work is supported in part by Grant-in-Aid for Scientific 
Research from the Ministry of Education, Culture, Sports, 
Science and Technology, Japan No.13640269 (NS and MN) 
and 16028203 for the priority area ``origin of mass'' 
(NS). 
The works of K.O. and M.N. are 
supported by Japan Society for the Promotion 
of Science under the Post-doctoral Research Program. 
One of the authors (Y.I.) gratefully acknowledges 
support from a 21st Century COE Program at 
Tokyo Tech ``Nanometer-Scale Quantum Physics" by the 
Ministry of Education, Culture, Sports, Science 
and Technology.

\appendix

\section{Positions of Walls} \label{position}
In this Appendix we present the method to evaluate the 
positions of walls.
With the infinite gauge coupling, 
det$\Omega $ can be calculated explicitly for a given 
moduli matrix $H_0$ as 
\begin{eqnarray}
 \det(\,\Omega) &=&
 \det\left(\frac{1}{c}H_0e^{2My}H_0^\dagger \right)
=\sum_{\langle A\rangle \in {\rm vacua}}
\Big|{\rm det}
\left(\frac{1}{c}H_0e^{My}H^\dagger_{0\langle A \rangle } 
\right)\Big|^2\nonumber\\
&=&\sum_{\langle A\rangle \in {\rm vacua}}
\Big|{\rm det}
\left(\frac{1}{c}H_0H^\dagger_{0\langle A \rangle }
\right)\Big|^2{\rm exp}
\left({2\sum_{r=1}^{N_{\rm C}} m_{A_r} y}\right)
,\label{detOmega} 
\end{eqnarray} 
where sum is taken over the whole supersymmetric vacua 
labeled by $\langle A \rangle = \langle A_1 A_2 
\cdots A_r \cdots A_{N_{\rm C}}  \rangle $.
Note that elements of the 
$N_{\rm C}\times N_{\rm C}$ matrix 
$H_0H^\dagger_{0\langle A \rangle }$ is given by 
$\left( H_0H^\dagger_{0\langle A \rangle } \right)^r{}_s
= \sqrt{c}H_0^{rA_s}$.
In vacua, the energy density vanishes 
$({c/2}){\partial_y^2}\log{\rm det}\Omega =0$ 
and the difference of $({c/2}){\partial_y}\log{\rm det}\Omega $
gives the topological charge of walls. 
We expect that det$\Omega $ exhibits 
a behaviour similar to the above formula (\ref{detOmega}) 
containing exponential factor of 
$ \sum_{r=1}^{N_{\rm C}} m_{A_r} y$, 
even if we consider the case of finite gauge couplings. 

\begin{figure}[htb]
\begin{center}
\includegraphics[width=6cm,clip]{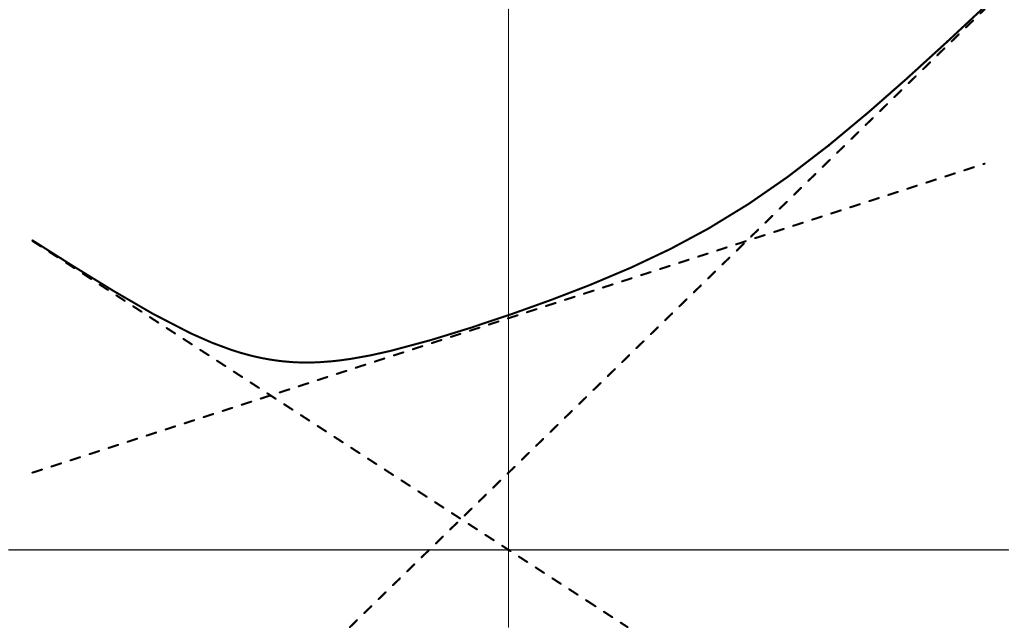}
\put(2,12){$y$}
\put(-88,120){$z$}
\put(-150,65){$\psi (y)$}
\put(-60,40){$f_1(y)$}
\put(-200,20){$f_2(y)$}
\put(-62,-10){$f_3(y)$}
\put(-105,0){$y_{12}$}
\put(-45,5){$y_1$}
\put(-130,5){$y_2$}
\put(-110,-30){a)~$y_2<y_1$}
\hspace{1cm}
\includegraphics[width=6cm,clip]{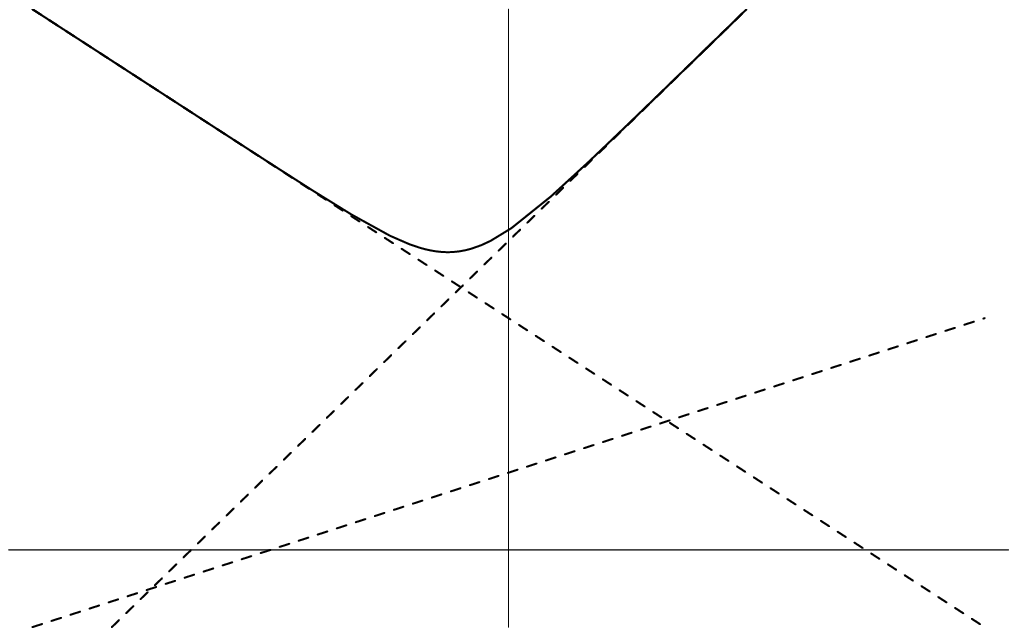}
\put(2,12){$y$}
\put(-150,110){$\psi (y)$}
\put(-145,40){$f_1(y)$}
\put(-0,55){$f_2(y)$}
\put(-0,-5){$f_3(y)$}
\put(-88,120){$z$}
\put(-105,0){$y_{12}$}
\put(-60,5){$y_2$}
\put(-150,20){$y_1$}
\put(-110,-30){b) $y_2>y_1$}
\end{center}
\caption{ Comparison of the profile of 
$z=\psi (y),f_1(y),f_2(y),f_3(y)$ as functions of $y$. 
Linear functions $f_i$ are good approximations in their 
respective dominant regions. 
}
\label{fig:position-of-wall}
\end{figure}

By use of the form (\ref{detOmega}), we can guess positions 
of walls more  accurately without calculating the energy 
density.  For 
simplicity, let us consider the following $N_{\rm F}=3$ case,
\begin{eqnarray}
 \psi (y)\equiv 
 \log{\rm det}\Omega 
 &=&\log\left(e^{2f_1(y)}+e^{2f_2(y)}+e^{2f_3(y)}\right),
\nonumber
\end{eqnarray}
with linear functions $f_i(y)\equiv m_iy-u_i, i=1,2,3$, 
and we assume 
$m_1>m_2>m_3$. 
Because of the exponential dependence in $y$, 
only one exponential factor in det$\Omega $ is dominant 
in each vacuum, 
thus $\psi (y)$ is close to the linear function of the 
dominant exponential  
\begin{eqnarray}
 \psi (y)\simeq 
 {\rm max}\left(f_1(y),\,f_2(y),\,f_3(y)\right).
 \label{eq:lin-approx}
\end{eqnarray}
As compared in Fig.\ref{fig:position-of-wall}, 
the linear approximation (\ref{eq:lin-approx}) 
for $\psi (y)$ is accurate except near the transition region. 
Since $\partial _y^2\psi (y)$ gives the energy density
of walls, position of walls are obtained as 
intersection points of the linear functions for each region 
defined by 
\begin{eqnarray}
 y_1&=&{u_1-u_2\over m_1-m_2},\quad y_2
 ={u_2-u_3\over m_2-m_3},
\quad y_{12}={u_1-u_3\over m_1-m_3},
\end{eqnarray}
if the two linear functions are really dominant in that 
region of $y$. 
Note that if other linear functions are dominant, 
the intersection point is hidden by 
contributions from other terms and 
there is no wall at that point.  
Thus we observe that in the case $y_1>y_2$ there exist 
two walls whose positions are represented by 
$y_1$ and $y_2$, while in the case $y_1<y_2$ 
the configuration of walls looks approximately like one wall 
whose position is given by $y_{12}$. 
If we take the limit of $y_2\rightarrow \infty $ with $y_{12}$
fixed, we obtain a compressed wall located at $y_{12}$.  

\section{The Standard Form of $H_0^1$}\label{TSFH01}
In this Appendix, we show that any moduli matrix $H_0^1$ can be 
transformed to the standard form 
\unitlength 0.8pt
 \vspace{1em}
 \begin{eqnarray}
 H_0^1=\left(
 \begin{array}{c}
\begin{picture}(380,90)(0,0)
 \put(10,-5){\scalebox{0.8}{\includegraphics{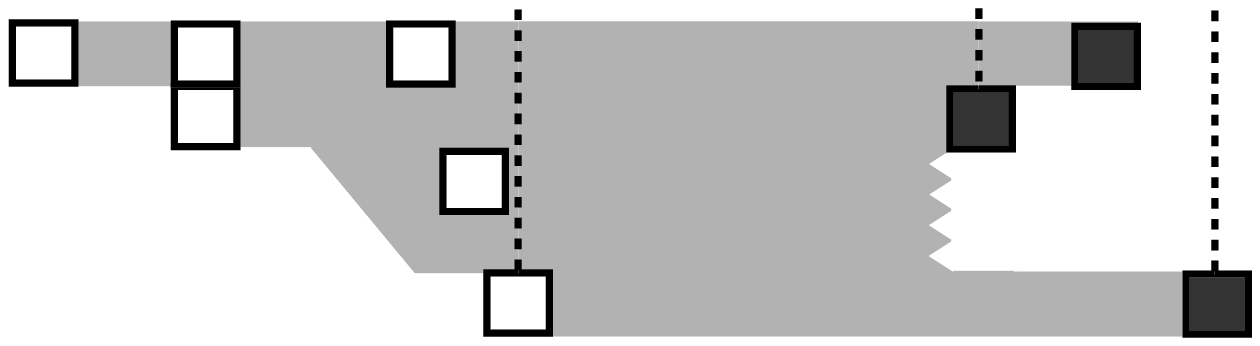}}} 
\end{picture}
\begin{picture}(0,0)(-25,-37)
 \put(-400,-25){{\Huge $0$}}
 \put(-385,65){$A_1$}
 \put(-338,65){$A_2$}
 \put(-305,65){$\cdots $}
 \put(-248,65){$A_{N_{\rm C}}$}
 \put(-115,65){$B_2$}
 \put(-82,65){$B_1  {\cdot} {\cdot}{\cdot } $}
 \put(-50,65){$B_{N_{\rm C}}$}
 \put(-385,39){1}
 \put(-338,39){0}
 \put(-275,39){0}
 \put(-338,21){1}
 \put(-260,2){0}
 \put(-248,-34){1}
 \put(0,-34){.}
 \put(-85,0){{\Huge $0$}} 
\end{picture}
 \end{array}
 \right)
 \label{sf}
\end{eqnarray}
Here, the matrix elements in gray color with flavor indices 
$A_r < A < B_r$ 
and in black color with flavor $B_r$ 
represent complex numbers 
which can and cannot vanish, respectively.  
Vanishing elements are fixed by the world-volume symmetry 
(\ref{art-sym}). 
We use the following simplified graphical representation 
\begin{eqnarray}
\begin{picture}(170,50)(-230,0)
\put(-230,0){\scalebox{0.333333333}{\includegraphics{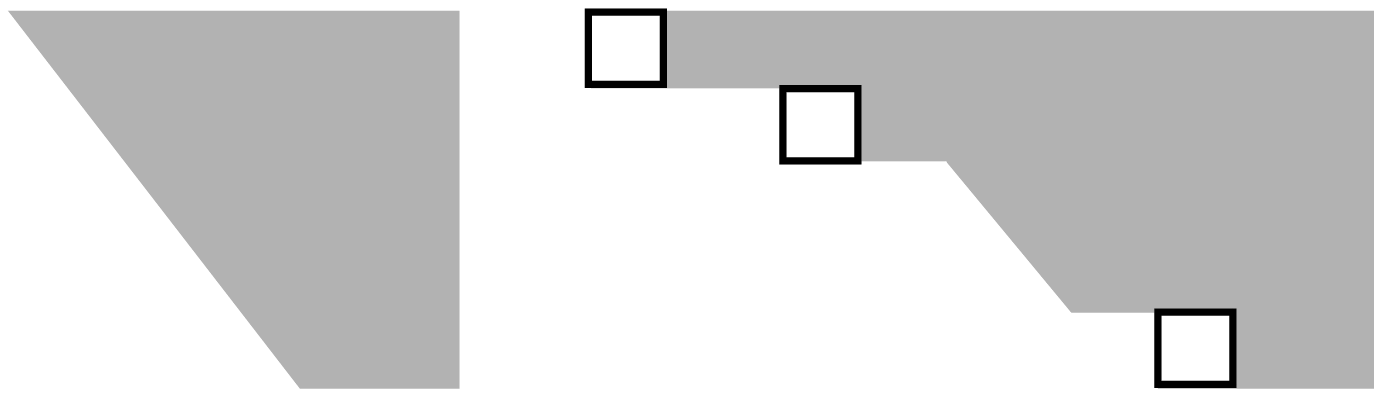}}}
\put(-157,40){{\scriptsize 1}}
\put(-133,31){{\scriptsize 1}}
\put(-88,4){{\scriptsize 1}}
\put(-170,20){$\equiv$} 
\end{picture},\qquad 
\begin{picture}(100,50)(0,0)
 \put(0,0){\scalebox{0.333333333}{\includegraphics{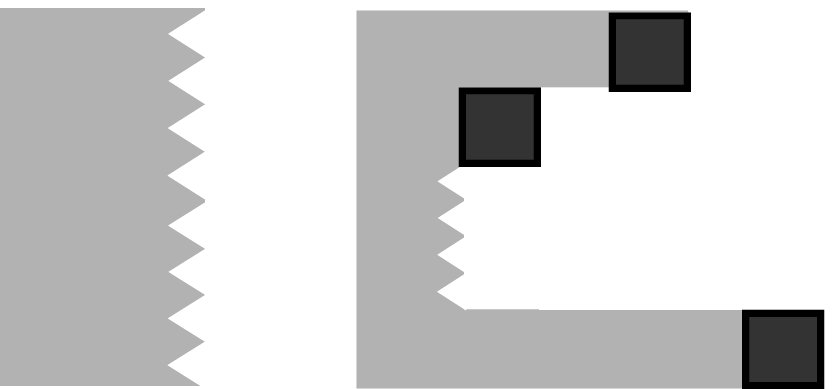}}}
\put(30,20){$\equiv$}
\end{picture}.
\end{eqnarray}

We now give a proof that an arbitrary moduli matrix $H_0^1$ 
can be brought to the standard form by the world-volume 
symmetry, and that no freedom of the world-volume 
symmetry remains once we have the standard form. 
Let us begin with arranging the left side of $H_0^1$. 
Let us assume that the first flavor index which has 
at least one non-vanishing element is $A_1$:
\unitlength 0.6pt
\vspace{1em}
\begin{eqnarray}
H_0^1=\left(
\begin{array}{c}
 \begin{picture}(140,90)(0,0)
\put(30,-5){\scalebox{0.3}{\includegraphics{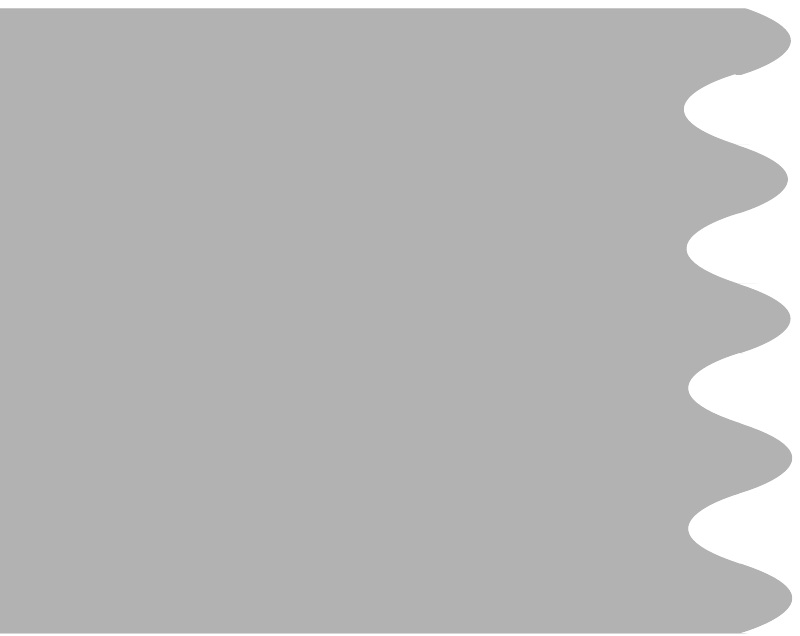}}} 
\end{picture}
\begin{picture}(0,0)(-65,-35)
\put(-170,60){$A_1$}
\put(-210,0){{\large$0$}} 
\end{picture}
\end{array}
\right),
\end{eqnarray}
In this equation, the right side of $H_0^1$ denoted 
by a wavy line indicates that the form on this side 
is not yet specified. 

By using a part of degree of freedom of the world-volume symmetry 
$V\in  GL(N_{\rm C},\mathbf{C})$ in (\ref{art-sym}),
it is possible to transform the $A_1$-th column 
to $(1,0,\cdots,0)^T$. 
Then the matrix is of the form 
\unitlength 0.67pt
\begin{eqnarray}
H_0^1=\left(
\begin{array}{c}
\begin{picture}(190,80)(0,0)
\put(20,-5){\scalebox{0.3}{\includegraphics{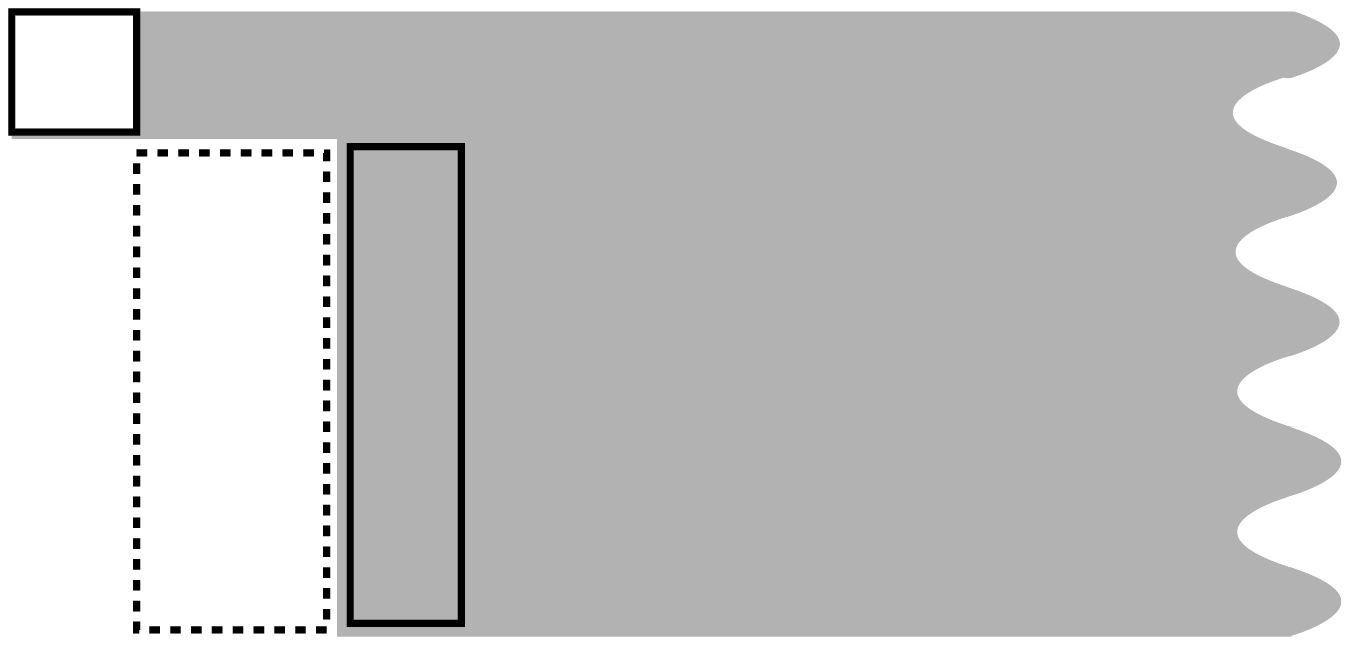}}} 
\end{picture}
\begin{picture}(0,0)(-45,-30)
\put(-165,55){$A_2$}
\put(-209,55){$A_1$}
\put(-206,35){1}
\put(-230,-10){{\Huge $0$}} 
\end{picture}
\end{array}
\right) ,
 \label{H0-0}
\end{eqnarray}
where 
we have assumed that elements enclosed by 
the dashed line happen to vanish 
and that the region enclosed by the solid line in 
the $A_2$-th column 
has at least one non-zero element. 
The following world-volume symmetry 
(\ref{art-sym}) remains after the fixing of (\ref{H0-0}): 
\begin{eqnarray}
V=\left(
\begin{array}{c}
\begin{picture}(80,80)(0,0)
\put(-3,-5){\scalebox{0.3}{\includegraphics{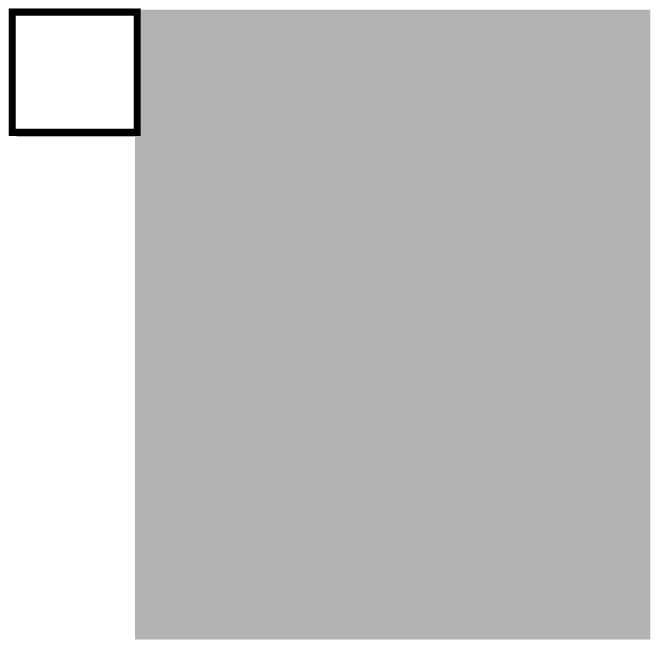}}} 
\end{picture}
\begin{picture}(0,0)(-30,-30)
\put(-108,35){1}
\put(-109,-10){{\large $0$}} 
\end{picture}
\end{array}
\right) .
  \label{V1}
\end{eqnarray}
Repeating these fixings $N_{\rm C}$ times, 
$H_0^1$ can be transformed to an echelon form
\vspace{1em}
\begin{eqnarray}
H_0^1=\left(
\begin{array}{c}
\begin{picture}(250,80)(0,0)
\put(10,-5){\scalebox{0.3}{\includegraphics{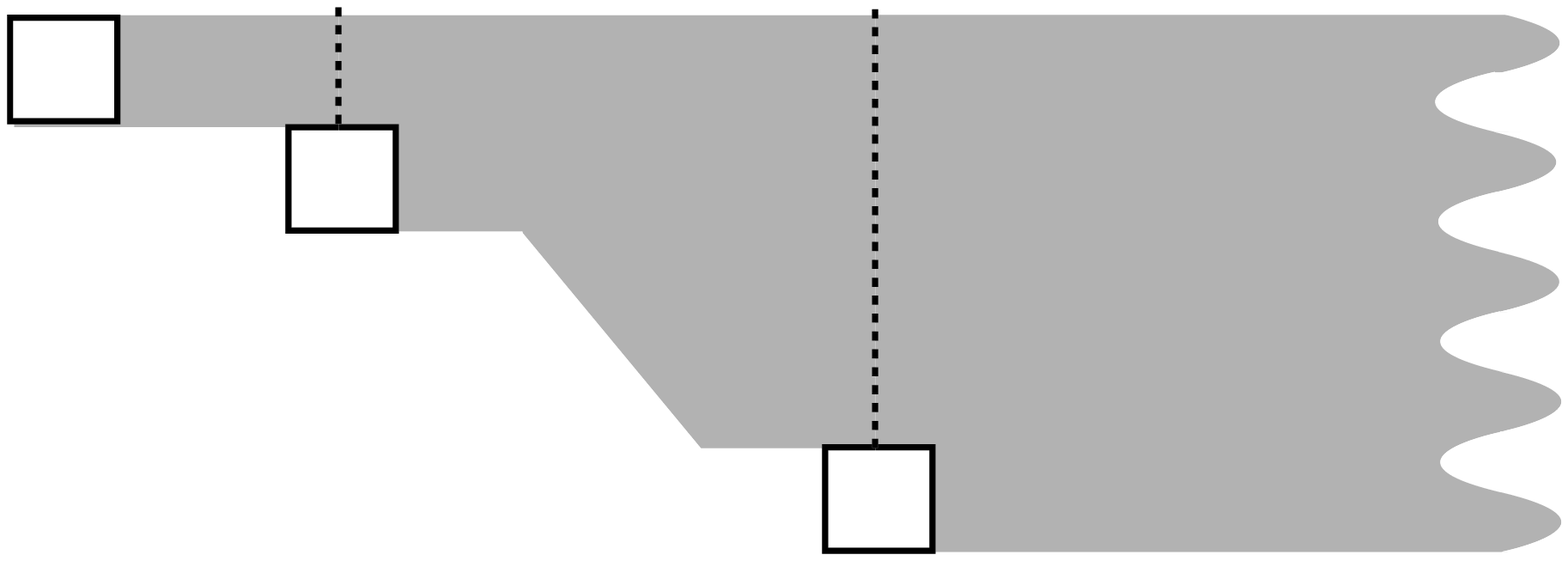}}} 
\end{picture}
\begin{picture}(0,0)(-30,-32)
 \put(-224,55){$A_2$}
\put(-267,55){$A_1$}
\put(-144,55){$A_{N_{\rm C}}$}
\put(-263,34){1}
\put(-221,18){1}
\put(-140,-30){1}
\put(-260,-20){{\Huge $0$}} 
\end{picture}
\end{array}
\right) .
\end{eqnarray}
Then the world-volume symmetry (\ref{V1}) reduces to
\begin{eqnarray}
V=\left(
\begin{array}{c}
\begin{picture}(80,80)(0,0)
\put(0,-5){\scalebox{0.3}{\includegraphics{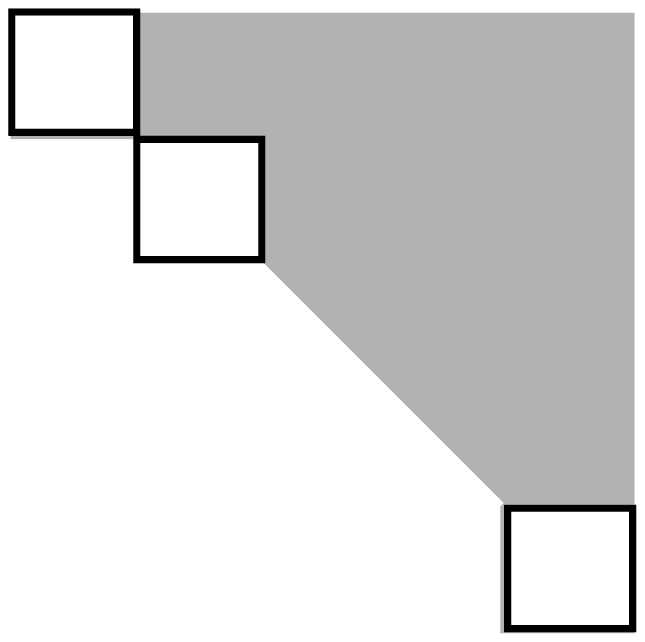}}} 
\end{picture}
\begin{picture}(0,0)(-35,-30)
 \put(-108,36){1}
\put(-92,20){1}
\put(-44,-28){1}
\put(-110,-20){{\Huge $0$}}
\end{picture}
\end{array}
\right) .
  \label{V2}
\end{eqnarray}

Next we fix the right side of $H_0^1$ by using the 
remaining world-volume symmetry (\ref{V2}).  
Let the first non-vanishing column from the right be
the $B_r$-th flavor and  
its lowest non-vanishing component be in the $r$-th row, like 
\begin{eqnarray}
H_0^1=\left(
\begin{array}{c}
\begin{picture}(180,80)(0,0)
\put(0,-5){\scalebox{0.3}{\includegraphics{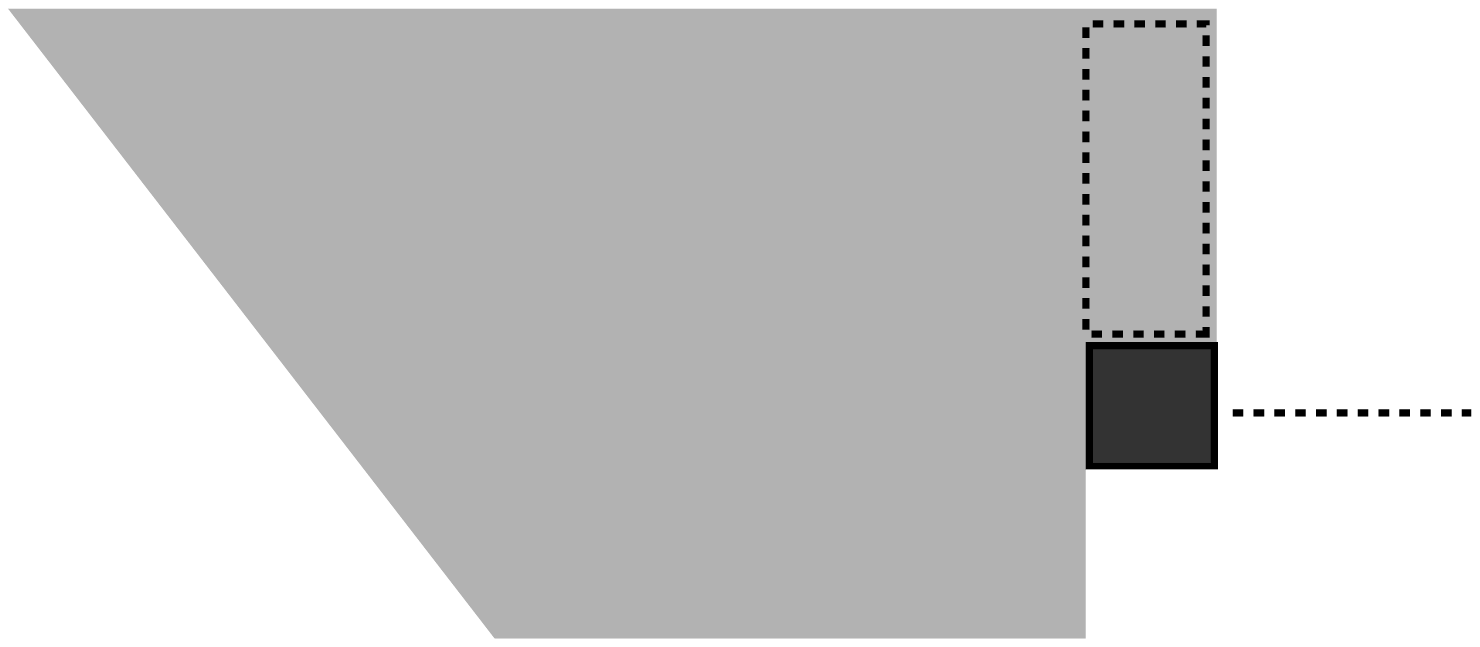}}} 
\end{picture}
\begin{picture}(0,0)(-50,-33)
 \put(-90,55){$B_r$}
\put(-66,-29){{\Large $0$}}
\put(-230,-20){{\Large $0$}}
\put(-28,-10){$r$} \label{H0-1}
\end{picture}
\end{array}
\right)\ \ \ .
\end{eqnarray}
It is possible to eliminate 
the region enclosed by the dashed line in (\ref{H0-1}) 
by using (\ref{V2}). 
Then $H_0^1$ and the world-volume symmetry become
\vspace{1em}
\begin{eqnarray}
H_0^1=\left(
\begin{array}{c}
\begin{picture}(165,90)(0,0)
\put(0,0){\scalebox{0.3}{\includegraphics{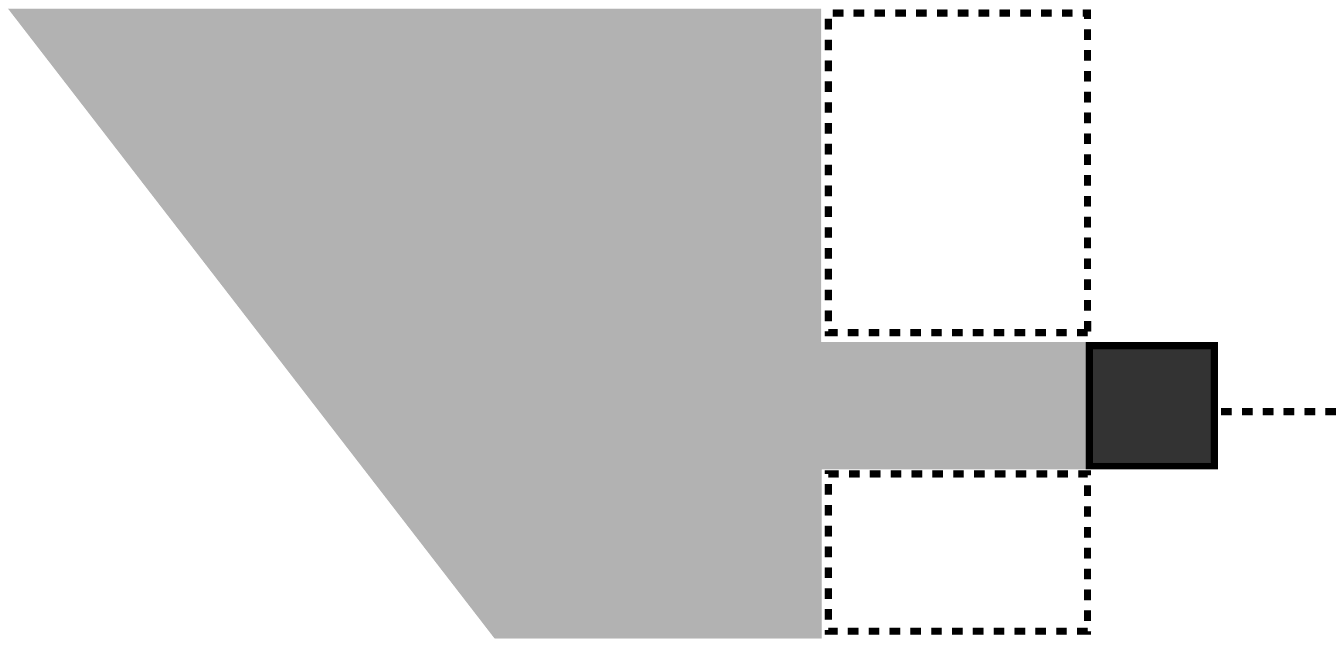}}} 		 
\put(142,95){$B_r$}
\put(10,20){{\Huge $0$}}
\put(190,30){$r$}
\end{picture}
\end{array}
\right)\ \ ,\quad 
V=\left(
\begin{array}{c}
\begin{picture}(80,90)(0,0) 
 \put(-2,0){\scalebox{0.3}{\includegraphics{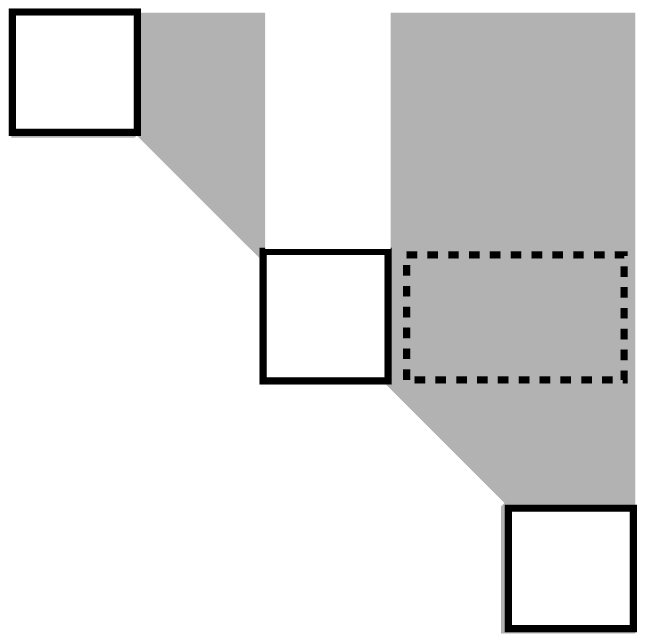}}}
\put(6,72){1}
\put(39,40){1}
\put(73,7){1}
\put(5,15){{\Huge $0$}}
\put(38,95){$r$} 
\put(105,40){$r$} 
\end{picture}
\end{array}
\right) \ \ \ , 
\label{H0-V}
\end{eqnarray}
where elements enclosed by the dashed line in $H_0^1$ 
vanish accidentally
and the $r$-th column in $V$ has been used for this fixing. 
By using elements enclosed by the dashed line 
in $V$ in Eq.~(\ref{H0-V}), it is possible to eliminate 
the 
$(r, \ A_{r+1}), \ (r, \ A_{r+2}), \ \cdots , \ (r, \ A_{N_{\rm F}})$ 
elements in $H_0^1$. 
Then $H_0^1$ and remaining world-volume symmetry become
\vspace{1em}
\unitlength 0.55pt
\begin{eqnarray}
H_0^1=\left(
\begin{array}{c}
\begin{picture}(335,110)(0,0)
\put(0,-30){\scalebox{0.4166666}{\includegraphics{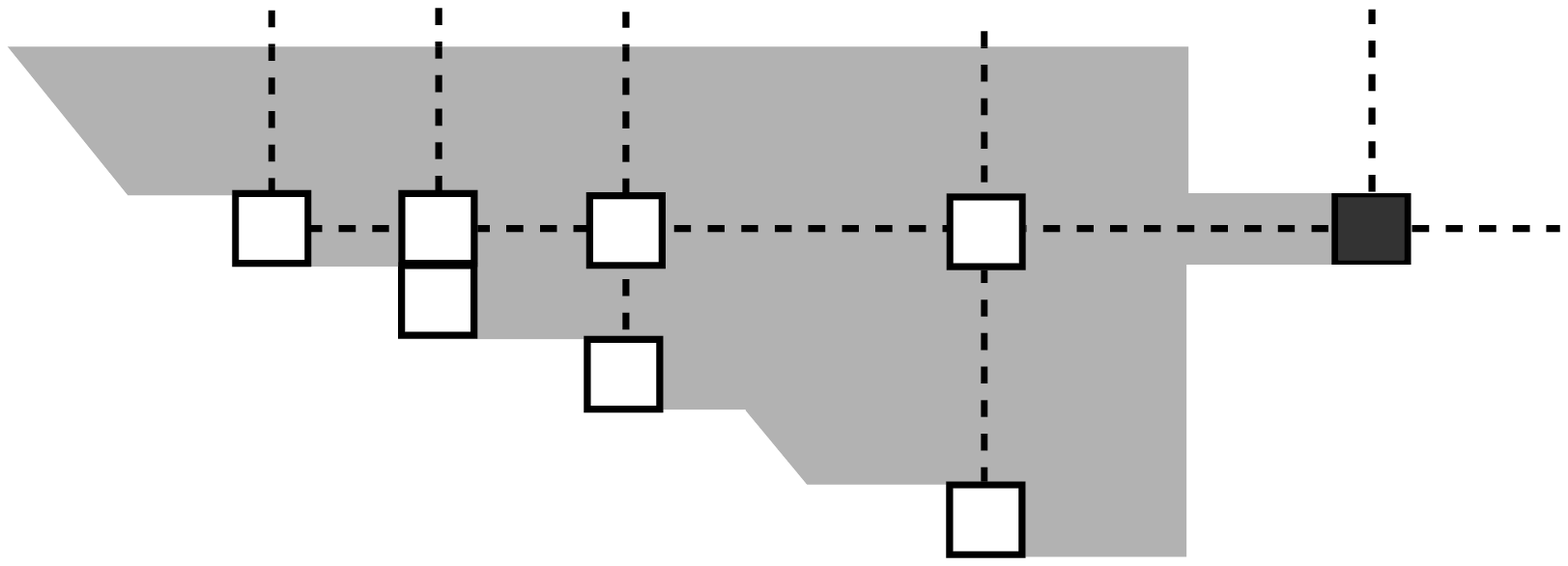}}} 
\put(60,120){$A_r$}
\put(98,120){$A_{r+1}$}
\put(140,120){$A_{r+2}  \ \cdots $}
\put(220,120){$A_{r+3}$}
\put(305,120){$B_r$}
\put(65,57){\small 1}
\put(102,57){\footnotesize 0}
\put(145,57){\footnotesize 0}
\put(226,57){\footnotesize 0}
\put(104,41){\footnotesize 1}
\put(145,25){\footnotesize 1}
\put(226,-8){\footnotesize 1}
\put(29,15){{\Huge $0$}}
\put(304,5){{\Huge $0$}}
\put(365,57){$r$}
\end{picture}
\end{array}
\right), \quad  
V=\left(
\begin{array}{c}
\unitlength 0.67pt
\begin{picture}(80,90)(0,0)
\put(0,0){\scalebox{0.3}{\includegraphics{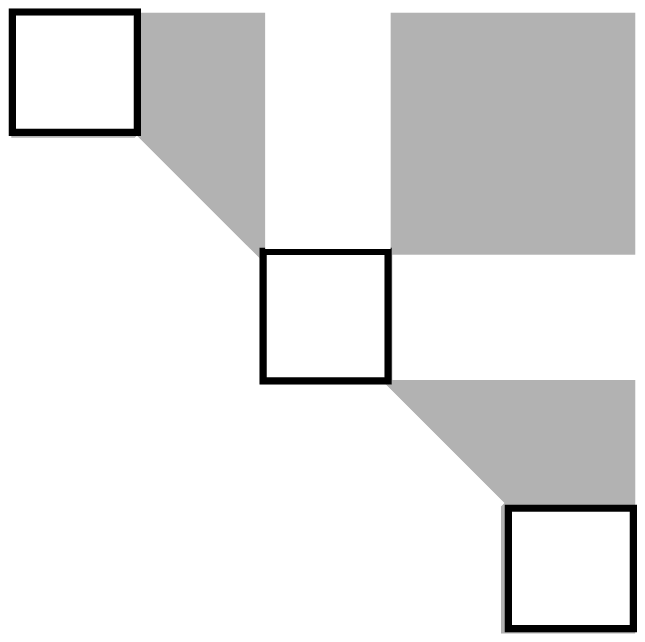}}} 
\put(10,72){1}
\put(43,40){1}
\put(75,7){1}
\put(9,7){{\Huge $0$}}
\put(43,89){$r$}
\put(101,40){$r$}
\end{picture}
\end{array}
\right) \ \ ,
\end{eqnarray}
respectively. 

Let us define $H_0^1[r]$ 
as 
the $(N_{\rm C}-1)\times N_{\rm F}$ submatrix 
removing the $r$-th row in the matrix $H_0$ 
and the $(N_{\rm C}-1)\times (N_{\rm C}-1)$ matrix $V[r]$ 
by removing the $r$-th row and the $r$-th column 
in the matrix $V$. 
Obviously, we can repeat the above procedure for $(H_0^1[r],V[r])$. 
Furthermore, the procedure to obtain $(H_0^1[r],V[r])$ from $(H_0^1,V)$ 
can be repeated for $(H_0^1[r],V[r])$ to obtain $(H_0^1[r,s], V[r,s])$. 
Continuing this process $N_{\rm C}$ times, 
all degrees of freedom in $V$ are 
finally used to fix $H^1_0$ to the standard form (\ref{sf}).

Let us give an alternative procedure to find the standard 
form which is equivalent to 
the above procedure. 
This procedure should be more practical if one wishes to list 
up all matrices in the standard form 
parametrizing the given topological sector labeled by 
$\langle A_1A_2\cdots A_{N_{\rm C}} 
\rangle \leftarrow \langle B_1B_2\cdots B_{N_{\rm C}} \rangle$. 
Let us first list up all possible orderings of $B_r$. 
Once $A_r$ and $B_r$ are chosen, 
we can find the vanishing elements 
between $A_r$ and $B_r$ in the $r$-th row 
in the following way. 
Let us illustrate the method using an example of $N_{\rm C}=6$ 
with general $N_{\rm F}$:
\unitlength 0.75pt
 \begin{eqnarray}
 H_0^1=\left(
 \begin{array}{c}
\begin{picture}(385,90)(0,0)
 \put(0,-3){\scalebox{0.5}{\includegraphics{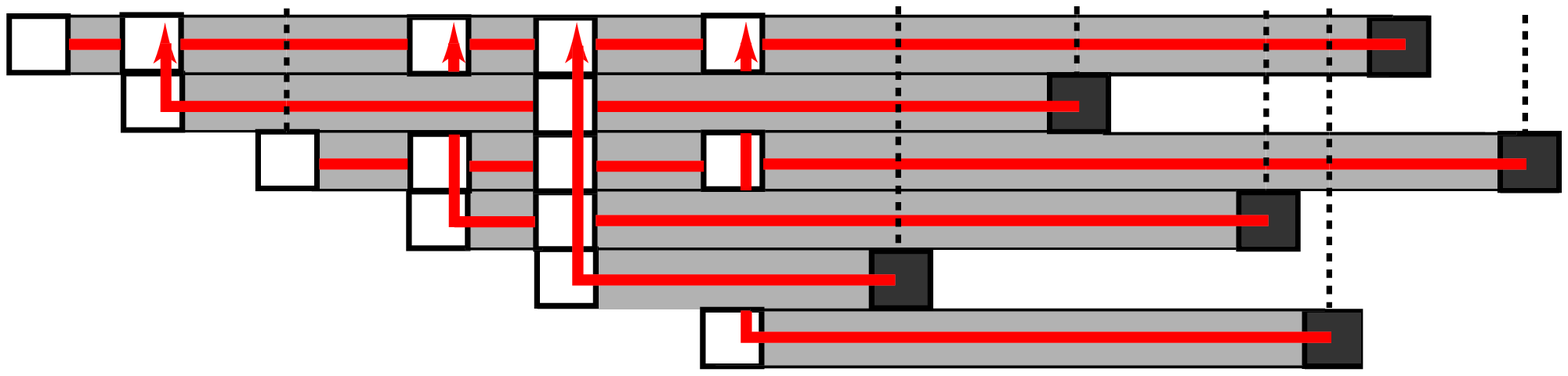}}} 
\end{picture}
 \end{array}
 \right) , 
 \put(-410,60){$A_1$}
 \put(-380,60){$A_2$}
 \put(-345,60){$A_3$}
 \put(-310,60){$A_4$}
 \put(-277,60){$A_5$}
 \put(-237,60){$A_6$}
 \put(-200,60){$B_5$}
 \put(-152,60){$B_2$}
 \put(-107,60){$B_4$}
 \put(-91,60){$B_6$}
 \put(-75,60){$B_1$}
 \put(-45,60){$B_3$}
 \put(-400,-30){{\Huge $0$}}
 \put(-50,-30){{\Huge $0$}}
 \put(-408,40){\footnotesize 1}
 \put(-380,40){\footnotesize 0}
 \put(-310,40){\footnotesize 0}
 \put(-278,40){\footnotesize 0}
 \put(-237,40){\footnotesize 0}
 \put(-380,26){\footnotesize 1}
 \put(-278,26){\footnotesize 0}
 \put(-345,12){\footnotesize 1}
 \put(-310,12){\footnotesize 0}
 \put(-278,12){\footnotesize 0}
 \put(-237,12){\footnotesize 0}
 \put(-310,-2){\footnotesize 1}
 \put(-278,-2){\footnotesize 0}
 \put(-278,-16){\footnotesize 1}
 \put(-237,-32){\footnotesize 1} \label{ex-NC=6}
\end{eqnarray}
1) Let $B_{r_1}$ the smallest among all the $B_r$. 
The $(A_{r_1},r_1)$ element should be unity, according 
to the rule of constructing the standard form. 
Then all the elements above this $(A_{r_1},r_1)$ element in 
the $A_{r_1}$-th column vanish. 
2) Remove the $r_1$-th row and the $A_{r_1}$-th column from $H_0^1$. 
3) Continue the same procedure $N_{\rm C}$ times. 
Then we obtain the standard form in Eq.~(\ref{ex-NC=6}). 

The generic region of the topological sector 
is covered by the generic moduli matrix with the ordering 
$B_1<B_2<\cdots<B_{N_{\rm C}}$. 
On the other hand the subspace with the smallest dimension 
is covered by 
the moduli matrix with the ordering 
$B_1>B_2>\cdots>B_{N_{\rm C}}$ 
which has $\1{2}N_{\rm C}(N_{\rm C}-1)$ zero elements 
by fixing of $V$. 
The other orderings are of intermediate dimensions 
between these two moduli matrices.

\section{A Proof of $H_0^2=0$}
\label{PH02=0}
In our wall configurations, 
$H^1$ is generated by the moduli matrix $H_0^1$, 
but $H^2$ always vanishes: 
$H_0^2=0$.
In this Appendix we give a proof of 
$H_0^2=0$. 
In the case of non-degenerate hypermultiplet masses,
this can be proved by requiring convergence of 
$H^2$ at $y\rightarrow \pm \infty$. 
The procedure of the proof is as follows.
First, using finiteness of the solution 
$H^1 = S^{-1} H_0^1 e^{My}$ 
at $y\rightarrow\pm \infty$, 
we will estimate the order of divergence in 
$y\rightarrow \pm \infty$ 
for elements in 
the $N_{\rm C}\times N_{\rm C}$ matrix $S$ 
defined in Eq.~(\ref{def-S}). 
Then, we study conditions for elements of $H_0^2$ 
imposed by convergence of $H^2 (=S^\dagger H_0^2 e^{-My})$ 
at $y\rightarrow \pm \infty$. 
At this stage, most elements of $H_0^2$ are proved to 
vanish. 
The remaining elements in $H_0^2$ are also proved to vanish 
by the orthogonality condition for 
the moduli matrices $H_0^1$ and $H_0^2$: 
$H_0^1 H_0^{2\dagger }=
0$ (\ref{cond-H01H02}).

First, let us investigate conditions for elements of $H_0^2$ 
imposed by convergence of 
$H^2 = S^\dagger H_0^2 e^{-My}$ at $y\rightarrow  +\infty$. 
To this end, introduce the notation $\mathcal{O}_A^{\pm}$ 
which represents the order of $e^{\pm m_A |y|}$ 
at $y\rightarrow  \infty$.
By using the standard form of $H_0^1$,
the order of the leading element in each row of 
$H_0^1 e^{My}$ at $y\rightarrow  +\infty$ 
is found to be 
\unitlength 0.67pt
\begin{eqnarray}
H_0^1 e^{My}=\left(
\begin{array}{c}
\begin{picture}(180,90)(0,0)
\put(0,0){\scalebox{0.3}{\includegraphics{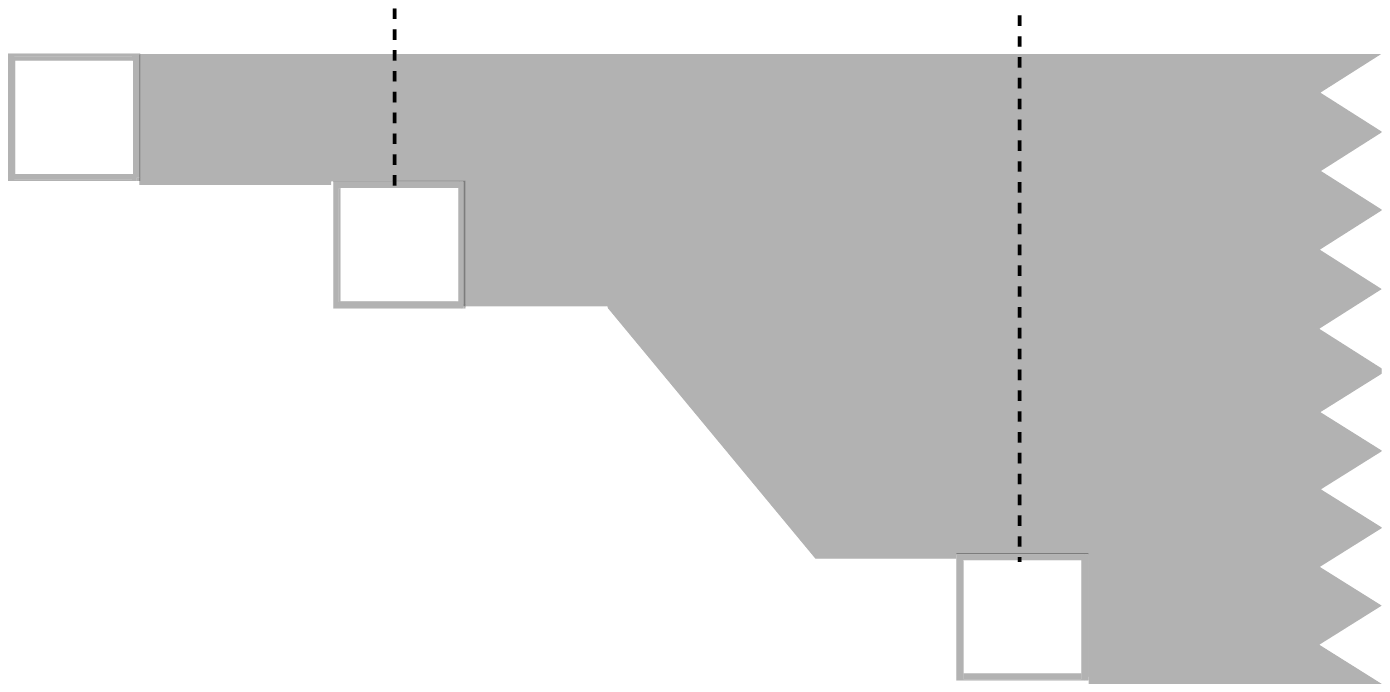}}} 
\end{picture}
\end{array}
\right) ,
\put(-207,55){$A_1$}
\put(-167,55){$A_2$}
\put(-88,55){$A_{N_{\rm C}}$}
\put(-209,34){{\scriptsize $\mathcal{O}_{A_1}^+$}}
\put(-167,18){{\scriptsize $\mathcal{O}_{A_2}^+$}}
\put(-86,-31){{\scriptsize $\mathcal{O}_{A_{N_{\rm C}}}^+$}}
\put(-210,-20){{\Huge $0$}}
\end{eqnarray}
where the order of divergence for subleading elements 
are less than the leading element in each row. 
Therefore, in order that $H^1$ converges at $y\rightarrow +\infty$, 
the orders of $S^{-1}$ and $S^\dagger$ should be 
\begin{eqnarray}
S^{-1}=\left(
\begin{array}{c}
\begin{picture}(80,90)(0,0)
\put(0,0){\scalebox{0.3}{\includegraphics{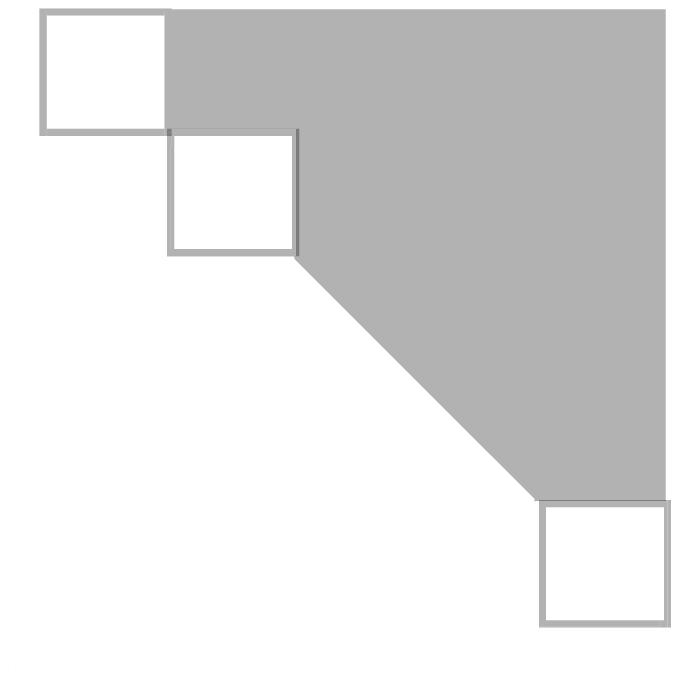}}} 
\put(9,74){{\scriptsize  $\mathcal{O}_{A_1}^-$}}
\put(25,58){{\scriptsize $\mathcal{O}_{A_2}^-$}}
\put(73,10){{\scriptsize $\mathcal{O}_{A_{N_{\rm C}}}^-$}}
\put(20,19){{\Huge $0$}}
\end{picture}
\end{array}
\right) 
\ , \ 
S^{\dagger}=\left(
\begin{array}{c}
\begin{picture}(80,90)(0,0)
 \put(-1,0){\scalebox{0.3}{\includegraphics{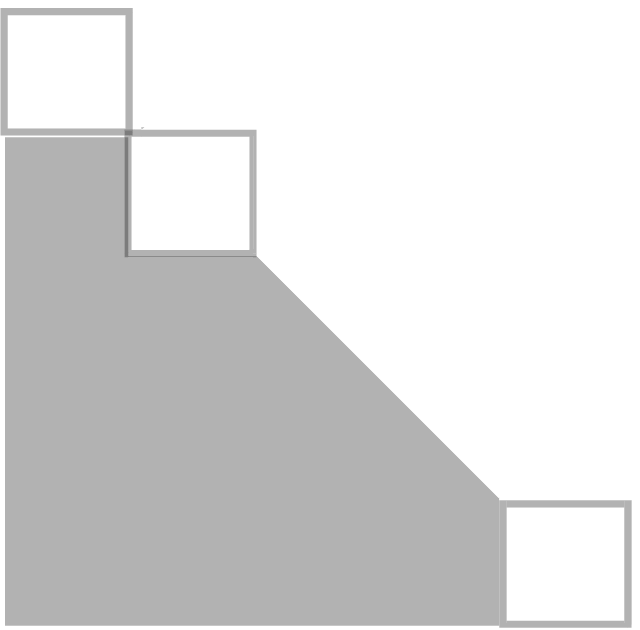}}}
\put(3,70){{\scriptsize  $\mathcal{O}_{A_1}^+$}}
\put(19,54){{\scriptsize $\mathcal{O}_{A_2}^+$}}
\put(67,6){{\scriptsize $\mathcal{O}_{A_{N_{\rm C}}}^+$}}
\put(60,50){{\Huge $0$}}
\end{picture}
\end{array}
\right) ,
\end{eqnarray}
where we have used $U(N_{\rm C})$ gauge symmetry 
to fix $S^{-1}$ as the upper triangular matrix 
with real diagonal elements. 
Since the order of $H_0^2 e^{-My}$ is 
\begin{eqnarray}
H_0^2 e^{-My}=\left(
\begin{array}{c}
\begin{picture}(100,70)(0,0)
\put(0,0){\scalebox{0.3}{\includegraphics{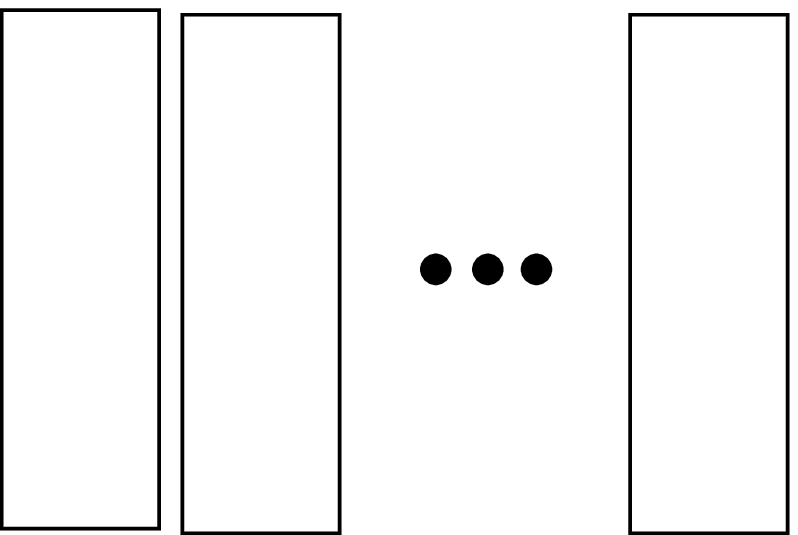}}} 
\end{picture}
\begin{picture}(0,0)(-30,-30)
 \put(-128,0){{\small $\mathcal{O}_{1}^-$}}
\put(-105,0){{\small $\mathcal{O}_{2}^-$}}
\put(-47,0){{\small $\mathcal{O}_{N_{\rm F}}^-$}}
\end{picture}
\end{array}
\right) ,
\end{eqnarray}
the order of the first row of 
$H^2 = S^\dagger H_0^2 e^{-My}$ becomes 
\begin{eqnarray}
H^2 \Bigl|_{\rm 1-st \ row}=(\mathcal{O}_{A_1}^+ \mathcal{O}_{1}^-, \ \mathcal{O}_{A_1}^+ \mathcal{O}_{2}^-, \ \cdots, \ \mathcal{O}_{A_1}^+ \mathcal{O}_{N_{\rm F}}^- ).
\end{eqnarray}
Therefore, convergence of $H^2$ at $y\rightarrow  + \infty$ 
requires
\begin{eqnarray}
H_0^2 \Bigl|_{\rm 1-st \ row}=\left(
\begin{array}{c}
\begin{picture}(200,20)(0,0)
\put(-1,0){\scalebox{0.3}{\includegraphics{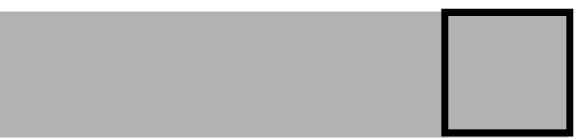}}}
\end{picture}
\begin{picture}(0,0)(-10,-3)
 \put(-162,20){ $A_1$}
\put(-135,0){ $ 0 \ \ \ \cdots \ \ \ \cdots \ \ \  0$}
\end{picture}
\end{array}
\right) .
\end{eqnarray}
Similarly, the $r$-th row of $H_0^2$ 
is of the form 
\begin{eqnarray}
H_0^2 \Bigl|_{\rm r-th \ row}=\left(
\begin{array}{c}
\begin{picture}(200,20)(0,0)
\put(0,0){\scalebox{0.3}{\includegraphics{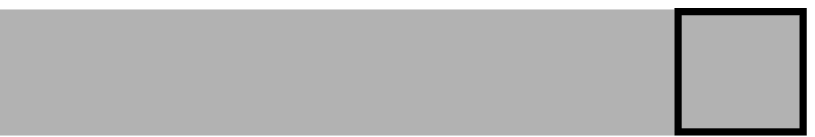}}} 
\end{picture}
\begin{picture}(0,0)(-15,-3)
\put(-133,20){ $A_r$}
\put(-110,0){ $ 0 \ \ \ \cdots \ \ \ \  0$}
\end{picture}
\end{array}
 \right) .
\end{eqnarray}
Then, convergence of $H^2$ at $y\rightarrow  +\infty$ 
requires that $H_0^2$ is in the form of 
\vspace{1em}
\begin{eqnarray}
H_0^2 =\left(
\begin{array}{c}
\begin{picture}(180,80)(0,0)
\put(0,-5){\scalebox{0.3}{\includegraphics{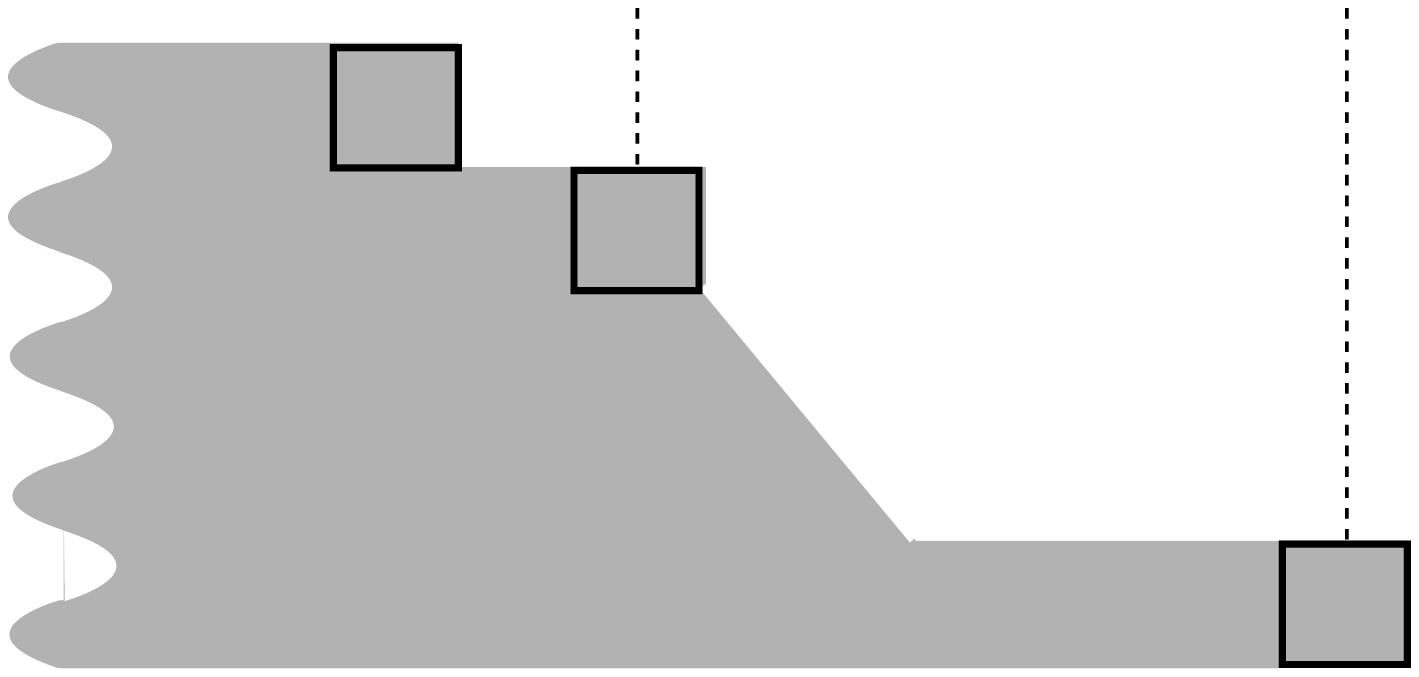}}} 
\end{picture}
\begin{picture}(0,0)(-40,-35)
\put(-90,15){{\Huge $0$}}
\put(-177,55){$A_1$}
\put(-145,55){$A_2 \ \ \cdots$}
\put(-53,55){$A_{N_{\rm C}}$} 
\end{picture}
\end{array}
\right) .
  \label{req1}
\end{eqnarray}

Next, let us investigate conditions on $H_0^2$ 
imposed by convergence at $y\rightarrow  -\infty$. 
The moduli matrix $H_0^1$ in the standard form 
can be transformed to the following form, 
by permuting its rows with a unitary matrix 
$V_{\rm IS}$ in the world-volume symmetry:
\vspace{1em}
\begin{eqnarray}
H_0^1 \ \rightarrow \ V_{\rm IS} H_0^1 =\left(
\begin{array}{c}
\begin{picture}(180,80)(0,0)
\put(0,-5){\scalebox{0.3}{\includegraphics{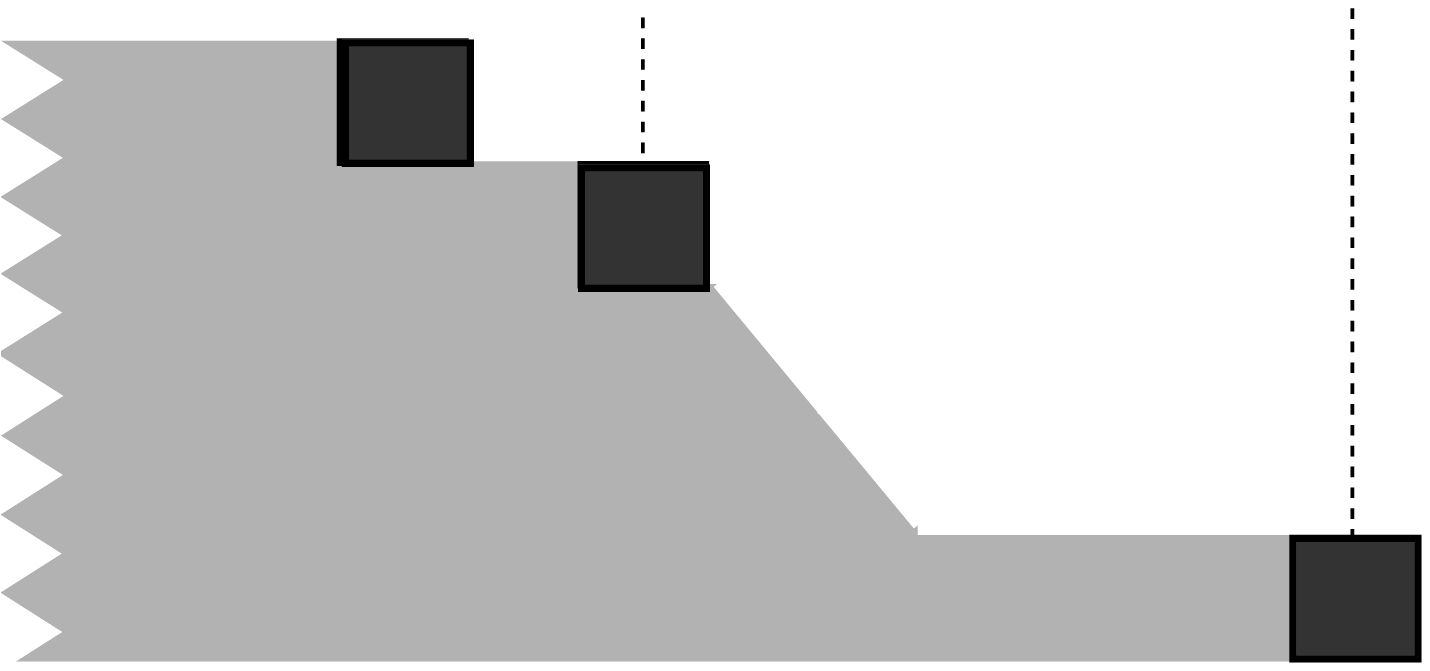}}} 
\end{picture}
\begin{picture}(0,0)(-40,-35)
 \put(-90,15){{\Huge $0$}}
\put(-179,55){\scriptsize $B_{V(1)}$}
\put(-145,55){\scriptsize $B_{V(2)} \ \ \cdots$}
\put(-53,55){\scriptsize $B_{V(N_{\rm C})}$}
\end{picture}
\end{array}
\right) ,
\label{is-h01}
\end{eqnarray}
where color indices $V(r)$ represent 
the permutation of $r$. 
In this equation, $B_{V(r)}$ is the right-most non-vanishing element 
in each column. 
The order of $V_{\rm IS} H_0^1 e^{My}$ at $y\rightarrow  -\infty$ is 
\vspace{1em}
\begin{eqnarray}
 V_{\rm IS} H_0^1 e^{My} =\left(
\begin{array}{c}
\begin{picture}(180,80)(0,0)
\put(0,-5){\scalebox{0.3}{\includegraphics{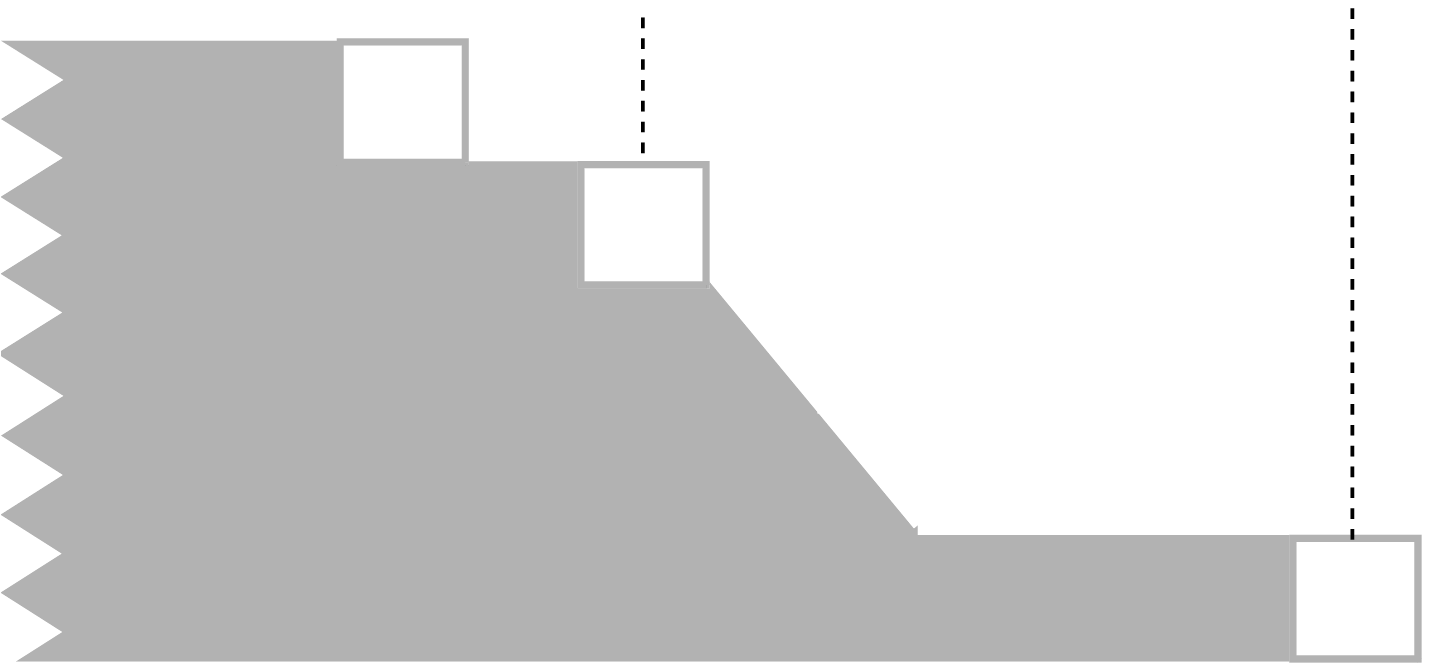}}} 
\end{picture}
\begin{picture}(0,0)(-40,-35)
\put(-90,15){{\Huge $0$}}
\put(-177,55){\scriptsize $B_{V(1)}$}
\put(-145,55){\scriptsize $B_{V(2)} \ \ \cdots$}
\put(-58,55){\scriptsize $B_{V(N_{\rm C})}$}
\put(-175,32){{\scriptsize $\mathcal{O}_{B_{V(1)}}^+$}}
\put(-145,16){{\scriptsize $\mathcal{O}_{B_{V(2)}}^+$}}
\put(-52,-33){{\scriptsize $\mathcal{O}_{B_{V(N_{\rm C})}}^+$}} 
\end{picture}
\end{array}
\right) .
\end{eqnarray}
Convergence of $H^1$ at $y\rightarrow -\infty$ 
requires that 
the orders of $V_{IS}S$ and $(V_{IS}S)^\dagger$ are
\begin{eqnarray}
(V_{\rm IS}S)^{-1}=\left(
\begin{array}{c}
\begin{picture}(80,80)(0,0)
\put(0,-5){\scalebox{0.3}{\includegraphics{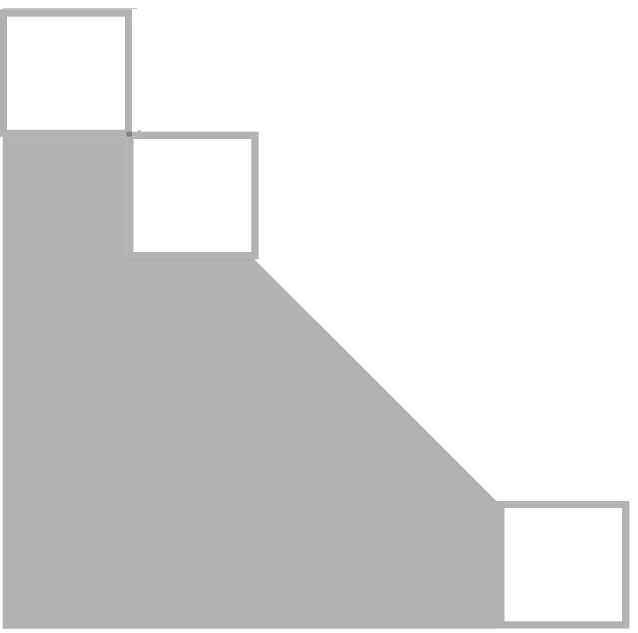}}} 
\end{picture}
\begin{picture}(0,0)(-215,-30)
\put(-291,33){{\scriptsize  $\mathcal{O}_{B_{V(1)}}^+$}}
\put(-274,17){{\scriptsize $\mathcal{O}_{B_{V(2)}}^+$}}
\put(-227,-32){{\scriptsize $\mathcal{O}_{B_{V(N_{\rm C})}}^+$}} 
\put(-230,10){{\Huge $0$}}
\end{picture}
\end{array}
\right) 
\ ,  
(V_{\rm IS}S)^{\dagger}=\left(
\begin{array}{c}
\begin{picture}(80,80)(0,0)
\put(0,-5){\scalebox{0.3}{\includegraphics{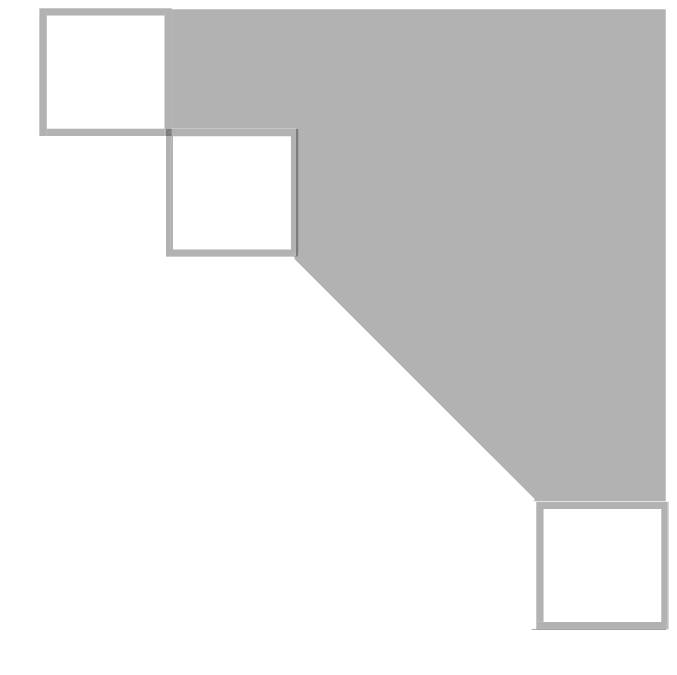}}} 
\end{picture}
\begin{picture}(0,0)(-30,-30)
 \put(-102,38){{\scriptsize  $\mathcal{O}_{B_{V(1)}}^-$}}
\put(-85,22){{\scriptsize $\mathcal{O}_{B_{V(2)}}^-$}}
\put(-37,-27){{\scriptsize $\mathcal{O}_{B_{V(N_{\rm C})}}^-$}}
\put(-90,-20){{\Huge $0$}}
\end{picture}
\end{array}
\right) .
\end{eqnarray}
where we have used $U(N_{\rm C})$ gauge symmetry 
to fix $(V_{\rm IS}S)^{-1}$ 
as the lower triangular matrix with real diagonal elements. 
Note that $H_0^{1,2}$ do not depend on the choice of gauge. 

Since the order of $H_0^2 e^{-My}$ is given by 
\begin{eqnarray}
(V_{\rm IS}^\dagger)^{-1}H_0^2 e^{-My}=\left(
\begin{array}{c}
 \begin{picture}(100,70)(0,0)
\put(-1,0){\scalebox{0.3}{\includegraphics{h02e-my.eps}}}  
 \end{picture}
\begin{picture}(0,0)(-30,-30)
\put(-129,0){{\small $\mathcal{O}_{1}^+$}}
\put(-106,0){{\small $\mathcal{O}_{2}^+$}}
\put(-49,0){{\small $\mathcal{O}_{N_{\rm F}}^+$}} 
\end{picture}
\end{array}
\right) ,
\end{eqnarray}
the order of the $N_{\rm C}$-th row of 
$H^2 (= S^\dagger H_0^2 e^{-My}) 
= (V_{\rm IS} S)^\dagger (V_{\rm IS}^\dagger)^{-1} H_0^2 e^{-My}$ 
at $y \to - \infty$ is found to be
\begin{eqnarray}
H^2 \Bigl|_{N_{\rm C}{\rm -th \ row}}=(\mathcal{O}_{B_{V(N_{\rm C})}}^- \mathcal{O}_{1}^+, \ \mathcal{O}_{B_{V(N_{\rm C})}}^- \mathcal{O}_{2}^+, \ \cdots, \ \mathcal{O}_{B_{V(N_{\rm C})}}^- \mathcal{O}_{N_{\rm F}}^+ ).
\end{eqnarray}
Therefore, convergence of $H^2$ at $y\rightarrow  -\infty$ 
requires 
\begin{eqnarray}
(V_{\rm IS}^\dagger )^{-1} H_0^2 \Bigl|_{\rm N_{\rm C}-th \ row}=\left(
\begin{array}{c}
\begin{picture}(200,10)(-25,5)
\put(0,0){\scalebox{0.3}{\includegraphics{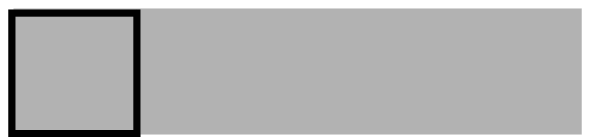}}} 
\end{picture}
\begin{picture}(0,0)(-10,0)
 \put(-102,20){ $B_{V(N_{\rm C})}$}
\put(-220,0){ $ 0 \ \ \ \cdots \ \ \ \cdots \ \ \ 0$}
\end{picture}
\end{array}
\right) .
\end{eqnarray}
Similarly, the $r$-th row of $H_0^2$ 
is found to be 
\begin{eqnarray}
(V_{\rm IS}^\dagger )^{-1} H_0^2 \Bigl|_{\rm r-th \ row}
=\left(
\begin{array}{c}
\begin{picture}(200,10)(0,0)
\put(20,-5){\scalebox{0.3}{\includegraphics{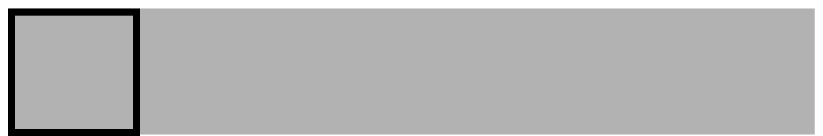}}} 
\end{picture}
 \begin{picture}(0,0)(-10,0)
  \put(-132,20){ $B_{V(r)}$}
\put(-220,0){ $ 0 \ \ \ \cdots \ \ \ \ 0$}
 \end{picture}
\end{array}
\right) .
\end{eqnarray}
In summary, convergence of $H^2$ at $y\rightarrow  -\infty$ 
requires
\vspace{1em}
\begin{eqnarray}
(V_{\rm IS}^\dagger)^{-1}H_0^2=\left(
\begin{array}{c}
 \begin{picture}(200,80)(0,0)
\put(10,-5){\scalebox{0.3}{\includegraphics{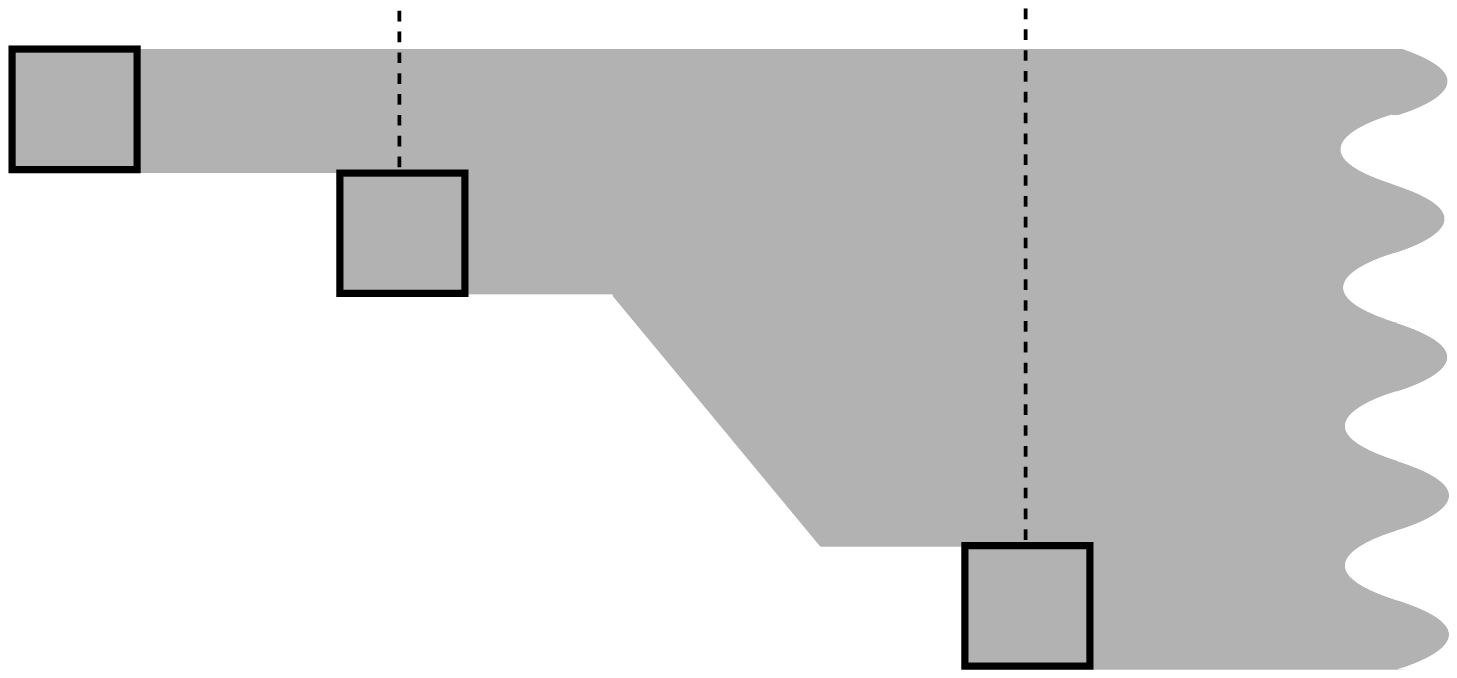}}}  
 \end{picture}
\begin{picture}(0,0)(-25,-35)
 \put(-165,55){$B_{V(2)} \ \cdots$}
\put(-210,55){$B_{V(1)}$}
\put(-87,55){$B_{V(N_{\rm C})}$}
\put(-210,-20){{\Huge $0$}}
\end{picture}
\end{array}
\right) .
\end{eqnarray}
By noting the unitarity of $V_{IS}$, 
$H_0^2$ is found to be of the form 
\vspace{1em} 
\begin{eqnarray}
H_0^2=\left(
\begin{array}{c}
\begin{picture}(200,80)(0,0)
\put(10,-5){\scalebox{0.3}{\includegraphics{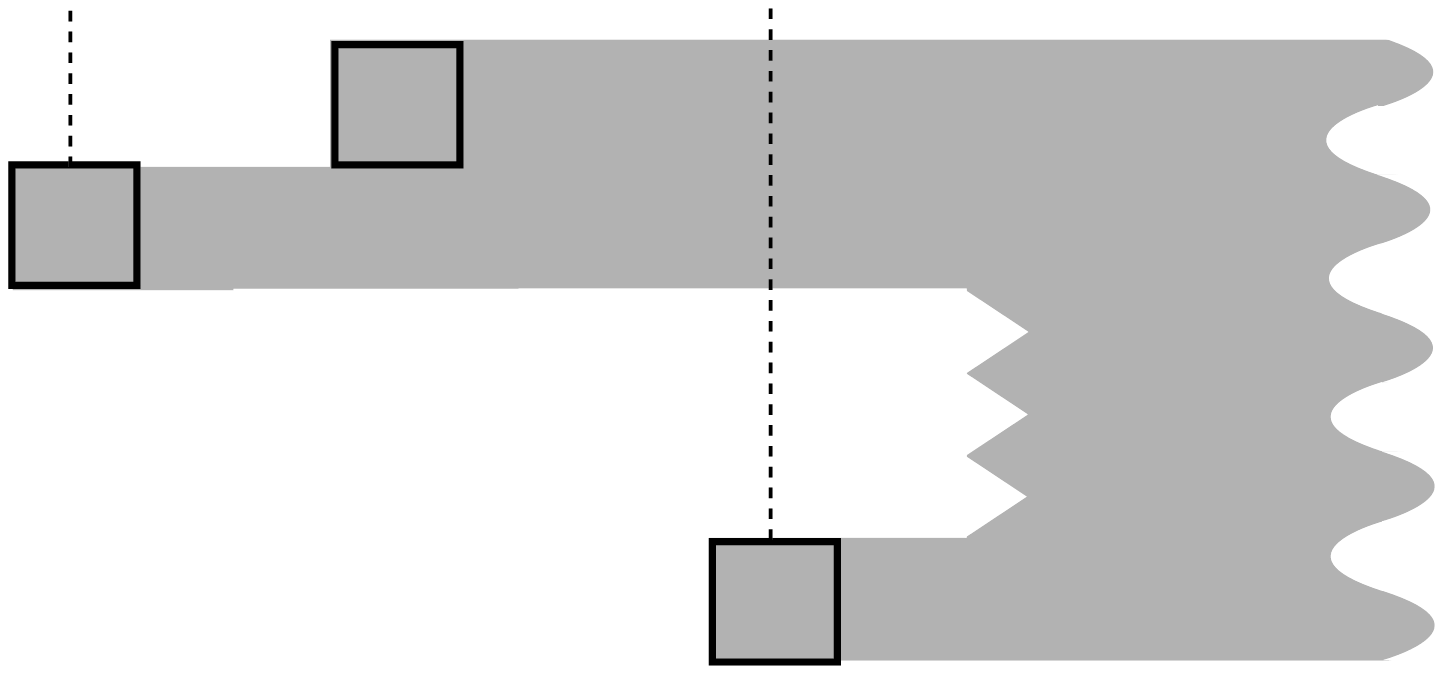}}}  
\end{picture}
\begin{picture}(0,0)(-25,-30)
\put(-170,55){$B_{1} \ \cdots$}
\put(-210,55){$B_{2}$}
\put(-120,55){$B_{N_{\rm C}}$}
\put(-210,-20){{\Huge $0$}}  
\end{picture}
\end{array}
\right) .
\label{req2}
\end{eqnarray}

Combining requirements (\ref{req1}) and (\ref{req2}) for $H_0^2$ 
and noting the relation $A_r \le B_r$, 
we find that if $A_r \neq B_r$ for some $r$
all elements in $A_r< A <B_r$ for each color $r$ have to vanish 
and that if $A_r=B_r$ for some $r$, 
the $(r,A_r)$ element is not required to vanish. 
Such remaining elements are also required to vanish 
by the orthogonality condition for moduli matrices 
$H_0^1$ and $H_0^2$: 
$H_0^1 H_0^{2\dagger }=0
$. 
The proof is completed.

\medskip
In the above proof we have assumed non-degenerate masses. 
If there exist some degenerate masses, 
$H_0^2 = 0$ 
needs not hold anymore as is shown below.
In this degenerate mass case the vacua are no longer 
discrete and there exist continuous degeneracy along  
(quasi-)Nambu-Goldstone directions 
in the moduli space of vacua.  
Then $H^2$ can be non-vanishing along non-compact directions 
corresponding to 
the quasi-Nambu-Goldstone (but not the Nambu-Goldstone) 
modes at both infinities. 
However this does not imply 
that $H^2$ includes additional moduli parameters, 
because such transformation to 
the quasi-Nambu-Goldstone directions 
does not have localized modes. 

We now show that non-vanishing components of $H_0^2$ 
can occur only for the degenerate mass flavor, 
and moreover only if all color components of $H_0^1$ 
with the same degenerate mass flavor 
combination are orthogonal to $H_0^2$. 
The proof given above holds until (\ref{req2})
by replacing each column (flavor) in the proof 
by a set of some columns (flavors) 
with degenerate masses. 
Then $A_r$ and $B_r$ represent 
row vectors $\vec{H_0^1}_{(s,A_r)}$ and 
$\vec{H_0^2}_{(s,B_r)}$ ($s=1,\cdots,N_{\rm C}$) respectively 
of the size $M_r$ of flavors with degenerate masses.  
In the case of $A_r \neq B_r$ 
all elements between $A_r$ and $B_r$ vanish 
in the same way with the degenerate case.
However if $A_r = B_r$ holds for some $r$ 
with degenerate masses, 
$\vec{H_0^2}_{(r,B_r)}$ does not vanish 
in general by the orthogonality condition 
between $H_0^1$ and $H_0^2{}^\dagger$ 
in contrast to the degenerate case.
Instead we have non-zero 
$\vec{H_0^2}_{(r,B_r)}$, say $|\vec{H_0^2}_{(r,B_r)}|^2 \neq 0$, 
with satisfying $\vec{H_0^1}_{(s,B_r)}\cdot 
\vec{H_0^2}_{(r,B_r)}^\dagger=0$ 
with all $s$-rows.
They can be written in the forms of 
\begin{eqnarray}
 H_0^1
= \left(
\begin{array}{ccc}
  *       & \vec{H_0^1}_{(1,B_r)}         & *          \\
  *       & \vdots                        & *          \\
 0\cdots 0& \vec{H_0^1}_{(r,B_r)}         & 0 \cdots 0 \\
  *       & \vdots                        & *          \\
  *       & \vec{H_0^1}_{(N_{\rm C},B_r)} & *          \\
\end{array}\right) , 
 H_0^2 
= \left(
\begin{array}{ccc}
  *       & \vec{0}               & *          \\
  *       & \vec{0}               & *          \\
 0\cdots 0& \vec{H_0^2}_{(r,B_r)} & 0 \cdots 0 \\
  *       & \vec{0}               & *          \\
  *       & \vec{0}               & *          \\
\end{array}\right).
\end{eqnarray} 
However $H^2$ generated by (some) 
non-zero $\vec{H_0^2}_{(r,B_r)}$ 
does not depend on the extra dimension $y$ 
but is fixed by the boundary condition. 
In addition the same row $\vec{H_0^1}_{(r,B_r)}$ 
do not generate any localized modes but 
are determined by the boundary condition. 
We thus have found that 
the non-vanishing $H_0^2$ components 
and the corresponding color components in 
$H_0^1$ are decoupled from the rest of the system 
and frozen to the vacuum value determined by boundary 
conditions. 

In the end we briefly make a comment on (non-)normalizability of 
non-Abelian flavor symmetry for degenerate masses.
We can fix $\vec{H_0^2}_{(r,B_r)}$ using the flavor symmetry $U(M_r)$ as 
$\vec{H_0^2}_{(r,B_r)} = (\alpha, 0, \cdots, 0)$ with $\alpha \in {\bf R}$. 
Here $\alpha$ is determined by the boundary condition and 
it is non-normalizable quasi-Nambu-Goldstone modes.
This breaks flavor symmetry to $U(M_r-1)$ and 
$\vec{H_0^1}_{(s,B_r)} = (0,*,\cdots,*)$ hold for all $s$.  
Other rows $\vec{H_0^1}_{(s,B_r)}$ ($s \neq r$) transform 
under the unbroken flavor symmetry $U(M_r-1)$ 
which is broken to the subgroup. 
Most of non-Abelian modes arose from this breaking 
are not localized but some with gauge symmetry transformation 
may be localized.  
\section{Duality between $U(N_{\rm C})$ and $U(\tilde N_{\rm C})$ 
at Infinite Coupling}
\label{BPSDBPSIC}
In this section, we discuss the dual relation between 
a $U(N_{\rm C})$ theory and a $U(N_{\rm F}-N_{\rm C}=\tilde N_{\rm C})$ theory
with fixed $N_{\rm F}$ appearing at the limit of infinite gauge coupling.
For simplicity, let us assume that $H^2=0$.
Under this assumption the constraint on $H^1$ (\ref{constraint2}) reduces to 
\begin{eqnarray}
 H^1H^1{}^\dagger =c{\bf 1}_{N_{\rm C}},
\end{eqnarray} 
and the components of vector multiplet which is composed by
hypermultiplets (\ref{constraint1}) are,   
\begin{eqnarray}
 \Sigma =c^{-1}H^1MH^1{}^\dagger ,\quad W_M=ic^{-1}(\partial_M H^1)H^1{}^\dagger .\label{constraintH1}
\end{eqnarray}
Thus the Lagrangian (\ref{reduced-L}) is also reduced to
\begin{eqnarray}
\mathcal{L}^{g\rightarrow\infty}&=&
{\rm Tr}_{\rm F}[({\cal D}_M H^1)^\dagger {\cal D}^M H^1] 
- {\rm Tr}_{\rm F}[(H^1{}^\dagger \Sigma - M H^1{}^\dagger)(\Sigma H^1 - H^1 M)],
\nonumber\\
&=&{\rm Tr}_{\rm F}[\left(
\partial _M H^1{}^\dagger \partial ^M H^1-MH^1{}^\dagger H^1M\right)
\left({\bf 1}_{N_{\rm F}}-c^{-1}H^1{}^\dagger H^1\right)]. 
\end{eqnarray}
In this form of the Lagrangian, an explicit duality relation can be easily
found as follows.
Let us introduce a normalized $\tilde N_{\rm C}\times N_{\rm F}$ matrix 
$\tilde H^1$ orthogonal to $H^1$,    
\begin{eqnarray}
 H^1\tilde H^1{}^\dagger =0, \quad 
\tilde H^1\tilde H^1{}^\dagger =c{\bf 1}_{\tilde N_{\rm C}}.\label{orthogonal}
\end{eqnarray}
These equations and the constraint (\ref{constraintH1}) make 
an $N_{\rm F}\times N_{\rm F}$ matrix 
$U^\dagger =c^{-\frac{1}{2}}(H^1{}^\dagger ,\tilde H^1{}^\dagger )$  unitary, 
$U U^\dagger ={\bf 1}_{N_{\rm F}}$. 
$U^\dagger U={\bf 1}_{N_{\rm F}}$  indicates the other expression of
(\ref{constraintH1}) and (\ref{orthogonal}),  
\begin{eqnarray}
 H^1{}^\dagger H^1+\tilde H^1{}^\dagger \tilde H^1=c {\bf 1}_{N_{\rm F}}.\label{unitaryH}
\end{eqnarray}
By use of this equation, the Lagrangian can be rewritten as
\begin{eqnarray}
 \mathcal{L}^{g\rightarrow\infty}&=&
c^{-1}{\rm Tr}_{\rm F}[\left(
\partial _M H^1{}^\dagger \partial ^M H^1-MH^1{}^\dagger H^1M\right)
\tilde H^1{}^\dagger \tilde H^1] \nonumber\\
&=&c^{-1}{\rm Tr}_{\rm F}[\left(
\partial _M \tilde H^1{}^\dagger \partial ^M \tilde H^1-M \tilde H^1{}^\dagger \tilde H^1M\right)
H^1{}^\dagger H^1],
\end{eqnarray}
where we used the orthogonality 
between $H^1$ and $\tilde H^1{}$ to show the second line.
Therefore we find that $\tilde H^1$ defined by Eq.~(\ref{orthogonal}) gives 
scalars of hypermultiplets 
in the dual theory, 
where components of composite $U(\tilde N_{\rm C})$ vector multiplets 
are given by,
\begin{eqnarray}
\tilde \Sigma =
c^{-1}\tilde H^1M\tilde H^1{}^\dagger ,\quad 
\tilde W_M=ic^{-1} (\partial_M \tilde H^1)\tilde H^1{}^\dagger .
\end{eqnarray}
Note that there is a direct relation between $\Sigma $ and $\tilde \Sigma $,    
\begin{eqnarray}
 {\rm Tr}_{\rm C}(\Sigma )+{\rm Tr}_{\tilde {\rm C}}(\tilde \Sigma )={\rm Tr}_{\rm F}(M),
\end{eqnarray}
which is obtained by multiplying the mass matrix $M$ to the both sides of
Eq.~(\ref{unitaryH}) and taking a trace.
 
An explicit dual relation in terms of the wall moduli manifolds 
can also be obtained.
The BPS equation for $H^1$ (\ref{BPSeq-H}) is rewritten as  
\begin{eqnarray}
 0=({\cal D}_y+\Sigma )H^1-H^1M=c^{-1}(\partial _yH^1-H^1M)\tilde H^1{}^\dagger \tilde H^1,
\end{eqnarray}
and by right-multiplication of $\tilde H^1$, we obtain a simple form of
the BPS equation, 
\begin{eqnarray}
 \partial _yH^1 \tilde H^1{}^\dagger =H^1M\tilde H^1{}^\dagger .
\end{eqnarray}
By use of Eq.~(\ref{orthogonal}), we obtain a dual equation for $\tilde H^1$,
\begin{eqnarray}
 \partial _y\tilde H^1 H^1{}^\dagger =-\tilde H^1M H^1{}^\dagger.
\end{eqnarray}
Thus if $H^1$ satisfies the BPS equation (\ref{BPSeq-H}), 
$\tilde H^1$ satisfies an anti-BPS equation, 
\begin{eqnarray}
 (\partial _y+i\tilde W_y)\tilde{H}^1=\tilde{\Sigma} \tilde{H}^1-\tilde{H}^1M,
\end{eqnarray}
which is also solved as 
\begin{eqnarray}
 \tilde H^1=\tilde S^{-1}\tilde H_0^1e^{-My},\quad \tilde \Sigma -i\tilde W_y=-
\tilde S^{-1}(\partial _y\tilde S),
\quad  \tilde S\tilde S^\dagger =c^{-1}\tilde H_0^1e^{-2My}\tilde H_0^1{}^\dagger 
\end{eqnarray}
with a dual moduli matrix $\tilde H_0^1$. 
The orthogonality (\ref{orthogonal}) is rewritten to the orthogonality
of the moduli matrices as, 
\begin{eqnarray}
 H_0^1\tilde H_0^1{}^\dagger =0.
\end{eqnarray}
This relation defines a one-to-one map from a point to a point 
on the Grassmann manifold.

\section{Proof of (\ref{eq:num-top-sec-BPS}): $N_{\rm BPS}$}
\label{PONNBPS}
We need a somewhat technical procedure 
to obtain the number of topological sectors with BPS saturated states 
(\ref{eq:num-top-sec-BPS}), 
\begin{eqnarray}
 N_{\rm BPS}={N_{\rm F}!\over N_{\rm C}!\tilde N_{\rm C}!}
{(N_{\rm F}+1)!\over (N_{\rm C}+1)!(\tilde N_{\rm C}+1)!}. 
\end{eqnarray}
Let us call ${\cal C}^r_{A_r,B_r}$ 
the number of sets 
of flavors $\{A_1,A_2,\cdots,A_{r-1}\}$ and
$\{B_1,B_2,\cdots,B_{r-1}\}$ which satisfy  
\begin{eqnarray}
\begin{array}{ccccccccc}
 1&\leq &A_1 &< &A_2&<\cdots<&A_{r-1}&<&A_r \\
&&\rotatebox[origin=c]{-90}{$\leq$}&&\rotatebox[origin=c]{-90}{$\leq$}&
&\rotatebox[origin=c]{-90}{$\leq$}&&\rotatebox[origin=c]{-90}{$\leq$} \\
&&B_1 &< &B_2&<\cdots<&B_{r-1}&<&B_r\\
&\multicolumn{7}{c}{\underbrace{\hspace{13em}}_{{\cal C}^r_{A_r,B_r}}}
\end{array}
\end{eqnarray}
with the $r$-th flavors $A_r$ $B_r$ fixed. 
We find that a recurrence formula 
for ${\cal C}^r_{A_r,B_r}$ and a initial value are given by
\begin{eqnarray}
 {\cal C}^{r+1}_{A_{r+1},B_{r+1}}=
\sum_{A_{r}=r}^{A_{r+1}-1}\sum_{B_r=A_r}^{B_{r+1}-1}{\cal C}^r_{A_r,B_r},
\quad {\cal C}^1_{A_1,B_1}=1.
\end{eqnarray}
Note that the indices $A_r$ are summed from a color $r$. 
By induction, we can prove the following
formula for ${\cal C}^r_{A_r,B_r}$ 
\begin{eqnarray}
 {\cal C}^r_{A_r,B_r}={(r B_r-(r-1)A_r)\over r!(r-1)!}
{(A_r-1)!\over (A_r-r)!}{(B_r-1)!\over (B_r-r+1)!} \, .
\end{eqnarray}
The number of BPS states $N_{\rm BPS}$ 
is obtained by summing over the 
$N_{\rm C}$-th flavor indices $A_{N_{\rm C}},B_{N_{\rm C}}$ of the 
${\cal C}^{N_{\rm C}}_{A_{N_{\rm C}},B_{N_{\rm C}}}$  
\begin{eqnarray}
\begin{array}{ccccc}
\multicolumn{4}{c}{\overbrace{\hspace{12em}}^{N_{\rm BPS}}}\\
\cdots<&A_{N_{\rm C}-1 }&<&A_{N_{\rm C}}&\\
&\rotatebox[origin=c]{-90}{$\leq$}&&\rotatebox[origin=c]{-90}{$\leq$}& \\
\cdots<&B_{N_{\rm C}-1} &<&B_{N_{\rm C}}&\leq N_{\rm F}.\\
\multicolumn{2}{c}{\underbrace{\hspace{7em}}_{{\cal C}^{N_{\rm C}}_{A_{N_{\rm C}},B_{N_{\rm C}}}}}
\end{array}
\end{eqnarray}
As a result, we find that $N_{\rm BPS}$ is given by  
\begin{eqnarray}
 N_{\rm BPS}=\sum_{A=N_{\rm C}}^{N_{\rm F}}
\sum_{B=A}^{N_{\rm F}}{\cal C}^{N_{\rm C}}_{A,B}
={\cal C}^{N_{\rm C}+1}_{N_{\rm F}+1,N_{\rm F}+1}.
\end{eqnarray}


\section{The Standard Forms for the 
$N_{\rm C} =2$ and $N_{\rm F}=4$ Case.} \label{STF}
We present the matrices in the standard forms in the case of 
$N_{\rm C} =2$ and $N_{\rm F}=4$. 
Following the way that we explained in Appendix~\ref{TSFH01}, 
the moduli matrices in this case are classified to 25 types of the
matrices in the standard form.  

First of all, this model contains six vacua 
which are determined by matrices in the standard form, given by
\begin{eqnarray}
 H_{0\langle12\rangle}&=&
\sqrt{c}
\left( 
\begin{array}{cccc}
1&0 &0 &0\\
0&1 &0&0
\end{array}
\right),
 \ \ 
 H_{0\langle 13\rangle}=
 \sqrt{c}
\left( 
\begin{array}{cccc}
1&0 &0 &0\\
0&0 &1&0
\end{array}
\right),
 \ \ 
 H_{0\langle 14\rangle}=
\sqrt{c}
\left( 
\begin{array}{cccc}
1&0 &0 &0\\
0&0 &0&1
\end{array}
\right),\nonumber \\
 H_{0\langle 23\rangle}&=&
 \sqrt{c}
\left( 
\begin{array}{cccc}
0&1 &0 &0\\
0&0 &1&0
\end{array}
\right),
 \ \ 
 H_{0\langle 24 \rangle}=
 \sqrt{c}
\left( 
\begin{array}{cccc}
0&1 &0 &0\\
0&0 &0&1
\end{array}
\right),
 \ \ 
 H_{0\langle 34\rangle}=
 \sqrt{c}
\left( 
\begin{array}{cccc}
0&0 &1 &0\\
0&0 &0&1
\end{array}
\right).\nonumber \\
\end{eqnarray}
Second, there exist six elementary walls generated by 
matrices in the standard form
\begin{eqnarray}
&& 
H_{0\langle 12\leftarrow 13\rangle } =
\sqrt{c}
 \left( 
 \begin{array}{cccc}
 1&0 &0 &0\\
 0&1 &e^{r_1}&0
 \end{array}
 \right),
 \ \ 
H_{0\langle 13\leftarrow 14\rangle }=
\sqrt{c}
 \left( 
 \begin{array}{cccc}
 1&0 &0 &0\\
 0&0 &1&e^{r_2}
 \end{array}
 \right), \nonumber\\
&&
H_{0\langle 13\leftarrow 23\rangle }=
\sqrt{c}
 \left( 
 \begin{array}{cccc}
 1&e^{r_3} &0 &0\\
 0&0 &1&0
 \end{array}
 \right),
H_{0\langle 14\leftarrow 24\rangle } =
\sqrt{c}
 \left( 
 \begin{array}{cccc}
 1& e^{r_3}&0 &0\\
 0&0 &0&1
 \end{array}
 \right), \nonumber\\
&&
H_{0\langle 23\leftarrow 24\rangle }=
\sqrt{c}
 \left( 
 \begin{array}{cccc}
 0&1 &0 &0\\
 0&0 &1&e^{r_2}
 \end{array}
 \right),
 \ \ 
H_{0\langle 24\leftarrow 34\rangle }=
\sqrt{c}
 \left( 
 \begin{array}{cccc}
 0&1 &e^{r_4} &0\\
 0&0 &0&1
 \end{array}
 \right),
\end{eqnarray}
as well as several compressed single walls which we have omitted. 

Third, the seven double wall configurations are given by 
\begin{eqnarray}
&&
H_{0\langle 12\leftarrow 14\rangle } =
\sqrt{c}
\left( 
\begin{array}{cccc}
1&0&0 &0\\
0&1 &e^{r_1}&e^{r_1+r_2}
\end{array}
\right),
 \ \ 
 H_{0\langle 12\leftarrow 23\rangle }=
 \sqrt{c}
\left( 
\begin{array}{cccc}
1&e^{r_3} &0 &0\\
0&1 &e^{r_1}&0
\end{array}
\right), \nonumber \\
&&
H_{0\langle 12\leftarrow 32\rangle }=
\sqrt{c}
 \left( 
 \begin{array}{cccc}
 1&0 &e^{r_5} &0\\
 0&1 &0&0
 \end{array}
 \right), \ \
H_{0\langle 13\leftarrow 24\rangle } =
\sqrt{c}
 \left( 
 \begin{array}{cccc}
 1&e^{r_3} &0 &0\\
 0&0 &1&e^{r_2}
 \end{array}
 \right),\nonumber\\
&&
H_{0\langle 14\leftarrow 34\rangle } =
\sqrt{c}
 \left( 
 \begin{array}{cccc}
 1&e^{r_3} &e^{r_3+r_4} &0\\
 0&0 &0&1
 \end{array}
 \right),
 \ \ 
H_{0\langle 23\leftarrow 34\rangle }=
\sqrt{c}
 \left( 
 \begin{array}{cccc}
 0&1 &e^{r_4} &0\\
 0&0 &1&e^{r_2}
 \end{array}
 \right), \nonumber \\
&& 
H_{0\langle 23\leftarrow 43\rangle }=
\sqrt{c}
 \left( 
 \begin{array}{cccc}
 0&1 &0 &e^{r_7}\\
 0&0 &1&0
 \end{array}
 \right),
\end{eqnarray}
where the third and the last matrices contain compressed walls. 

The triple wall configurations are generated by
\begin{eqnarray}
H_{0\langle 12\leftarrow 24\rangle }&=&
\sqrt{c}
\left( 
\begin{array}{cccc}
1&e^{r_3} &0 &0\\
0&1 &e^{r_1}&e^{r_1+r_2}
\end{array}
\right),
 \ \ 
 H_{0\langle 12\leftarrow 42\rangle }=
 \sqrt{c}
\left( 
\begin{array}{cccc}
1&0 &e^{r_5} &e^{r_5+r_6}\\
0&1 &0&0
\end{array}
\right),\nonumber \\
H_{0\langle 13\leftarrow 34\rangle }&=&
\sqrt{c}
\left( 
\begin{array}{cccc}
1&e^{r_3} &e^{r_3+r_4} &0\\
0&0 &1&e^{r_2}
\end{array}
\right),
 \ \ 
 H_{0\langle 13\leftarrow 43\rangle }=
 \sqrt{c}
\left( 
\begin{array}{cccc}
1& e^{r_3}&0 &e^{r_3+r_7}\\
0&0 &1&0
\end{array}
\right).    \label{triple-}
\end{eqnarray}
The second and the last matrices represent 
compressed triple walls.  

In the end, four walls and 
a compressed triple wall are given by
\begin{eqnarray}
H_{0\langle 12\leftarrow 34\rangle }&=&
\sqrt{c}
\left( 
\begin{array}{cccc}
1&e^{r_3} &e^{r_3+r_4} &0\\
0&1 &e^{r_1}&e^{r_1+r_2}
\end{array}
\right),
 H_{0\langle 12\leftarrow 43\rangle }=
 \sqrt{c}
\left( 
\begin{array}{cccc}
1&0 &e^{r_5} &e^{r_5+r_6}\\
0&1 &e^{r_1}&0
\end{array}
\right), \label{4-wall-st}
\end{eqnarray}
respectively. 

As we explained in Sec.~\ref{NC2NF3Case}, one can discuss relations
between parameters of moduli matrices by using world-volume symmetry
and taking appropriate limit.
The complex parameters $r_5,r_6$ 
in, for instance $H_{0\langle 12\leftarrow 43\rangle }$ are 
related to the complex parameters $r_1,r_2,r_3,r_4$ 
which parametrize generic part of the moduli as 
\begin{eqnarray}
 r_5&=&r_1+r_3+\log(e^{r_4-r_1}-1),\quad 
 r_6=r_2-\log(e^{r_4-r_1}-1)+\pi i,
\end{eqnarray}
with limits $ r_2\rightarrow -\infty ,\quad r_3\rightarrow \infty $ 
and $r_4\rightarrow r_1$, which can be shown by considering a row-reduced 
echelon
form of $H_{0\langle 12\leftarrow 34\rangle }$.
With a row-reduced echelon form of $H_{0\langle 13\leftarrow 34\rangle }$, 
the parameter 
$r_7$ in $H_{0\langle 13\leftarrow 43\rangle }$ is obtained in the limit
\begin{eqnarray}
  r_7&=&r_2+r_4+\pi i,\quad r_2\rightarrow -\infty ,\quad 
  r_1\rightarrow \infty .
\end{eqnarray}

\newcommand{\J}[4]{{\sl #1} {\bf #2} (#3) #4}
\newcommand{\andJ}[3]{{\bf #1} (#2) #3}
\newcommand{\AP}{Ann.\ Phys.\ (N.Y.)}
\newcommand{\MPL}{Mod.\ Phys.\ Lett.}
\newcommand{\NP}{Nucl.\ Phys.}
\newcommand{\PL}{Phys.\ Lett.}
\newcommand{\PR}{ Phys.\ Rev.}
\newcommand{\PRL}{Phys.\ Rev.\ Lett.}
\newcommand{\PTP}{Prog.\ Theor.\ Phys.}
\newcommand{\hep}[1]{{\tt hep-th/{#1}}}


\begin{thebibliography}{100}

  \bibitem{HoravaWitten}     
    P.~Horava and E.~Witten, 
     Nucl.\ Phys.\ {\bf B460}, 506 (1996) [arXiv:hep-th/9510209]. 
     
 \bibitem{LED}N.~Arkani-Hamed, S.~Dimopoulos and G.~Dvali, 
             Phys.\ Lett.\ {\bf B429}, 263  (1998) 
             [arXiv:hep-ph/9803315]; 
             I.~Antoniadis, N.~Arkani-Hamed, S.~Dimopoulos 
             and G.~Dvali, 
             Phys.\ Lett.\ {\bf B436}, 257  (1998) 
             [arXiv:hep-ph/9804398]. 
 \bibitem{RandallSundrum}L.~Randall and R.~Sundrum, 
             Phys.\ Rev.\ Lett.\ 
             {\bf 83}, 3370 (1999)  [arXiv:hep-ph/9905221]; 
             Phys.\ Rev.\ Lett.\ {\bf 83}, 4690  (1999) 
             [arXiv:hep-th/9906064].

 \bibitem{WittenOlive} E.~Witten and D.~Olive, 
             Phys.\ Lett.\  {\bf B78}, 97 (1978).

\bibitem{Cvetic:1991vp}
M.~Cvetic, F.~Quevedo and S.~J.~Rey,
Phys.\ Rev.\ Lett.\  {\bf 67}, 1836 (1991);
M.~Cvetic, S.~Griffies and S.~J.~Rey,
Nucl.\ Phys.\ B {\bf 381}, 301 (1992)
[arXiv:hep-th/9201007];
M.~Cvetic, S.~Griffies and H.~H.~Soleng,
Phys.\ Rev.\ D {\bf 48}, 2613 (1993)
[arXiv:gr-qc/9306005].

\bibitem{AT} E.~Abraham and P.~K.~Townsend, 
             Phys.\ Lett.\ {\bf B 291}, 85 (1992).


  \bibitem{DGSW}
    S.~Dimopoulos and H. Georgi, 
     Nucl.\ Phys.\ {\bf B193}, 150 (1981); 
    N.~Sakai, 
     Z.\ f.\ Phys.\ {\bf C11}, 153 (1981);
    E.~Witten, 
     Nucl.\ Phys.\ {\bf B188}, 513 (1981);
    S.~Dimopoulos, S.~Raby and F.~Wilczek, 
     Phys.\ Rev.\ {\bf D24}, 1681 (1981).

  \bibitem{Rubakov}
 V.~A.~Rubakov, 
 Phys.~Usp.~{\bf 44}, 871  (2001) 
 [arXiv:hep-ph/0104152]. 

\bibitem{DvaliShifman}  
     G.~Dvali and M.~Shifman, 
      Phys.\ Lett.\  {\bf  B396}, 64 (1997) 
      [arXiv:hep-th/9612128]. 

\bibitem{Akhmedov}  E.~K.~Akhmedov, 
         Phys.\ Lett.\  {\bf  B521}, 79 (2001) 
         [arXiv:hep-th/0107223]. 


\bibitem{DubovskyRubakov} S.~L.~Dubovsky and V.A.~Rubakov,  
         Int.\ J.\ Mod.\ Phys. {\bf A16}, 4331 (2001) 
        [arXiv:hep-ph/0105243]. 

\bibitem{ShifmanYung}  M.~Shifman and A.~Yung, 
 Phys.\ Rev.\ {\bf D67}, 125007 (2003)  
               [arXiv:hep-th/02122293]. 


\bibitem{MaruSakai}   N.~Maru and N.~Sakai, 
     Prog.~Theor.~Phys.~{\bf 111}, 907 
     (2004) 
         [arXiv:hep-th/0305222]. 
\bibitem{IOS1} Y.~Isozumi, K.~Ohashi, and N.~Sakai, 
  JHEP\ {\bf 11}, 060 (2003) 
  [arXiv:hep-th/0310189]. 

\bibitem{IOS2} Y.~Isozumi, K.~Ohashi, and N.~Sakai, 
 JHEP\ {\bf 11}, 061 (2003)
  [arXiv:hep-th/0310130]. 

\bibitem{SY2}
M.~Shifman and A.~Yung,
 Phys.\ Rev.\ {\bf D70}, 025013 (2004)  
[arXiv:hep-th/0312257].

%
\bibitem{ANS}
M.~Arai, M.~Nitta and N.~Sakai, to appear in Prog.Theor.Phys. 
[arXiv:hep-th/0307274];
to appear in the Proceedings of the 3rd International Symposium on Quantum Theory and Symmetries (QTS3), September 10-14, 2003, 
[arXiv:hep-th/0401084];
to appear in the Proceedings of the International Conference on ``Symmetry Methods in Physics (SYM-PHYS10)'' held at Yerevan, Armenia, 13-19 Aug. 2003
[arXiv:hep-th/0401102]; 
to appear in the Proceedings of  
SUSY 2003 held at the University of Arizona, Tucson, AZ, June 5-10, 2003
[arXiv:hep-th/0402065].


\bibitem{INOS}
Y.~Isozumi, M.~Nitta, K.~Ohashi and N.~Sakai, 
Phys.Rev.Lett.{\bf 93}, 161601 (2004) 
[arXiv:hep-th/0404198].

%
%
\bibitem{Wi}
E.~Witten,
Nucl.\ Phys.\  {\bf B460}, 541 (1996)
[arXiv:hep-th/9511030].

\bibitem{brane-monopole}
M.~B.~Green and M.~Gutperle,
Phys.\ Lett.\  {\bf B377}, 28 (1996)
[arXiv:hep-th/9602077]; 
D.~E.~Diaconescu,
Nucl.\ Phys.\ {\bf B503}, 220 (1997)
[arXiv:hep-th/9608163].

\bibitem{HT}
A.~Hanany and D.~Tong,
JHEP {\bf 0307}, 037 (2003)
[arXiv:hep-th/0306150];
[arXiv:hep-th/0403158].

\bibitem{ADHM} 
M.~F.~Atiyah, N.~J.~Hitchin, V.~G.~Drinfeld and Yu.~I.~Manin,
Phys.\ Lett.\ {\bf A65}, 185 (1978).

\bibitem{Nahm}  
W.~Nahm,
Phys.\ Lett.\ {\bf B90}, 413 (1980).

%
%
\bibitem{GTT2} J.~P.~Gauntlett, D.~Tong and P.~K.~Townsend,  
               Phys.\ Rev.\ {\bf D64}, 025010 (2001)  
               [arXiv:hep-th/0012178]. 
%
\bibitem{To}  D.~Tong, 
               Phys.\ Rev.\ {\bf D66}, 025013 (2002)  
               [arXiv:hep-th/0202012].
%
\bibitem{To2}  D.~Tong, 
JHEP {\bf 0304}, 031 (2003) 
               [arXiv:hep-th/0303151].  

\bibitem{gravity2}
M.~Eto, S.~Fujita, M.~Naganuma and N.~Sakai, 
Phys. Rev. {\bf D69}, 025007 (2004) 
[arXiv:hep-th/0306198]. 
%
\bibitem{Lee} K.~S.~M.~Lee, 
               Phys.\ Rev.\ {\bf D67}, 045009 (2003) 
               [arXiv:hep-th/0211058]. 
%

\bibitem{HNOO}
K.~Higashijima, M.~Nitta, K.~Ohta and N.~Ohta,
Prog.\ Theor.\ Phys.\ {\bf 98}, 1165 (1997) 
[arXiv:hep-th/9706219].

\bibitem{Ni}
M.~Nitta,
Int.\ J.\ Mod.\ Phys.\ {\bf A14}, 2397 (1999)
[arXiv:hep-th/9805038].

%
\bibitem{Zu}  B.~Zumino,
               Phys.\ Lett.\ {\bf B87}, 203 (1979); 
   L.~Alvarez-Gaum\'{e} and D.~Z.~Freedman, 
   Commun.\ Math.\ Phys.\ {\bf 80},  443 (1981). 
%
\bibitem{AF2} 
L.~Alvarez-Gaum\'{e} and D.~Z.~Freedman, 
             Commun.\ Math.\ Phys.\ {\bf 91}, 87 (1983).
%
\bibitem{APS} P.~C.~Argyres, M.~R.~Plesser and N.~Seiberg, 
              Nucl.\ Phys.\ {\bf B471}, 159 (1996)
             [arXiv:hep-th/9603042].
%
\bibitem{AP} I.~Antoniadis and B.~Pioline, 
             Int.\ J.\ Mod.\ Phys.\ {\bf A12}, 4907 (1997) 
             [arXiv:hep-th/9607058].

%
\bibitem{LR} U.~Lindstr\"{o}m and M.~Ro\v{c}ek, 
               Nucl.\ Phys.\ {\bf B222}, 285 (1983).
%
\bibitem{HKLR}
     N.~J.~Hitchin, A.~Karlhede, U.~Lindstr\"{o}m and M.~Ro\v{c}ek,
     Commun.\ Math.\ Phys.\ {\bf 108}, 535 (1987). 

%
\bibitem{ANNS} M.~Arai, M.~Naganuma, M.~Nitta, and N.~Sakai, 
   Nucl.\ Phys.\ {\bf B652},  35 (2003) [arXiv:hep-th/0211103]; 
``BPS Wall in N=2 SUSY Nonlinear Sigma Model with Eguchi-Hanson Manifold''
in Garden of Quanta - In honor of Hiroshi Ezawa, 
Eds. by J.~Arafune et al. 
(World Scientific Publishing Co. Pte. Ltd. Singapore, 2003) 
pp 299-325, [arXiv:hep-th/0302028].
%
\bibitem{AIN}
M. Arai, E.~Ivanov and J.~Niederle, 
              Nucl.\ Phys.\ {\bf B680},  23 (2004) 
              [arXiv:hep-th/0312037]. 
%
\bibitem{KS} K.~Kakimoto and N.~Sakai, 
          Phys.\ Rev.\ {\bf D68}, 065005 (2003)  
          [arXiv:hep-th/0306077]. 
%
\bibitem{GTT1}J.~P.~Gauntlett, D.~Tong, and P.K.~Townsend, 
               Phys.\ Rev.\ {\bf D63}, 085001 (2001)  
               [arXiv:hep-th/0007124].           
%
\bibitem{NNS1}M.~Naganuma, M.~Nitta, and N.~Sakai,
               Grav.\ Cosmol.\ {\bf 8}, 129 (2002) 
               [arXiv:hep-th/0108133].
%
\bibitem{PT} R.~Portugues and P.~K.~Townsend, 
               JHEP\ {\bf 0204}, 039 (2002) [arXiv:hep-th/0203181]. 
%
\bibitem{GPTT}
              J.~P.~Gauntlett, R.~Portugues, D.~Tong, and P.K.~Townsend, 
               Phys.\ Rev.\ {\bf D63}, 085002 (2001)  
               [arXiv:hep-th/0008221].
%
\bibitem{INOS2}
Y.~Isozumi, M.~Nitta, K.~Ohashi and N.~Sakai,
[arXiv:hep-th/0405129].
%

\bibitem{T*CPN}T.~L.~Curtright and D.~Z.~Freedman,
               Phys.\ Lett.\ {\bf B90}, 71 (1980); 
               L.~Alvarez-Gaum\'{e} and D.~Z.~Freedman, 
               Phys.\ Lett.\ {\bf B94}, 171 (1980);
              M.~Ro\v{c}ek and P.~K.~Townsend, 
               Phys.\ Lett.\ {\bf B96}, 72 (1980).
%
\bibitem{Ca}  E.~Calabi, Ann. Scient. Ec. Norm. Sup. {\bf 12}, 269 (1979).
%
\bibitem{EH}  T.~Eguchi and A.~J.~Hanson,  
               Phys.\ Lett.\ {\bf B74}, 24 (1978);
               Ann.\ Phys.\ {\bf 120},  82 (1979). 
%
\bibitem{Ma}
N.~S.~Manton, 
Phys.\ Lett.\ {\bf B110}, 54 (1982). 
%
\bibitem{HN} K.~Higashijima and M.~Nitta, 
             Prog.\ Theor.\ Phys.\ {\bf 103}, 635 (2000)  
             [arXiv:hep-th/9911139].


\bibitem{Eto:2004ii}
M.~Eto, M.~Nitta and N.~Sakai, 
Nucl.Phys.{\bf B701}, 247 (2004)
[arXiv:hep-th/0405161].

\bibitem{Sakai:2002wc}
N.~Sakai and R.~Sugisaka,
Phys.\ Rev.\ D {\bf 66}, 045010 (2002)
[arXiv:hep-th/0203142].


%
\bibitem{CNV}
T.~E.~Clark, M.~Nitta and T.~ter Veldhuis,
Phys.\ Rev.\  {\bf D67}, 085026 (2003)
[arXiv:hep-th/0208184];
Phys.\ Rev.\  {\bf D69}, 047701 (2004)
[arXiv:hep-th/0209142]; 
Phys.\ Rev.\  {\bf D70}, 105005 (2004)
[arXiv:hep-th/0401163].


%
\bibitem{gravity1}
M.~Arai, S.~Fujita, M.~Naganuma and N.~Sakai, 
Phys. Lett. {\bf B556}, 192 (2003) [arXiv:hep-th/0212175]; 
in the proceedings of International Seminar on 
Supersymmetries and Quantum Symmetries SQS 03, 
Dubna, Russia, 24-29 Jul 2003,  
[arXiv:hep-th/0311210]. 

\bibitem{KugoOhashi} T.~Kugo and K.~Ohashi,
                Prog.\ Theor.\ Phys.\ {\bf 105}, 323 (2001) 
                [arXiv:hep-ph/0010288];
                 T.~Fujita and K.~Ohashi, 
                Prog.\ Theor.\ Phys.\ {\bf 106},  221 (2001)
                [arXiv:hep-th/0104130];
T.~Fujita, T.~Kugo and K.~Ohashi,
                Prog.\ Theor.\ Phys.\ {\bf 106}, 671 (2001) 
                [arXiv:hep-th/0106051]. 

\end{thebibliography}
\end{document}